\documentclass[twocolumn]{aastex631}
\usepackage[version=4]{mhchem}
\usepackage{mathrsfs}
\usepackage{xcolor}
\usepackage{rotating}

\begin{document}

\title{The JDISC Survey: Linking the physics and chemistry of inner \& outer protoplanetary disk zones} 

\author[0000-0003-2631-5265]{Nicole Arulanantham}
\affiliation{Astrophysics \& Space Institute, Schmidt Sciences, New York, NY 10011, USA}

\author[0000-0003-3682-6632]{Colette Salyk}
\affiliation{Vassar College, 124 Raymond Avenue, Poughkeepsie, NY 12604, USA}

\author[0000-0001-7552-1562]{Klaus Pontoppidan} 
\affiliation{Jet Propulsion Laboratory, California Institute of Technology, 4800 Oak Grove Drive, Pasadena, CA 91109, USA}

\author[0000-0003-4335-0900]{Andrea Banzatti}
\affiliation{Department of Physics, Texas State University, 749 N Comanche Street, San Marcos, TX 78666, USA}

\author[0000-0002-0661-7517]{Ke Zhang}
\affil{Department of Astronomy, University of Wisconsin-Madison, Madison, WI 53706, USA}

\author[0000-0001-8798-1347]{Karin \"{O}berg}
\affil{Center for Astrophysics | Harvard \& Smithsonian, 60 Garden St., Cambridge, MA 02138, USA}

\author[0000-0002-7607-719X]{Feng Long}
\affiliation{Lunar and Planetary Laboratory, University of Arizona, Tucson, AZ 85721, USA}
\affiliation{NASA Hubble Fellowship Program Sagan Fellow}

\author[0000-0002-6695-3977]{John Carr}
\affiliation{Department of Astronomy, University of Maryland, College Park, MD 20742, USA}

\author[0000-0002-5758-150X]{Joan Najita}
\affiliation{NSF’s NOIRLab, 950 N. Cherry Avenue, Tucson, AZ 85719, USA}

\author[0000-0001-7962-1683]{Ilaria Pascucci}
\affil{Department of Planetary Sciences, University of Arizona, 1629 East University Boulevard, Tucson, AZ 85721, USA}

\author[0000-0002-5296-6232]{Mar\'{i}a Jos\'{e} Colmenares}
\affiliation{Department of Astronomy, University of Michigan, Ann Arbor, MI 48109, USA}

\author[0000-0001-8184-5547]{Chengyan Xie}
\affiliation{Lunar and Planetary Laboratory, University of Arizona, Tucson, AZ 85721, USA}

\author[0000-0001-6947-6072]{Jane Huang}
\affiliation{Department of Astronomy, Columbia University, 538 W. 120th Street, Pupin Hall, New York, NY 10027, USA}

\author[0000-0003-1665-5709]{Joel Green}
\affiliation{Space Telescope Science Institute, 3700 San Martin Drive, Baltimore, MD 21218, USA}

\author[0000-0003-2253-2270]{Sean M. Andrews}
\affiliation{Center for Astrophysics \textbar~Harvard \& Smithsonian, 60 Garden St, Cambridge, MA 02138, USA}

\author[0000-0003-0787-1610]{Geoffrey A. Blake}
\affiliation{Division of Geological and Planetary Sciences, California Institute of Technology, MC 150-21, Pasadena, CA 91125, USA}

\author[0000-0003-4179-6394]{Edwin A. Bergin}
\affiliation{Department of Astronomy, University of Michigan, 1085 S. University, Ann Arbor, MI 48109, USA}

\author[0000-0001-8764-1780]{Paola Pinilla}
\affiliation{Mullard Space Science Laboratory, University College London, Holmbury St Mary, Dorking, Surrey RH5 6NT, UK}

\author[0000-0002-4147-3846]{Miguel Vioque}
\affiliation{European Southern Observatory, Karl-Schwarzschild-Str. 2, 85748 Garching bei München, Germany}

\author[0000-0003-2985-1514]{Emma Dahl}
\affiliation{Division of Geological and Planetary Sciences, California Institute of Technology, MC 150-21, Pasadena, CA 91125, USA}

\author[0009-0002-2380-6683]{Eshan Raul}
\affil{Department of Astronomy, University of Wisconsin-Madison, Madison, WI 53706, USA}

\author[0000-0002-3291-6887]{Sebastiaan Krijt}
\affil{Department of Physics and Astronomy, University of Exeter, Exeter, EX4 4QL, UK}

\author{the JDISCS Collaboration}

\begin{abstract}

Mid-infrared spectroscopy of protoplanetary disks provides a chemical inventory of gas within a few au, where planets are readily detected around older stars. With the \emph{JWST} Disk Infrared Spectral Chemistry Survey (JDISCS), we explore demographic trends among 31 disks observed with MIRI (MRS) and with previous ALMA millimeter continuum imaging at high angular resolution (5-10 au). With these S/N $\sim$200-450 spectra, we report emission from H$_2$O, OH, CO, C$_2$H$_2$, HCN, CO$_2$, [Ne II], [Ne III], and [Ar II]. Emission from H$_2$O, OH and CO is nearly ubiquitous for low-mass stars, and detection rates of all molecules are higher than for similar disks observed with Spitzer-IRS. Slab model fits to the molecular emission lines demonstrate that emission from C$_2$H$_2$, HCN, and possibly CO$_2$ is optically thin; thus since column densities and emitting radii are degenerate, observations are actually sensitive to the total molecular mass.  C$_2$H$_2$ and HCN emission also typically originate in a hotter region ($920^{+70}_{-130}$, $820^{+70}_{-130}$ K, respectively) than CO$_2$ ($600^{+200}_{-160}$ K). The HCN to cold H$_2$O luminosity ratios are generally smaller in smooth disks, consistent with more efficient water delivery via icy pebbles in the absence of large dust substructures. The molecular emission line luminosities are also correlated with mass accretion rates and infrared spectral indices, similar to trends reported from \emph{Spitzer-IRS} surveys. This work demonstrates the power of combining multi-wavelength observations to explore inner disk chemistry as a function of outer disk and stellar properties, which will continue to grow as the sample of observed Class II systems expands in the coming \emph{JWST} observation cycles.     

\end{abstract}


\section{Introduction} \label{sec:intro}

In the classical solar nebula model \citep[e.g.,][]{grossman72}, a well-mixed solar-composition gas condenses into solid planetary building blocks, with the local temperature determining which materials condense into the solid phase versus remaining in the gas phase.  This results in planets close to the Sun that are more refractory rich, and planets (or planetary cores) that are far from the Sun being more volatile rich.  In addition, the condensation of water ice at the Solar System’s so-called ``snow line’’ may have facilitated the formation of more massive planetary cores, seeding the runaway gas accretion needed to form gas giants \citep[e.g.,][]{hayashi81,pollack96, drazkowska17}.  This classical story is likely incomplete, however, for many reasons.  For example, planet-forming regions may inherit minimally processed materials from their parent clouds \citep[e.g.][]{Visser11}, radial temperature zones are blurred by mixing \citep[e.g.][]{Brownlee06}, and disk heating may be highly stochastic \citep[e.g.][]{zhu09,vorobyov10}.  The classical story may also require modifications for non-sun-like stars.

Direct measurements of disk chemistry in planet-forming regions are the means to test theories regarding the origin of planetary compositions.  This field of study blossomed with the launch of the Spitzer-InfraRed Spectrograph \citep{houck04}, from which mid-IR spectra probed disk atmospheres in the terrestrial planet-forming zone (see e.g., \citealt{henning13, pontoppidan14} and references therein). Spitzer-IRS showed that emission from simple molecules, including H$_2$O, OH, HCN, C$_2$H$_2$ and CO$_2$ was nearly ubiquitous around low-mass stars \citep{pontoppidan10b,carr11}, demonstrating that the study of inner disk chemistry was possible.  It was discovered, in contrast, that higher mass Herbig Ae/Be stars only rarely show molecular emission lines, possibly from colder water vapor \citep{pontoppidan10b,fedele12}.  Disks with large inner dust cavities also had minimal molecular emission lines besides CO \citep{salyk09, pontoppidan10b, banzatti2017}, though weaker emission lines, including from photodissociation-produced OH, could be detected with sufficiently high dynamic range \citep{najita10}.  For full disks around low-mass stars, chemical differences were more difficult to extract, likely due to Spitzer's relatively low spectral resolution \citep{salyk11}.  Nevertheless, intriguing trends in HCN/H$_2$O line strengths were discovered \citep{najita13,najita18}, which later, with larger samples, expanded into trends between water luminosity and the outer disk radius and were potentially linked to inner disk water enrichment by pebble drift \citep{banzatti2020}.

The James Webb Space Telescope (JWST) Mid InfraRed Instrument Medium Resolution Spectrometer (MIRI-MRS; \citealp{rieke15,wells15})  provides increased sensitivity and spectral resolution compared to Spitzer-IRS, and has already demonstrated a greatly improved ability to tease out chemical details and differences.  In particular, MRS has revealed a variety of line strength ratios between H$_2$O and C-bearing species, including CO$_2$, HCN and C$_2$H$_2$ \citep[e.g.][]{tabone23,banzatti23b,grant23,xie23,gasman23, gasman25, long25}. It is also possible to ``map'' molecular emitting regions via modeling of level populations \citep[e.g.][]{temmink24b,romero-mirza24b} and from the observed line broadening \citep{banzatti24, grant2024}. Evidence for pebble migration followed by water sublimation is now observed in drift-dominated disks \citep{banzatti23b}, and trace molecules and isotopologues are detected for the first time \citep{grant23,perotti23,salyk25}.

These observations have resulted in the emergence of several new theoretical frameworks.  One, which we term ``peeling back the onion,’’ suggests that all disks have an onion-like chemical structure with temperature as the determining factor, but the presence of gaps and rings, known to be nearly ubiquitous from ALMA imaging \citep{andrews18,long18}, preferentially reveals different parts of the ``onion".  For example, \citet{grant23} and \citet{vlasblom24} suggest that the stronger CO$_2$ emission observed in GW Lup relative to water may be caused by an inner disk cavity that is preferentially revealing gas between the H$_2$O and CO$_2$ snow lines.  A second emerging theoretical idea is that radial pebble drift, or its inhibition, leads to changes in inner disk chemistry. In the inner disk, evidence for pebble drift is arguably strongest in water vapor \citep{kalyaan21,banzatti23b,banzatti24,romero-mirza24b, gasman25}, but pebble drift is also expected to change the inner disk carbon-to-oxygen elemental ratios and, thus, the ratios of C-bearing to O-bearing molecular abundances \citep[e.g.][]{najita11,booth17,booth19,mah23}.  A third framework involves the influence of stellar mass on inner disk chemistry \citep[e.g.][]{pascucci09, pascucci13, tabone23,xie23,colmenares24}, with inner disks around lower-mass stars typically, though not always, displaying higher abundances of C-bearing species at ages as old as 30 Myr \citep{kanwar24, long25}.  The cause of these differences is still being debated, with leading candidate theories including variations in radial drift rates \citep{pinilla12,mah23}, radiation field differences \citep{walsh15}, and carbon-grain destruction inside of the so-called soot line \citep{kress10, tabone23,colmenares24}. 
Note that some of these theoretical frameworks predict links between disk dust structure and inner disk chemistry that are potentially observable by combining mid-infrared spectroscopy with dust imaging, especially from the Atacama Large Millimeter/submillimeter Array (ALMA).

To date, many MRS studies of disk chemistry have focused on individual disks, but such studies may not be conducive to revealing global trends related to planet formation chemistry. To more comprehensively characterize the large variety of inner disk molecular spectra observed to date, and to relate this variety to other disk or stellar properties within unified theoretical frameworks, we must begin to analyze larger samples of disk spectra, ideally samples with ancillary disk imaging (see e.g., \citealt{banzatti23b, banzatti24, henning24, gasman25}).  The \emph{JWST} Disk Infrared Spectral Chemistry Survey \citep[JDISCS;][and this work]{pontoppidan24} was designed to accelerate this process by building a large sample of MRS observations of protoplanetary disks with high-quality ancillary ALMA imaging.  This paper presents a first JDISCS program analysis of all of the disks from our Cycle 1 programs, representing primarily K and M stars of a few Myr age.  Section \ref{sec:obs} provides an overview of the first sample of JDISCS data from JWST Cycle 1. In Section \ref{sec:results}, we describe the basic observables of the sample spectra, including detection statistics of molecular and atomic lines, and retrievals of physical parameters using slab models. Finally, in Sections \ref{sec:analysis} and \ref{sec:discussion}, we begin to make connections between MRS spectra and dust substructures observed in ALMA data.  We conclude with discussions of the next steps needed in this field, notably an expansion towards wider ranges and more complete sampling of the parameter space of star and disk properties.

\section{The JWST Disk Infrared Spectral Chemistry Survey (JDISCS)} \label{sec:obs}

\subsection{Description of the Sample}

JDISCS comprises MIRI-MRS data from several GO programs, collectively designed to provide a legacy dataset from which to explore outstanding questions about inner disk chemistry and its link to planet formation. Targets whose outer disks have been well-studied at sub-mm wavelengths are critical to this effort, particularly those that were systematically observed as part of large programs with ALMA. One sample of fundamental importance comes from the ALMA ``Disk Substructures at High Angular Resolution Project" (DSHARP; \citealt{andrews18}), which was optimized to search for regions of enhanced sub-mm continuum emission in protoplanetary disks at angular resolutions of $\sim 0.035\arcsec$ (5 au at $d = 150$ pc). Such substructures, including rings, gaps, spirals, and azimuthal asymmetries, reveal reservoirs of dust grains that may be trapped in local pressure maxima where they can eventually grow into larger planetesimals (although \citealt{dullemond18} show that not all substructures are necessarily dust traps). Meanwhile, the chemical conditions of the outer disk have been mapped at the highest sensitivity and spatial resolution to date within a sample of five sources in the ``Molecules with ALMA at Planet-Forming Scales" (MAPS; \citealt{Oberg2021}) survey. The observations provided by these two ALMA programs enable synergic studies between outer disk dust substructures and chemical evolution. 

\begin{figure*}
\epsscale{1.2}
\plotone{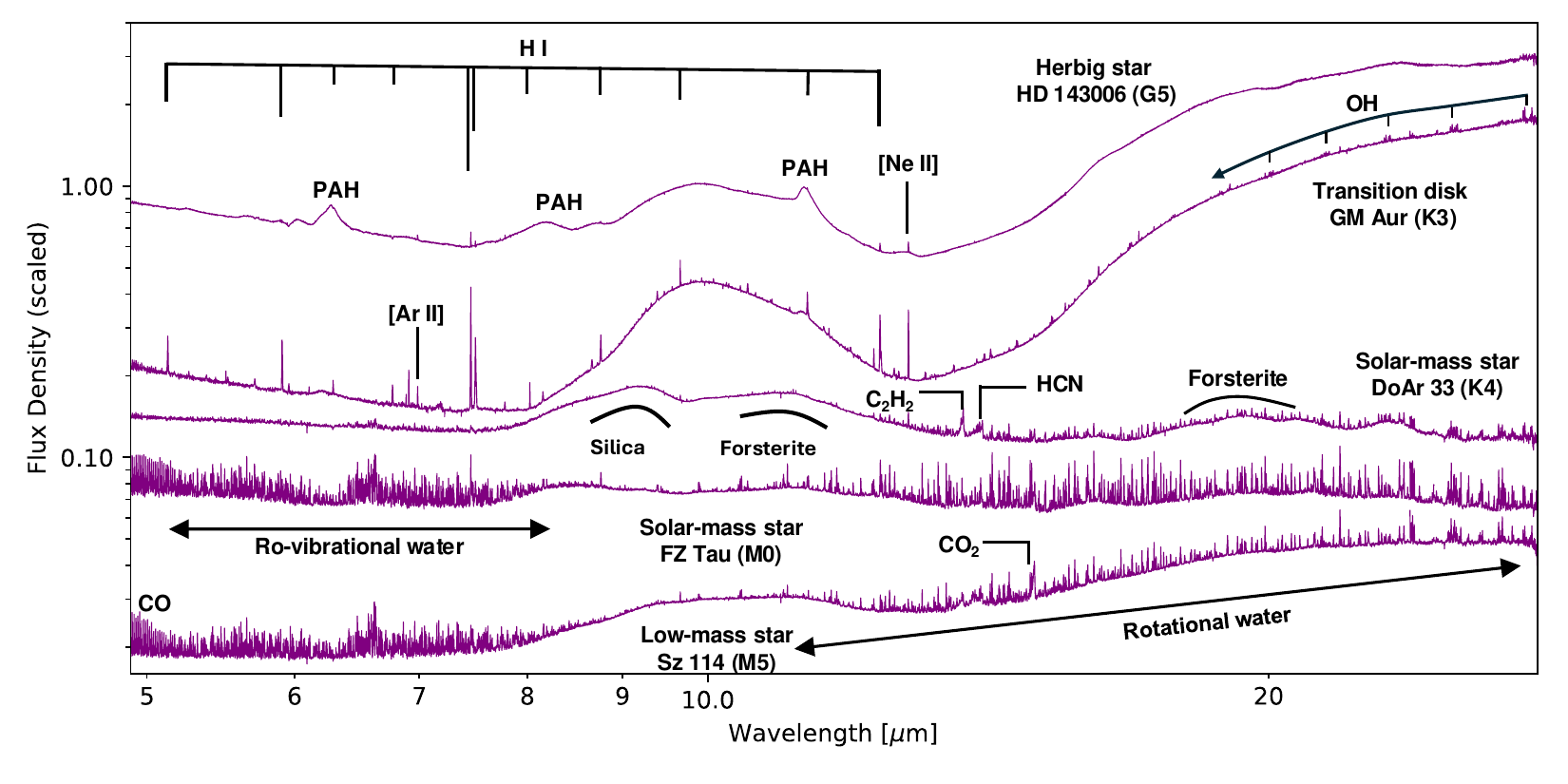}
\caption{MIRI spectra of select targets from the JDISCS sample, representing broad categories of sources, namely: intermediate mass, transition disk (with a large dust cavity), solar-mass with weak water emission, solar-mass with strong water emission, and a low-mass star. Emission lines from prominent atomic and molecular gas, and solid state features, are highlighted with labels.}
\label{fig:jdiscs_bytype}
\end{figure*}

JDISCS adds observations of the chemistry of the terrestrial planet-forming regions to such ALMA data sets. Specifically, the Cycle 1 GO program PID 1584 (PI: C. Salyk; co-PI: K. Pontoppidan) acquired \emph{JWST} MIRI MRS spectra of 17/20 disks that were included in DSHARP \citep{andrews18}. Another three disks are included in the MAPS survey as part of PID 2025 (PI: K. \"{O}berg; \citealt{romero-mirza24a}). \footnote{Three remaining disks from DSHARP (GW Lup, IM Lup, and Wa Oph 6) are available as archival data from the GTO program PID 1282 (PI: T. Henning). These sources are not analyzed in this work.} Additional disks are included from PID 1549 (3 disks, PI: K. Pontoppidan; \citealt{pontoppidan24}) and PID 1640 (8 disks, PI: A. Banzatti; \cite{banzatti23b,romero-mirza24b}). These additional disks have moderate-resolution ALMA data ($0.12"$, or 16 au at $d = 140$ pc) available from \cite{long19,Hendler20} and new unpublished data (Long et al. 2025, in prep). In total, the JDISCS Cycle 1 sample analyzed in this work includes 30 systems (31 disks, since two spectra can be extracted from the binary system AS 205 N + S). Figure \ref{fig:jdiscs_bytype} presents an overview of the MIRI spectra for different categories of sources observed by JDISCS: the K- and M-type T Tauri stars, the intermediate mass systems, and ``transition" disks with large mm cavities.
While this first paper includes only Cycle 1 targets, the JDISC Survey now includes targets from Cycles 2 and 3 for a total sample of $\sim$100 disks, including PIDs 3034 (PI: K. Zhang), 3153 (PI: F. Long), and 3228 (PI: I. Cleeves); these additional targets will be analyzed in future works. The JWST data presented in this article were obtained from the Mikulski Archive for Space Telescopes (MAST) at the Space Telescope Science Institute. The specific observations analyzed can be accessed via \dataset[doi: 10.17909/hx6h-qw97]{https://doi.org/10.17909/hx6h-qw97}.

\begin{figure}
\epsscale{1.2}
\plotone{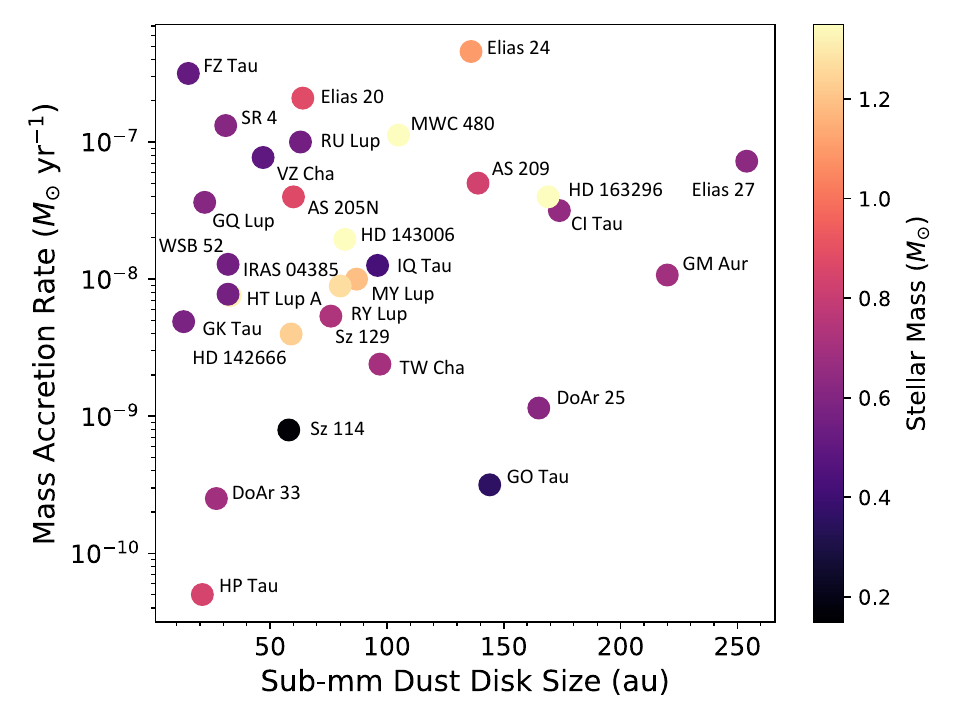}
\caption{Summary of stellar and disk properties for 30/31 JDISCS targets included in this work, including mass accretion rates (y-axis), sub-mm dust disk radii enclosing 90-95\% of the flux at $\sim$1.3 mm (x-axis), and stellar masses (marker colors). The sample spans $\sim$4 orders of magnitude in mass accretion rate, $\sim$2 orders of magnitude in dust disk sizes, and a narrow range in stellar masses (27 K- and M-type T Tauri stars, one G-type star, and three Herbig Ae/Be systems). AS 205 S is excluded from the figure, as it does not yet have a measured accretion rate.}
\label{fig:jdiscs_sample}
\end{figure}

Table \ref{tab:targprops} and Table \ref{tab:diskprops} list the stellar and outer disk properties for all 31 sources included in this study. Representative spectra for broad categories of sources are highlighted in Figure \ref{fig:jdiscs_bytype}. Figure \ref{fig:jdiscs_sample} shows a graphical representation of our coverage of stellar mass, disk radius, and accretion rate.  Distances for all targets were obtained by inverting the parallaxes reported in Gaia Data Release 3 \citep{prusti2016, brown2021}. Where possible, stellar masses, spectral types, luminosities, mass accretion rates, and sub-mm dust disk masses were taken from the table constructed by \citet{manara2023}, which provides such properties derived from a self-consistent analysis of data from large surveys of star-forming regions (see e.g., \citealt{manara14, manara16, manara17, manara21, herczeg14, pascucci16, ansdell16, alcala17, long18, long19, testi22}). This includes stellar masses ($0.16 \, M_{\odot} < M_{\ast} < 3.58 \, M_{\odot}$) and spectral types for all 31 sources (27 K- and M-type T Tauri stars, one G-type star, and three Herbig Ae/Be systems), mass accretion rates for 30/31 $\left(10^{-9.1} \,  M_{\odot} \, \rm{yr}^{-1} < \dot{M}_{\rm{acc}} < 10^{-6.225} \, M_{\odot} \, \rm{yr}^{-1} \right)$, and sub-mm dust disk masses for 23/31 $\left(2.43 \, M_{\oplus} < M_{\rm{dust}} < 210.52 \, M_{\oplus} \right)$. Stellar properties for the remaining targets were taken from other near- and mid-infrared surveys that are referenced in Table \ref{tab:targprops} (see e.g., \citealt{eisner05, salyk13, fairlamb15, mcclure19, donati24}). With the sample spanning only a small range in stellar and dust disk masses, the survey is effectively optimized to characterize how the broad diversity of mid-infrared spectra depends on outer disk substructures and mass accretion rates (which span $\sim$4 orders of magnitude; see Figure \ref{fig:jdiscs_sample}).

We also report the disk inclinations and position angles derived from the sub-mm dust distributions in Table \ref{tab:diskprops} (\citealt{banzatti2017, wu2017, huang2018, huang2020a, kurtovic2018, liu2019, long2019, long20, francis2020, andrews21}; Long et al. 2025, in prep). The sample includes inclination angles from $18^{\circ}-67.4^{\circ}$, in addition to one highly inclined target (MY Lup; $i_\mathrm{disk}=73.2^{\circ}$). Dust rings are resolved in most disks, while four show spiral arms (AS 205 N, HT Lup A, Elias 27, HD 143006; \citealt{kurtovic2018, huang2018c, andrews21}) and two have smooth distributions down to $\sim$0.12$\arcsec$, or 16 au (GK Tau, HP Tau; \citealt{long2019}). We note that inner disk substructures at radii that would overlap with the mid-infrared emitting regions are not spatially resolved with ALMA; instead, our dataset can be used to assess the impact of outer disk evolution on inner disk chemistry (see e.g., \citealt{najita13}).  

\subsection{MIRI-MRS Observations and Data Reduction}

The JDISCS data reduction follows that initially described in \cite{pontoppidan24}. Minor updates since then include reprocessing using the JWST Calibration Pipeline version 1.15.0 and reference file context 1253 (corresponding to the internal JDISCS version 8.0). JDISCS sources observed before 1 January 2024 use the observation of asteroid 526 Jena obtained on 21 September 2023 as part of PID 1549 for MRS channels 2-4, and the observation of early-type star HD 163466 obtained as part of calibration program PID 4499 on 5 July 2023 for channel 1. Extraction aperture radii scale with wavelength as 1.4, 1.3, 1.2, and 1.1 times $1.22\lambda/D$ for MRS channels 1 through 4, respectively. Apertures are kept the same for the source and calibrator, such that any PSF complexities should cancel out during division by the calibrator, improving spectro-photometric precision. Independent observations of HD 163466 suggest that the absolute spectrophotometric precision is a few percent in MRS channels 1-3 and up to 10\% at the long wavelength end of channel 4. Future versions of the JDISCS pipeline may achieve higher precision by a combination of multiple calibrators and improved models of the asteroid spectra. The signal-to-noise ratio (S/N) achieved near 17~$\mu$m is 200--450 in all spectra except for HD~163296, which is affected by saturation and fringe residuals.

\subsection{Continuum Subtraction}
For continuum estimation, we use the automatic iterative algorithm described in \cite{pontoppidan24}.\footnote{The continuum subtraction routine is available at \url{https://github.com/pontoppi/ctool}.} This method works well in estimating the continuum across all four MIRI channels in conditions of narrow gas emission on top of broad dust emission; however, the continuum level can be uncertain in regions where dense clustering of molecular gas emission lines produce a pseudo-continuum \citep{carr11, pascucci13, tabone23, kanwar24, arabhavi24, kaeufer24, kaeufer2024b}. For this reason, we exclude the organic region at 13.4--14.1~$\mu$m from the continuum fit to avoid subtracting part of the emission from HCN and \ce{C2H2}. In case of strong emission from \ce{CO2} (as in MY~Lup, HT~Lup, Sz~114), we also exclude the 14.9--15~$\mu$m region for the same reason. In case of gas absorption in a wind or a stellar photosphere, an additional adjustment of the continuum is necessary using line-free regions, as described in \citet{banzatti24}. We discuss the impact of this continuum subtraction method on our analysis and interpretation of the molecular gas emission lines in Section 5.2. 

\begin{deluxetable*}{lccccccc}
\tablecaption{JDISCS Stellar Properties \label{tab:targprops}}
\tablehead{
\colhead{Target} & \colhead{Distance} & \colhead{SpT} & \colhead{$L_{\ast}$} & \colhead{$M_{\ast}$} & \colhead{$\log \dot{M}_{\rm{acc}}$} & \colhead{Radial Velocity} \\
\colhead{} & \colhead{(pc)} & & \colhead{($L_{\odot}$)} & \colhead{($M_{\odot}$)} & \colhead{($M_{\odot}$ yr$^{-1}$)} & \colhead{(km s$^{-1}$)}}
\startdata
AS 205 N & 142 & K5 & 1.3 & 0.87 & -7.4 & -5.4 \\
AS 205 S & 142 & K7+M0 & 0.7 & 1.28 & \nodata & -5.4 \\
AS 209 & 121 & K5 & 1.4 & 0.83 & -7.3 & -9.1 \\
CI Tau & 160 & K7 & 0.8 & 0.65 & -7.5 & 19.9 \\
DoAr 25 & 138 & K5 & 0.9 & 0.62 & -8.9 & -12.6 \\
DoAr 33 & 142 & K4 & 1.5 & 0.69 & -9.6 & -6.6 \\
Elias 20 & 138 & M0 & 2.6 & 0.88 & -6.7 & -3.3 \\
Elias 24 & 139 & K5 & 6.8 & 1.10 & -6.3 & -7.3 \\
Elias 27 & 110 & M0 & 1.5 & 0.63 & -7.1 & -7.7 \\
FZ Tau & 129 & M0 & 1.0 & 0.51 & -6.5 & 15.8 \\
GK Tau & 129 & K7 & 0.9 & 0.58 & -8.3 & 17.0 \\
GM Aur & 158 & K7 & 1.0 & 0.69 & -8.0 & 16.5 \\
GO Tau & 140 & K5 & 0.2 & 0.36 & -9.5 & 17.1 \\
GQ Lup & 154 & K7 & 1.4 & 0.61 & -7.4 & -2.1 \\
HD 142666 & 146 & A8 & 9.1 & 1.23 & -8.4 & -13.1 \\
HD 143006 & 167 & G7 & 3.9 & 1.48 & -7.7 & -0.2 \\
HD 163296 & 101 & A1 & 17.0 & 2.04 & -7.4 & -4.0 \\
HP Tau & 171 & K0 & 1.1 & 0.84 & -10.3 & 17.7 \\
HT Lup A+B & 153 & K2 & 5.1 & 1.32 & -8.1 & -1.4 \\
IQ Tau & 132 & M0.5 & 1.0 & 0.42 & -7.9 & 15.3 \\
IRAS 04385+2550 & 160 &  M0.5 & 0.5 & 0.56 & -8.1 & 17.0 \\
MWC 480 & 156 & A2 & 22.0 & 3.58 & -7.0 & 27.7 \\
MY Lup & 157 & K0 & 0.9 & 1.20 & -8.0 & 4.4 \\
RU Lup & 158 & K7 & 1.5 & 0.55 & -7.0 & -0.8 \\
RY Lup & 153 & K2 & 1.9 & 1.27 & -8.1 & -0.4 \\
SR 4 & 135 & K7 & 1.2 & 0.61 & -6.9 & -4.5 \\
Sz 114 & 157 & M5 & 0.2 & 0.16 & -9.1 & 4.0 \\
Sz 129 & 160 & K7 & 0.4 & 0.73 & -8.3 & 3.2 \\
TW Cha & 183 & K7 & 0.4 & 0.70 & -8.6 & 17.8 \\
VZ Cha & 191 & K7 & 0.5 & 0.50 & -7.1 & 16.3 \\
WSB 52 & 135 & M1 & 1.7 & 0.55 & -7.9 & -5.5 \\
\enddata
\tablecomments{All distances are calculated from Gaia DR3 parallaxes \citealt{GaiaDR3}. Stellar properties are taken from the compiled table in \citealt{manara2023}, and heliocentric radial velocities are from \citealt{banzatti19}, \citealt{Fang18} and references therein (see below). Significant digits across the sample are rounded to match the targets with the least precise published measurements.}
\tablerefs{\citealt{eisner05} (AS 205N, AS 205S); \citealt{salyk13} (AS 205S, AS 209, CI Tau); \citealt{donati24} (CI Tau); \citealt{gaiaDR2} (DoAr 25); \citealt{cieza2010} (DoAr 33); \citealt{GaiaESO23} (DoAr 33); \citealt{testi22} (Elias 20, Elias 24, Elias 27, HD 143006, IRAS 04385); \citealt{sullivan2019} (Elias 20, WSB 52); \citealt{APOGEE20} (Elias 24, Elias 27, IRAS 04385, SR 4); \citealt{mcclure19} (FZ Tau); \citealt{manara14} (GM Aur); \citealt{alcala17} (GQ Lup, HT Lup A+B, RU Lup, RY Lup); \citealt{fairlamb15} (HD 142666); \citealt{miretroig22} (HD 142666); \citealt{gontcharov06} (HD 163296); \citealt{kounkel2019} (IQ Tau); \citealt{white04} (IRAS 04385); \citealt{najita09, mendigutia13} (MWC 480); \citealt{alcala19} (MY Lup); \citealt{frasca2017} (MY Lup, Sz 114, Sz 129); \citealt{hourihane23} (TW Cha); \citealt{banzatti2017} (FZ Tau, TW Cha, VZ Cha); \citealt{nguyen12} (VZ Cha)}
\end{deluxetable*}

\begin{deluxetable*}{cccccclc}
\tablecaption{Inner and Outer Dust Disk Properties \label{tab:diskprops}}
\tablehead{
\colhead{Target} & \colhead{$n_{13-26}$\tablenotemark{a}} & \colhead{$M_{\rm{dust}}$\tablenotemark{b}} & \colhead{$r_{\rm{dust}}$\tablenotemark{c}} & \colhead{$i_{\rm{disk}}$} & \colhead{PA\tablenotemark{d}} & \colhead{Sub-mm} & \colhead{Marker\tablenotemark{e}} \\
\colhead{} & \colhead{} & \colhead{($M_{\oplus}$)} & \colhead{(au)} & \colhead{($^{\circ}$)} & \colhead{($^{\circ}$)} & \colhead{Substructures} & \colhead{}}
\startdata
AS 205 N & -0.20 & 192.9 & 60 & $20$ & $114$ & spirals & $\mathcal{S}$ \\
AS 205 S & 0.50 & \nodata & 34 & $66$ & $1102$ & cavity, ring (34 au) & $\odot$ \\
AS 209 & -0.24 & \nodata & $139$ & $35$ & $86$ & rings (14-141 au) & $\odot$ \\
CI Tau & -0.38 & 103.4 & $174$ & $50$ & $11$ & rings (28-153 au) & $\odot$ \\
DoAr 25 & 0.23 & 138.8 & $165$ & $67$ & $111$ & rings (86-137 au) & $\odot$ \\
DoAr 33 & -1.08 & 20.3 & $27$ & $67$ & $111$ & ring (17 au) & $\odot$ \\
Elias 20 & -0.90 & 54.9 & $64$ & $49$ & $153$ & rings (29, 36 au) & $\odot$ \\
Elias 24 & -0.83 & 210.5 & $136$ & $29$ & $46$ & rings (77, 123 au) & $\odot$ \\
Elias 27 & -0.66 & 113.8 & $254$ & $56$ & $119$ & ring (86 au), spirals & $\mathcal{S} + \odot$ \\
FZ Tau & -1.07& 5.3 & $\sim$15 & $\sim$26 & $30$ & smooth & $\bullet$ \\
GK Tau & -0.30 & 2.4 & $13$ & $40$ & $120$ & smooth & $\bullet$ \\
GM Aur & 2.08 & 95.9 & 220 & 53 & 57 & cavity, rings (40, 84, 168 au) & $\odot$ \\
GO Tau & -0.03 & \nodata & $144$ & $54$ & $21$ & rings (73, 109 au) & $\odot$ \\
GQ Lup & -0.39 & 25.6 & 22 & $61$ & $346$ & gap (10 au) & $\bigcirc$ \\
HD 142666 & -0.54 & \nodata & $59$ & $62$ & $162$ & rings (6-58 au) & $\odot$ \\
HD 143006 & 1.33 & 49.7 & $82$ & $19$ & $169$ & cavity, rings (6-65 au), spirals & $\mathcal{S}+\odot$ \\
HD 163296 & -0.99 & \nodata & $169$ & $47$ & $133$ & rings (14-155 au) & $\odot$ \\
HP Tau & 0.14 & 27.3 & 21 & $18$ & $57$ & smooth & $\bullet$ \\
HT Lup A+B & -0.38 & 49.8 & 33, 5 & $48$ & $166$ & spirals (A); smooth (B) & $\mathcal{S}$ \\
IQ Tau & -0.63 & 35.5 & $96$ & $62$ & $42$ & rings (48-83 au) & $\odot$ \\
IRAS 04385+2550 & 0.53 & \nodata & $\sim 32$ & $\sim 60$ & $\sim 162$ & smooth & $\bullet$ \\
MWC 480 & \nodata & 184.7 & $105$ & $36$ & $148$ & ring (98 au) & $\odot$ \\
MY Lup & 0.19 & 50.4 & $87$ & $73$ & $59$ & rings (20, 40 au) & $\odot$ \\
RU Lup & -0.14 & 125.2 & $63$ & $19$ & $121$ & rings (17-50 au) & $\odot$ \\
RY Lup & 0.45 & 64.6 & 80 & 67 & 109 & cavity (69 au) & $\bigcirc$ \\
SR 4 & 0.56 & 38.5 & $31$ & $22$ & $18$ & ring (18 au) & $\odot$ \\
Sz 114 & -0.16 & 30.6 & $58$ & $21$ & $165$ & ring (45 au) & $\odot$ \\
Sz 129 & 0.47 & 58.5 & $76$ & $34$ & $151$ & cavity, rings (10-69 au) & $\odot$ \\
TW Cha & -0.03 & \nodata & $\sim 97$ & $\sim 27$ & $121$ & cavity ($\sim 30$~au) & $\bigcirc$ \\
VZ Cha & -1.22 & \nodata & $\sim 47$ & $\sim 19$ & $30$ & gap ($<0.1''$) & $\odot$ \\
WSB 52 & -0.43 & 36.6 & $32$ & $54$ & $138$ & ring (25 au) & $\odot$ \\
\enddata
\tablenotetext{a}{$n_{13-26}$ is the infrared spectral index measured in this work, using the continuum flux at 13 and 26~$\mu$m (see Section 3.1 for details).}
\tablenotetext{b}{$M_{\rm{dust}}$ are obtained from mm fluxes (see references below). Significant digits across the sample are rounded to match the targets with the least precise published measurements.}
\tablenotetext{c}{$r_{\rm{dust}}$ values represent the boundaries containing 90\% \citep{long18} to 95\% \citep{huang2018, long19} of the flux at $\sim$1.3 mm.}
\tablenotetext{d}{PA is the disk position angle.}
\tablenotetext{e}{Symbols indicate the markers used to represent each disk in Figures \ref{fig:detection_limits}, \ref{fig:allmols_T}, \ref{fig:Lemit_vs_T}, \ref{fig:Lemit_vs_Memit}, \ref{fig:T_vs_Memit}, \ref{fig:correlations}, and \ref{fig:organics_vs_rdust}.}
\tablerefs{\citealt{kurtovic2018} (A205N, AS 205S, HT Lup A+B); \citealt{huang2018} (AS 209, DoAr 25, DoAr 33, Elias 20, Elias 24, Elias 27, HD 142666, HD 163296, MY Lup, RU Lup, SR 4, Sz 114, Sz 129, WSB 52); \citealt{long19} (CI Tau, GK Tau, HP Tau, IQ Tau); \citealt{huang2020a} (GM Aur); \citealt{francis2020} (RY Lup); \citealt{long20} (GQ Lup; see also \citealt{wu2017}); \citealt{andrews21} (HD 143006); Long et al., private communication (FZ Tau, IRAS 04385+2550, TW Cha, VZ Cha)}
\end{deluxetable*}

\begin{deluxetable*}{lcccc}
\tablecaption{Description of \emph{JWST} MIRI-MRS Observations \label{tab:obstable}}
\tablehead{
\colhead{Target} & \colhead{Start Date} & \colhead{End Date}  & \colhead{Visit ID} & \colhead{Exposure Time\tablenotemark{$\ast$}}} 
\startdata
AS 205 N & 2023-04-04T16:37:46.419 & 2023-04-04T17:21:20.598 & 01584010001 & 388.504 \\
AS 205 S\tablenotemark{a} & 2023-04-04T16:37:46.419 & 2023-04-04T17:21:20.598 & 01584010001 & 388.504 \\
AS 209 & 2022-08-02T02:37:53.904 & 2022-08-02T04:51:31.158 & 2025001001 & 2242.232 \\
CI Tau & 2023-02-27T14:03:00.277 & 2023-02-27T15:56:38.677 & 1640005001 & 1776.024 \\
DoAr 25 & 2023-08-16T04:19:04.241 & 2023-08-16T05:19:50.954 & 1584013001 & 677.108 \\
DoAr 33 & 2023-03-31T01:26:39.326 & 2023-03-31T03:15:56.870 & 1584016001 & 1687.224 \\
Elias 20 & 2023-03-31T03:46:55.594 & 2023-03-31T04:48:07.616 & 1584012001 & 721.512 \\
Elias 24 & 2023-08-16T05:41:51.706 & 2023-08-16T06:41:15.174 & 1584014001 & 654.908 \\
Elias 27 & 2024-03-16T16:34:36.646 & 2024-03-16T17:35:15.936 & 1584021001 & 677.108 \\
FZ Tau & 2023-02-28T03:01:11.403 & 2023-02-28T04:14:35.498 & 1549001001 & 987.916 \\
GK Tau & 2023-02-28T04:42:09.445 & 2023-02-28T05:58:14.500 & 1640003001 & 987.916 \\
GM Aur & 2023-10-14T11:18:58.286 & 2023-10-14T14:19:51.433 & 2025007001 & 3057.908 \\
GO Tau & 2023-10-09T18:51:59.611 & 2023-10-09T21:14:13.058 & 1640002001 & 2319.932 \\
GQ Lup & 2023-08-13T14:46:28.511 & 2023-08-13T16:40:12.236 & 1640009001 & 1776.024 \\
HD 142666 & 2023-04-04T23:57:26.617 & 2023-04-05T00:40:58.557 & 1584008001 & 388.504 \\
HD 143006 & 2023-04-04T19:35:31.188 & 2023-04-04T20:35:39.403 & 1584009001 & 721.512 \\
HD 163296 & 2022-08-10T00:42:14.512 & 2022-08-10T02:56:31.662 & 2025004001 & 2253.332 \\
HP Tau & 2023-02-27T16:23:53.232 & 2023-02-27T17:37:00.687 & 1640001001 & 987.916 \\
HT Lup A+B & 2023-04-04T12:35:49.858 & 2023-04-04T13:29:51.204 & 1584001001 & 588.308 \\
IQ Tau & 2023-02-27T11:42:46.236 & 2023-02-27T13:35:12.413 & 1640004001 & 1764.924 \\
IRAS 04385+2550 & 2023-10-15T22:09:24.933 & 2023-10-15T23:25:13.620 & 1640011001 & 987.916 \\
MWC 480 & 2023-10-13T07:30:50.504 & 2023-10-13T09:58:45.508 & 02025006001 & 2253.332 \\
MY Lup & 2023-08-13T18:58:02.052 & 2023-08-13T19:50:07.041 & 1584007001 & 555.008 \\
RU Lup & 2023-08-13T20:11:26.148 & 2023-08-13T21:02:19.004 & 1584004001 & 521.708 \\
RY Lup & 2023-08-13T17:10:16.071 & 2023-08-13T18:24:32.902 & 1640010001 & 987.916 \\
SR 4 & 2023-08-16T02:48:38.943 & 2023-08-16T03:50:32.240 & 1584011001 & 688.208 \\
Sz 114 & 2023-03-31T12:15:41.650 & 2023-03-31T13:22:01.166 & 1584005001 & 832.512 \\
Sz 129 & 2023-03-31T07:49:47.701 & 2023-03-31T09:25:18.832 & 1584006001 & 1110.016 \\
TW Cha & 2023-07-24T11:32:55.271 & 2023-07-24T13:57:17.689 & 1549003001 & 2386.536 \\
VZ Cha & 2023-07-24T14:29:51.363 & 2023-07-24T16:23:03.536 & 1549004001 & 1764.924 \\
WSB 52 & 2023-08-28T04:47:16.633 & 2023-08-28T05:30:46.089 & 1584017001 & 333.004 \\
\enddata
\tablenotetext{\ast}{Exposure times are provided in seconds per sub-band.}
\end{deluxetable*}

\section{Results} \label{sec:results}

\subsection{Establishing the Inner and Outer Dust Disk Contexts}

All JDISCS targets included in this work were observed with ALMA at high angular resolution through DSHARP ($\sim$0.035'' or 5 au; \citealt{andrews18}), surveys of the Taurus star-forming region (0.12'' or 16 au; \citealt{long18, long19}), and a program targeting the remaining disks observed in \emph{JWST} Cycle 1 that lacked high-resolution mm images (0.05" or 7 au; Long et al., in prep; ALMA Project Code \#2021.1.00854.S); we refer the reader to these works for further details about the ALMA observations. While the four programs differ in spatial resolution by a factor of $\sim$4, the closest resolved structures at $\sim$5 au are still outside the expected emitting radii for the mid-infrared molecular gas emission (see e.g., \citealt{walsh15,woitke18,anderson21, kanwar24a}). We use the published sub-mm datasets to provide some outer disk context for the interpretation of inner disk emission lines, with the caveat that we are not able to discern the presence of inner disk dust substructure from the ALMA observations themselves. 

As a complementary tracer of inner disk dust evolution, we measure the infrared spectral index $n_{13-26}$ as in \cite{banzatti23b} by taking the MIRI continuum flux measured at 13 and 26~$\mu$m. A similar index was previously introduced with Spitzer-IRS spectra at 13 and 30~$\mu$m and was used to infer the presence of an inner dust cavity from positive values of the infrared index $n_{13-30} > 0$ \citep{brown07,furlan09,banzatti2020}. Negative values of $n_{13-30}$ instead indicate emission from abundant small grains in the inner disk region, although inclination effects also play a role \citep[see Appendix D in][]{banzatti2020}. The specific 13 and 26 $\mu$m wavelength regions used in this work were selected using the analysis in \citet{banzatti24} from those that are most free from detectable line emission: at 13.095--13.113~$\mu$m and 26.3--26.4~$\mu$m. At these wavelengths, molecular emission is as weak as the noise on the continuum even in a strong-emission case as CI~Tau, which is used for reference in \citet{banzatti24}. We use 26~$\mu$m instead of 30~$\mu$m due to the different wavelength coverage of the MRS and the S/N that decreases at longer wavelengths. The measured $n_{13-26}$ values for the whole sample are provided in Table \ref{tab:diskprops}.

\begin{figure*}
\epsscale{1.17}
\centering 
\plotone{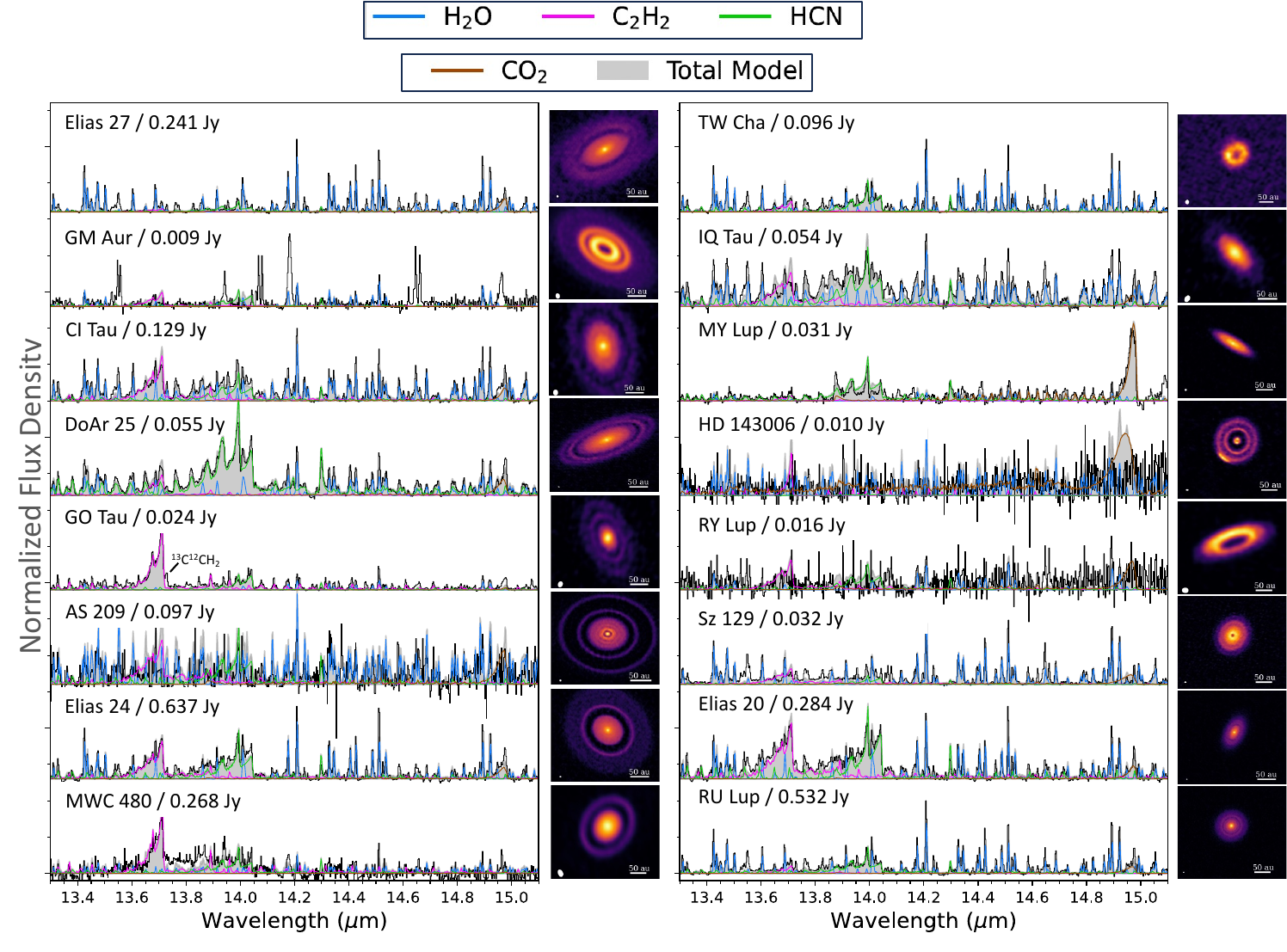}
\caption{MRS continuum-subtracted spectra of C$_2$H$_2$ (magenta), HCN (green), H$_2$O (blue), and CO$_2$ (brown) emission lines between 13.6-15.3 $\mu$m from JDISCS sources with sub-mm dust sizes $r > 63$ au. ALMA images are shown to the right of the spectra (\citealt{andrews18, long18, long19}, Long et al. 2025, in prep), in order of sub-mm dust disk size from largest (top left; Elias 27, $r \sim 250$ au) to smallest (bottom right; RU Lup, $r \sim 63$ au). Slab model fits to the molecular emission lines are overplotted on the spectra, corresponding to the best-fit parameters and upper limits reported in Tables \ref{tab:slabparams_C2H2}, \ref{tab:slabparams_HCN}, \ref{tab:slabparams_CO2}, and \ref{tab:slabparams_H2O}. $^{13}$C$^{12}$CH$_2$ emission is also detected in the spectrum of GO Tau, along with multiple CO$_2$ isotopologues in MY Lup \citep{salyk25}, and an OH-rich spectrum with no CO$_2$ emission from GM Aur (Romero-Mirza et al., submitted).}
\label{fig:organics_gallery_large}
\end{figure*}

\begin{figure*}
\epsscale{1.17}
\centering
\plotone{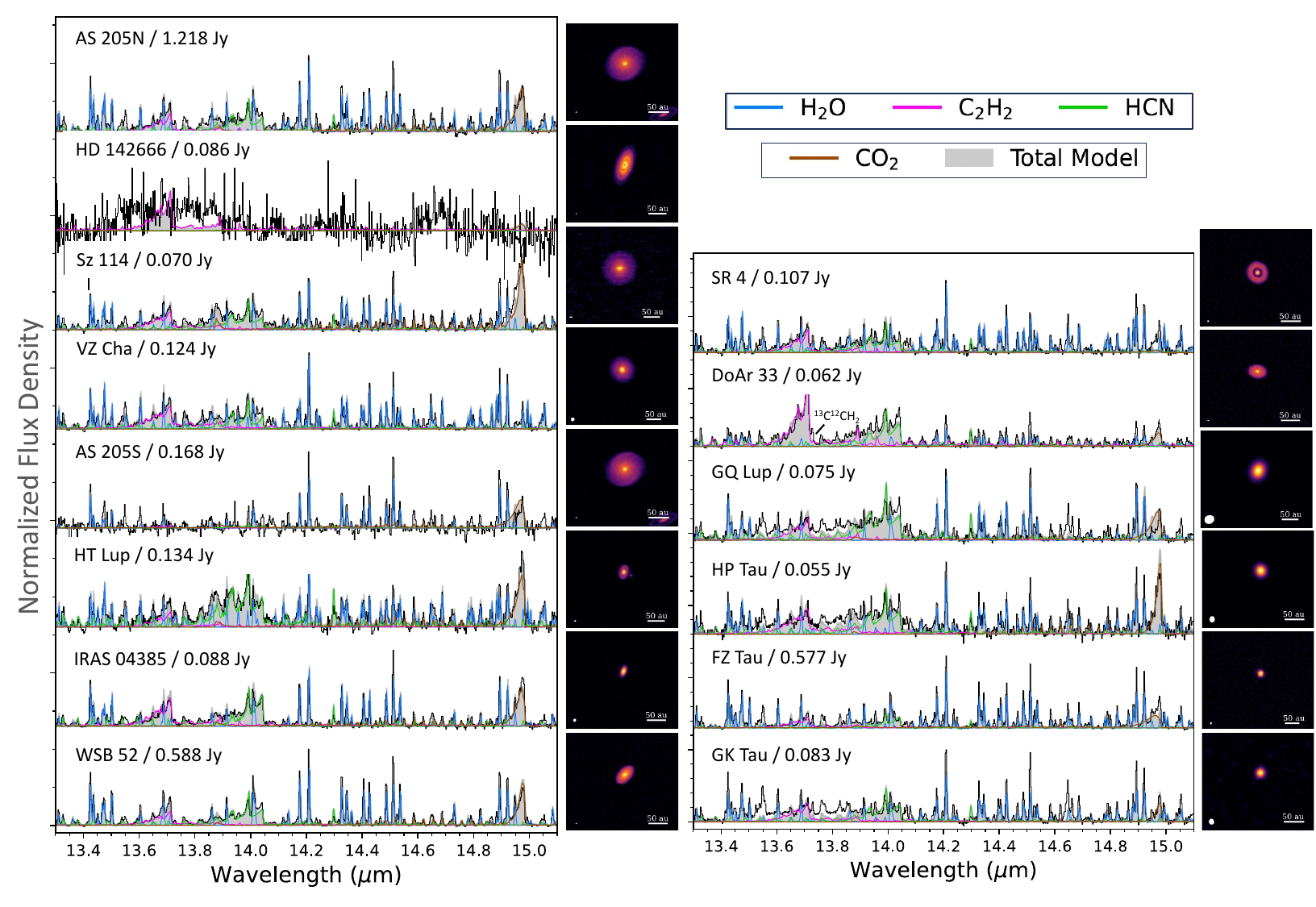}
\caption{MRS continuum-subtracted spectra of C$_2$H$_2$ (magenta), HCN (green), H$_2$O (blue), and CO$_2$ (brown) emission lines between 13.6-15.3 $\mu$m from JDISCS sources with sub-mm dust sizes $r < 60$ au. ALMA images are shown at right (\citealt{andrews18, long18, long19}, Long et al. 2025, in prep), in order of sub-mm dust disk size from largest (top left; AS 205N, $r \sim 60$ au) to smallest (bottom right; GK Tau, $r \sim 13$ au). Slab model fits to the molecular emission lines are overplotted on the spectra, corresponding to the best-fit parameters and upper limits reported in Tables \ref{tab:slabparams_C2H2}, \ref{tab:slabparams_HCN}, \ref{tab:slabparams_CO2}, and \ref{tab:slabparams_H2O}. $^{13}$C$^{12}$CH$_2$ emission is also detected in the spectrum of DoAr 33 \citep{colmenares24}.}
\label{fig:organics_gallery_small}
\end{figure*}

Figures \ref{fig:organics_gallery_large} and \ref{fig:organics_gallery_small} zooms in on the $Q$ branches of C$_2$H$_2$, HCN, and CO$_2$, along with overlapping H$_2$O transitions, from all JDISCS sources included in this work, ordered by mm dust disk radius from largest (Elias 27; $r \sim 250$ au; \citealt{huang2018}) to smallest (GK Tau; $r \sim 13$ au; \citealt{long19}). As also observed with \emph{Spitzer}, the MIRI spectra of the organics are diverse, with clear visual variations in both the shapes and strengths of emission lines across the sample. With the increased spectral resolution of MRS relative to IRS, it is readily apparent where $P$ and $R$ branch transitions of all three molecules overlap, particularly in the case of C$_2$H$_2$ and HCN. This makes it challenging to model molecules individually, as the line fluxes at wavelengths corresponding to overlapping transitions can not easily be separated. 

The sample presented in Figures \ref{fig:organics_gallery_large} and \ref{fig:organics_gallery_small} includes five disks with brighter atomic emission lines from H I, [Ne II], [Ne III], and sometimes [Ar II] relative to the molecules: GM Aur, MY Lup, HD 143006, RY Lup, and HD 142666. A set of rotational H$_2$O emission lines from these systems, where detected, is shown in Figure \ref{fig:herbig_cavity_water}. The group includes one Herbig Ae/Be star (HD 142666), one G-type star (HD 143006), and three T Tauri stars (RY Lup, GM Aur, MY Lup); however, the S/N across the brighter disks (HD 142666, HD 143006, RY Lup) is so high that residual fringing becomes more apparent in the spectra. Since all three targets are brighter than the asteroid calibrators at these wavelengths (along with HD 163296, which is excluded from both Figures \ref{fig:organics_gallery_large} and \ref{fig:herbig_cavity_water} for poor data quality), the JDISCS reduction pipeline cannot yet fully correct for the residual non-linearity in the MIRI detectors. Since this subset of disks is not molecule-rich, the effect does not impact the results presented here.

\subsection{Detection Rates of H$_2$O, OH, and CO}

\begin{figure*}
\centering
\epsscale{1.2}
\plotone{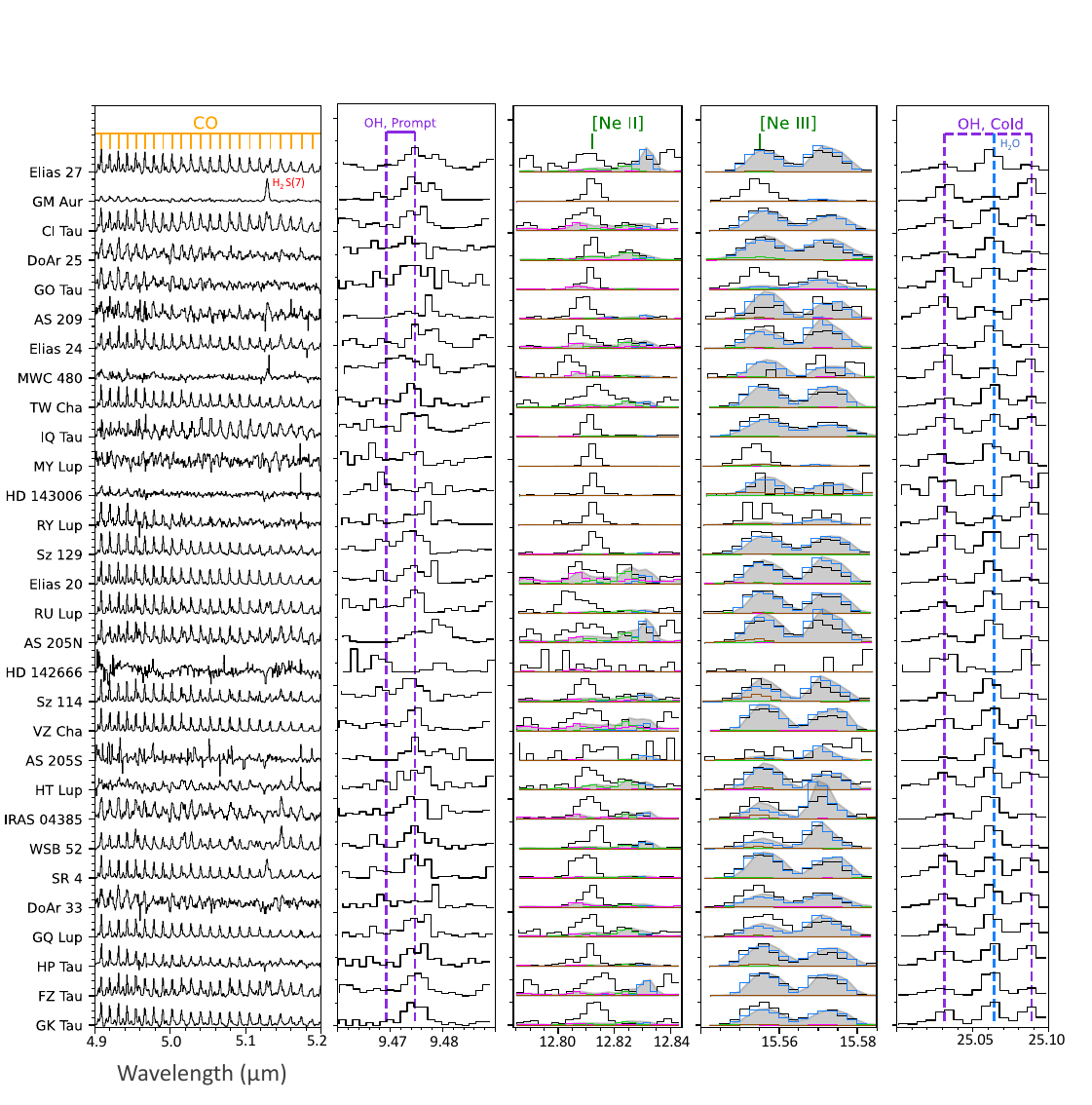}
\caption{Gallery of ro-vibrational CO, OH, [Ne II], and [Ne III] emission lines from 30/31 JDISCS sources (HD 163296 is excluded due to poor data quality). The spectra are in order from largest to smallest sub-mm disk size, with representative slab models of H$_2$O overplotted on the data surrounding the [Ne II] and [Ne III] emission lines \citep[following][]{banzatti24}. C$_2$H$_2$ and HCN slab models are shown as well, with the same color scheme as in Figure \ref{fig:organics_gallery_large}.}
\label{fig:CO_OH_NeII_NeIII_gallery}
\end{figure*}

We find that emission lines from H$_2$O, OH, and CO are nearly ubiquitous at the 2-$\sigma$ threshold across the JDISC sample (see Figures \ref{fig:CO_OH_NeII_NeIII_gallery} and \ref{fig:histograms}), consistent with previous findings from \emph{Spitzer} and ground-based spectroscopy surveys \citep{najita03, pontoppidan10b, salyk11, brown13, banzatti22, banzatti23a}. Detailed modeling of these species is required to confirm the fainter lines, but our initial detection rates are reported in Table \ref{tab:detectionrates}, for the full sample, classical K- and M-type T Tauri stars, transitional disks with infrared spectral indices $>0$, and an ``intermediate mass" category that includes all three Herbigs and the G-type star HD 143006. The detection rate for rotational transitions of H$_2$O is 94\%, with HD 142666 and HD 143006 being the only disks where water emission is not detected \citep{banzatti24}. Notably, mid-infrared water emission is detected for the first time towards the transition disks RY Lup \citep{banzatti24} and GM Aur (see Figure \ref{fig:herbig_cavity_water}; Romero-Mirza et al. 2025, submitted). While less ubiquitous than the rotational H$_2$O transitions, ro-vibrational water emission lines are detected in 77\% of sources. Non-detections include all three Herbig sources (HD 142666, HD 143006, HD 163296) and the transition disk GM Aur, along with AS 205S, GO Tau, and MY Lup. Ro-vibrational $v=1-0$ CO emission from is observed in 87\% of disks with MRS, with the exception of AS 205S, HD 142666, HD 163296 and MY Lup.  HD 163296 does have detectable rovibrational CO in ground-based high resolution spectra \citep{salyk11b} as well as water and OH previously detected with Spitzer and Herschel \citep{fedele12}. 

In the case of OH, we check for detections both in very high energy levels at the short wavelengths and in low energy levels at long wavelengths. We focus on five lines with $E_u \sim 27,000 - 36,000$~K emitting between 9.4 and 10.3~$\mu$m, and six lines with $E_u \sim 3000 - 4000$~K emitting between 23 and 28~$\mu$m. The lines we consider in these ranges are free from water contamination, as determined from the analysis presented in \cite{banzatti24}. Figure \ref{fig:CO_OH_NeII_NeIII_gallery} shows examples of these lines near 9.5~$\mu$m and 25~$\mu$m. We detect the high-energy OH lines in 19/30 disks (63\% detection rate) and the low-energy lines in 25/30 disks (83\% detection rate). 

For all of the high-energy OH detections, the lines show the typical asymmetry produced by prompt emission following water photodissociation by UV radiation \citep[e.g.][]{carr14, tabone24}, suggesting that this may be common in T~Tauri disks of a few Myr old. Recent models by \cite{tabone24} propose the prompt emission would produce prominent OH lines with a strong asymmetry at $< 12$~$\mu$m, but both prominence and asymmetry would become weaker at longer wavelengths due to the increasing separation of transitions and possibly chemical pumping affecting the excitation of OH lines. We note that, contrary to their models, all the OH spectra in our sample show a strong increase in line flux with wavelength, suggesting that either an additional OH reservoir or other excitation processes dominate the observed populations of low-energy lines. We leave a more detailed modeling of water and OH to future work, which will likely require multiple temperature components or a temperature gradient to fully reproduce the emission lines (see e.g., \citealt{romero-mirza24b}) along with a dedicated treatment of OH emission from disks around intermediate T Tauri stars and Herbig systems.

\begin{deluxetable*}{cccccccc}
\tablecaption{Molecular Detection Rates \label{tab:detectionrates}}
\tablehead{\colhead{Disk Type\tablenotemark{a}} & \colhead{N\tablenotemark{b}}&\colhead{H$_2$O} & \colhead{OH\tablenotemark{c}} & \colhead{CO} & \colhead{CO$_2$} &\colhead{HCN} &\colhead{C$_2$H$_2$}  }
\startdata
Classical & 18 & 100 & 94 & 100 & 72 & 94 & 89 \\
Transitional & 9 & 100 & 89 & 78 & 56 & 56 & 56 \\
Intermediate Mass & 4 & 50 & 50 & 50 & 0 & 0 & 25 \\
\hline
Full Sample & 31 & 94 & 87 & 87 & 58 & 71 & 71
\enddata
\tablenotetext{a}{The ``Intermediate Mass" category includes all three Herbig disks (HD 142666, HD 163296, and MWC 480) and the G-type star HD 143006. For consistency with \emph{Spitzer}-IRS surveys, the ``classical" and ``transitional" groups include all remaining disks with $n_{13-26} < 0$ or $n_{13-26} > 0$, respectively. However, we note that not all disks with positive infrared spectral indices have resolved sub-mm dust cavities (see Table \ref{tab:diskprops}).}
\tablenotetext{b}{Number of disks}
\tablenotetext{c}{The overall detection rate of OH is higher than reported in the text, since not all disks show emission from both high and low-energy transitions.}
\end{deluxetable*}

\subsection{A MIRI-MRS Inventory of Molecular Gas Emission Lines} 

Figures \ref{fig:organics_gallery_large} and \ref{fig:organics_gallery_small} highlights the $Q$ branch emission lines from C$_2$H$_2$, HCN, and CO$_2$ from JDISCS targets included in this work. At the exquisite sensitivity and spectral resolution of MIRI across the Channel 3 detector ($R \sim 2190-3160$ between 11.55-17.98 $\mu$m; \citealt{pontoppidan24}), the $P$ and $R$ branch transitions are also now readily detected. These emission lines can remain optically thin even when the lower energy $Q$ branch lines are optically thick, critically breaking the degeneracy between gas temperatures and column densities within the inner disk molecular layer. However, as identified in \emph{Spitzer} spectra, the emission lines from organic molecules overlap with each other, with rotational H$_2$O emission lines, and with strong atomic features from H I, [Ne II], and [Ne III]. Even with the increased spectral resolution of MRS, it is challenging to isolate each individual species and characterize temperature and density stratifications within the warm molecular disk surface layers. In fact, previous analyses of MRS spectra have often taken the approach of sequential fits where one molecule is fitted and subtracted before the next molecule is fitted \citep{grant23,vlasblom24}, although more recently molecules are fit together \citep{temmink24b,grant2024}.

Instead of measuring line fluxes and reporting detection rates directly from the data, we use local-thermodynamic-equilibrium (LTE) slab models made with \texttt{spectools-ir} \citep{salykspectoolsir} to reproduce the observed spectrum from each disk, measure the total emission line luminosity for each modeled species, and retrieve the underlying physical parameters that describe the molecular gas. 

\subsubsection{Description of LTE Slab Models}
In brief, \texttt{spectools-ir} produces synthetic emission spectra from a slab of gas with three properties: temperature ($T$), column density ($N_{\rm{col}}$), and projected emitting area ($A_\mathrm{proj}$), which we convert to an emitting radius ($r_{\rm{slab}}$) assuming A$_\mathrm{proj}=\pi r_{\rm{slab}}^2 \cos{i_\mathrm{disk}}$.

\texttt{spectools-ir} accounts for two fundamental features that are essential for successfully reproducing the observed organic emission lines: line broadening (which is a combination of thermal, instrumental, and Keplerian) and line opacity overlap, which can be particularly significant in the densely clustered spectral regions around the $Q$-branches \citep{tabone23} and for water ortho-para line pairs that overlap in wavelength \citep{banzatti24}. In previous work with spectrally unresolved \emph{IRS} data, the thermal broadening of emission lines was treated as a fixed parameter assumed from the sound speed of H$_2$ at 1000~K \citep{salyk09}. With MRS, emission lines originating from gas populations with different temperatures are now spectrally separated; for example, a warm water emission component ($T \sim 400$ K) has been readily detected against hotter water emission ($T \sim 850$ K) \citep{banzatti23b, pontoppidan24, romero-mirza24b, grant2024, temmink24b, banzatti24}. These two temperatures correspond to thermal widths (standard deviation) of 0.4 km s$^{-1}$ and 0.6 km s$^{-1}$, respectively, which differ by a factor of 1.5. This can lead to a significant under- or over-prediction of the model optical depth and a corresponding over- or under-prediction of the column density in order to reproduce the data. Rather than fixing the thermal broadening to a single value across all molecules as assumed in previous IRS analyses and recently in some analyses of MRS spectra \citep{grant23, tabone23, xie23, banzatti23b}, we compute it from the temperature used to generate each individual slab model as in \citet{pontoppidan24} and \citet{romero-mirza24b}. This approach accounts for differences in local line broadening between emission from radially or vertically separated layers of the disk, allowing for better constraints on the column densities retrieved from optically thin features.    

At the temperatures and rotation velocities found in inner disks, Keplerian broadening dominates the observed emission line widths as observed at high resolution from the ground \citep[e.g.][]{najita03,salyk11,brown13,banzatti22}. This effect has now been detected also in molecular emission lines observed with MRS, which are broader than the instrument resolution and show, in some disks observed at high inclination, an increase in FWHM as a function of upper level energy \citep{banzatti24}. Although the broadening is more challenging to detect in the spectrally unresolved $Q$ branch emission lines from the organic molecules, convolving the water slab models with the instrumental widths alone leads to an under-prediction of emission in the wings of the water line profiles. To account for this, we convolve the slab models to a FWHM measured from the rotational water lines within the same wavelength range as the organics, which represents the contribution from both Keplerian and instrumental broadening \citep{banzatti24}. 

Further saturation and blending of individual emission lines may be seen when the emitting layer is optically thick \citep{carr11, salyk11, tabone23, banzatti24}. \texttt{spectools-ir} accounts for this line opacity overlap by calculating the optical depth as a function of wavelength and summing optical depth over all overlapping transitions before computing flux (see also, \citealt{tabone23}). If this effect is not included, the slab models will consistently over-predict the observed line fluxes and force an under-prediction of the column densities \citep[for the case of water, see Figure 5 in][]{banzatti24}. This is particularly important for identifying optically thick populations of gas in the absence of spectrally resolved, optically thin $Q$ branch emission lines \citep{tabone23}.

We use slab models generated with \texttt{spectools-ir} to provide fits first to H$_2$O and subsequently C$_2$H$_2$, HCN, and CO$_2$ emission lines between 12-16 $\mu$m for each disk in our sample. We first use a Markov chain Monte Carlo (MCMC) ensemble sampler \citep{emcee} to fit the water lines, by exploring the temperature, column density, and slab emitting area as variable parameters. The sampler minimizes the L2 loss function, such that 
\begin{equation}
\mathcal{L}(\theta) = \sum_{i=1}^N \parallel y_i - f(x_i, \theta) \parallel
\end{equation}
approaches 0. Here $y_i$ represent the observed flux at each wavelength $x_i$, and $f(x_i, \theta)$ is the slab model generated for each $T$, $N_{\rm{col}}$, and $r_{\rm{slab}}$. A statistical analysis that includes the error measurements on the observed fluxes will be included in future works, particularly those searching for emission from isotopologues and rare species, but this simple method sufficiently retrieves best-fit parameters for the more abundant molecules presented here. We also note that the uncertainties produced by degeneracies across the model parameter space far exceed the uncertainties on the flux measurements.  

We subtract the best-fit water model from the data to more clearly identify the remaining molecular emission lines and then use a second MCMC ensemble sampler to simultaneously fit all C$_2$H$_2$, HCN, and CO$_2$ emission lines within the same wavelength range. The second sampler also minimizes the L2 loss function; we do not include RMS noise in the loss function, as it is minimal in comparison to the intrinsic model degeneracies. We discuss the impacts of using a single temperature to describe the water emission, excluding fits to low S/N isotopologue emission, and the challenges associated with identifying the true dust continuum in Appendix \ref{appendix:waterremoval} and Section \ref{sec:analysis}.

The slab models can also be used to place upper limits on the emitting mass and emission line luminosity from molecules that are not detected in the spectra (see e.g., \citealt{salyk11}). In other analyses, this has been done molecule-by-molecule (see e.g., \citealt{grant23, xie23}). After subtracting a slab model fit to a single species, the residuals are examined in the wavelength region around the expected $Q$ branch transitions. Marginal and non-detections are identified when an additional molecule does not lead to a significant reduction in the Akaike Information Criterion \citep{romero-mirza24a}, when the RMS noise is equal in strength to the slab model fit \citep{xie23, schwarz24}, or when the excess emission is not significant relative to a MRS spectrum in which the molecule is detected \citep{gasman25}. While quantitatively robust, iterative fits such as those described in \citet{romero-mirza24a} require significant computational time, making them best suited for carefully characterizing all detected species in a single spectrum and thus prohibitive for this study of 31 sources. For the overview of C$_2$H$_2$, HCN, and CO$_2$ presented here, we set the detection thresholds across the sample by comparing the integrated emission line luminosities from the best-fit slab models and peak-to-continuum ratios calculated as the maximum flux from the same slab models divided by the continuum flux at the corresponding wavelength. We outline this method and discuss its limitations in Section 3.3.3.    

\subsubsection{Characterizing Rotational H$_2$O Emission Lines}

By fitting transitions between 12--16 $\mu$m only, the slab model fits to water emission lines from 24/31 disks in the JDISCS sample all return similar column densities, with a median value of $\log N_\mathrm{col} = 18.4$ cm$^{-2}$ and standard deviation across the sample of 0.3 dex. This similarity was also identified in the \emph{IRS} data by \citet{carr11} when the fit was limited to the same spectral range as is done in this work and when all transitions between 10--35 $\mu$m were considered \citep{salyk11}. The median temperature of 710 K derived from the JDISCS sample, with a standard deviation of 100 K, is consistent with those reported in \citealt{carr11} for disks around T Tauri stars, which fit over a similar wavelength region. This result is also generally consistent with the hot ($T \sim 800$--900~K) temperature component found in a sub-set of the JDISCS sample in \cite{romero-mirza24b} and \cite{banzatti23b}, consistent with the fact that higher-energy lines dominate the emission at wavelengths $<$17$\mu$m \citep{banzatti24}. Finally, the retrieved emitting areas across the sample correspond to a median slab radius of 0.4 au, with a standard deviation of 0.6 au. 

The best-fit H$_2$O model for each disk was subtracted from the spectrum, leaving water-free residuals which are used to fit the organics and measure atomic emission lines. In Appendix \ref{appendix:waterremoval}, we discuss the impact of weak residual water vapor emission on the retrieval of best-fit parameters for the organic molecules. The residual water vapor emission includes signatures of non-LTE excitation in the rovibrational versus rotational states, making them more difficult to reproduce \citep{Meijerink09,bosman22,banzatti23b,banzatti24}.
  
The seven remaining disks (GM Aur, MWC 480, AS 209, HD 142666, HD 143006, MY Lup, and RY Lup) show much stronger residual fringing that overlaps with the water emission lines between 12--16 $\mu$m, making it challenging to identify the best-fit slab model parameters. Instead of fitting this region directly, we fit the slab models to water emission lines between 16.6--17.6 $\mu$m where detected (all these disks except for HD143006, HD142666, MWC 480, and HD 163296) and use those column densities and temperatures to predict a water model at wavelengths that overlap with the organics between 12--16 $\mu$m (see Appendix \ref{appendix:waterremoval}, Figure \ref{fig:herbig_cavity_water}). As previously discussed, the disk of HD 163296 is excluded from this analysis due to saturation effects that led to poor data quality. 

Appendix \ref{appendix:waterremoval} highlights the spectra of two disks around Herbig stars with resolved dust rings (HD 142666 and HD 143006), one disk with a large dust cavity (RY Lup; \citealt{vandermarel18, francis2020, ribas24}), and one disk with an inner dust gap near $r \sim 8$ au (MY Lup; \citealt{huang2018}) in Figure \ref{fig:herbig_cavity_water}. At the sensitivity of \emph{Spitzer-IRS}, water emission lines were generally found to be absent in transition disks and were only tentatively detected in some Herbig disks \citep{pontoppidan10b}, with a firm detection only in HD~163296 \citep{fedele12}. A few transition disks (DoAr~44, TW~Hya, SR~9), however, did show water emission preferably at longer IR wavelengths (but in the case of DoAr~44 at short IR wavelengths too), indicative of colder water in comparison to what is typically found in T~Tauri disks \citep{salyk15,banzatti2017}. However, we identify water emission lines in the spectrum of RY Lup, with $\log N_\mathrm{col} \sim 17.6$ cm$^{-2}$ and $T \sim 520$ K. The water emission in MY Lup, instead, can be reproduced with best-fit model indicating $\log N_\mathrm{col} \sim 18$ cm$^{-2}$ and $T \sim 330$ K \citep{salyk25}. The analysis of the sub-sample of disks with dust cavities will be presented in a forthcoming paper (Mallaney et al. 2025 in prep). 

\subsubsection{A carbon carrier: C$_2$H$_2$}

\begin{figure*}
\epsscale{1.2}
\plotone{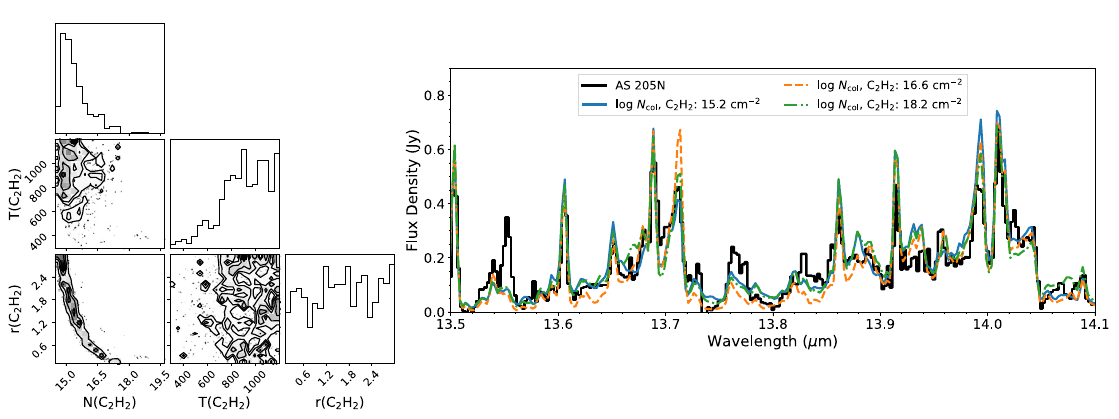}
\caption{\emph{Left:} Example corner plot, showing the strong degeneracy between column density and slab radius in LTE slab model fits to the C$_2$H$_2$ emission from AS 205N. The effect is still apparent when HCN and CO$_2$ emission lines are included in the fit. \emph{Right:} Degenerate slab model solutions overlaid on JDISCS spectrum of AS 205N.}
\label{fig:C2H2_degeneracy} 
\end{figure*}

We use the \emph{IRS} and published MRS results to inform the parameter space for slab model retrievals, using uniform priors of C$_2$H$_2$ column densities between $10^{13}$ cm$^{-2} < N_\mathrm{col} < 10^{24}$ cm$^{-2}$, temperatures between 400 K $< T < 1500$ K, and slab radii between 0.1 au $< r_\mathrm{slab} < 3$ au (see e.g., \citealt{salyk11, anderson21, tabone23}). While cooler, optically thick C$_2$H$_2$ emission has been observed in disks around very low mass stars \citep{tabone23}, we keep the lower limit on the temperature consistent with what was observed from K- and M-type stars with \emph{Spitzer}. However, the slab model fits across the 12--16 $\mu$m wavelength range show a strong degeneracy between the column densities and the slab emitting radii. At the same time, the temperatures remain roughly constant with respect to variations in the column densities  --- see Figure \ref{fig:C2H2_degeneracy}. This effect indicates that the emitting gas is optically thin (see e.g., \citealt{carr11, grant2024}), making it challenging to quantitatively converge on a set of best-fit parameters for all disks in the sample without additional prior information. 

Within the parameter space identified with \emph{IRS} observations, we detect clusters of ``best-fit" slab model parameters that can reproduce the C$_2$H$_2$ emission from each disk, with $<5\%$ changes in the minimum $L2$-norm test statistics. As an example, three possible solutions for AS 205N, shown in Figure \ref{fig:C2H2_degeneracy}, are: $\log N_\mathrm{col} \sim 15.2$ cm$^{-2}$ and $r_\mathrm{slab} \sim 2.5$ au, $\log N_\mathrm{col} \sim 16.7$ cm$^{-2}$ and $r_\mathrm{slab} \sim 1.5$ au, and $\log N_\mathrm{col} \sim 18.2$ and $r_\mathrm{slab} \sim 0.6$ au. This effect was accounted for in analyses of \emph{IRS} spectra by scaling the slab models to match the observed flux, leaving only the temperatures and column densities as variable parameters in the retrievals (see e.g., \citealt{carr08, carr11, salyk11}).

Additional solutions appear when the parameter space is extended to include smaller or larger slab radii (see also, \citealt{carr11}). \citet{anderson21} find that CO$_2$ and C$_2$H$_2$ emission can come from radii as large as 3--5 au, making all degenerate slab model fits consistent with the physical-chemical models presented in that work. However, the low $\log N_\mathrm{col}$ solutions are close to the column densities used to place upper limits on the C$_2$H$_2$ emission from \emph{IRS} ($N_\mathrm{col} \sim 10^{13}-10^{14}$ cm$^{-2}$; \citealt{carr11, salyk11}). The slab emitting areas were treated as scale factors in that work, requiring smaller values to reproduce the non-detections than the large slab radii that are consistent with the detected emission lines in the JDISCS spectra. Meanwhile, the optically thick $N_\mathrm{col} > 10^{18}$ cm$^{-2}$ solutions should have resulted in ubiquitous detections of the $^{13}$C$^{12}$CH$_2$ isotopologue assuming $^{12}$C/$^{13}$C $\approx 70$ \citep{tabone23, kanwar24, arabhavi24}, which is only unambiguously detected in two JDISCS sources (DoAr 33 and GO Tau; \citealt{colmenares24}). 

\begin{figure*}[h!]
\epsscale{1.0}
\plotone{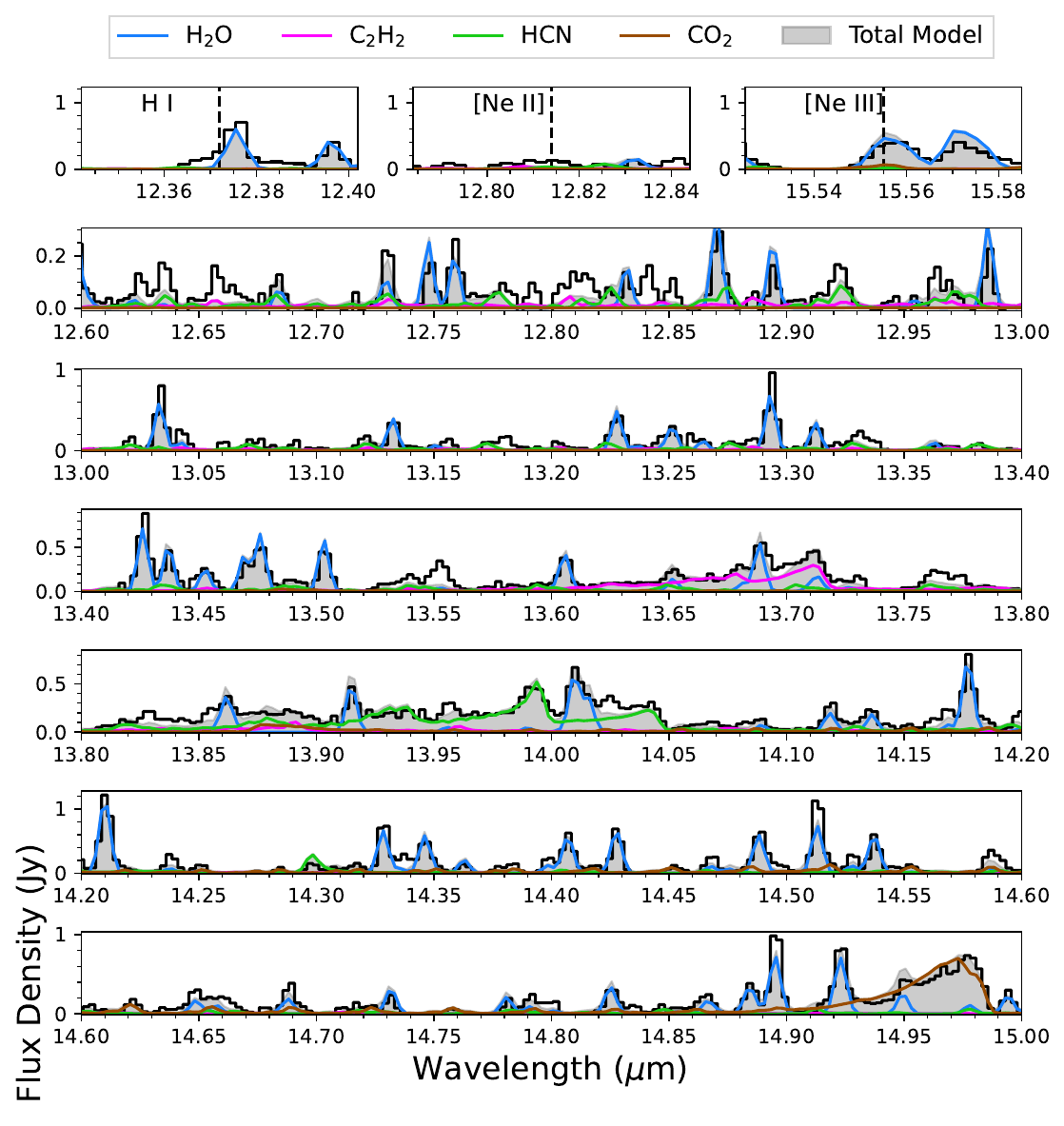}
\caption{MIRI-MRS spectrum of the brightest disk between 12-16 $\mu$m in our sample, AS 205N (black), with slab model fits to emission lines from H$_2$O (blue), C$_2$H$_2$ (magenta), HCN (green), and CO$_2$ (brown), as an example of a JDISCS target with prominent emission lines from all four molecules. The total model is shaded in gray. The complete set of similar figures for all targets in our sample is available in the online journal and \url{https://github.com/narulanantham/JDISCS_Cyc1_slabmodelfits.git}.}
\label{fig:example_retrieval}
\end{figure*}

For all other disks where the molecules are detected but the isotopologues are not, we plot the slab models generated from the intermediate solutions.  An example retrieved model is shown for AS 205N in Figure \ref{fig:example_retrieval}, which has bright emission lines from H$_2$O, C$_2$H$_2$, HCN, and CO$_2$. Rather than interpreting the slab model parameters themselves, we carry out our analysis using the total number of molecules (measured from the column density and emitting area to eliminate the degeneracy, and converted to units of $M_{\oplus}$), gas temperatures, and integrated model luminosities, as all three quantities are generally better preserved across the degenerate solutions. This means that model fits with identical $L2$-norm test statistics have temperatures, emitting masses, and luminosities that are similar to each other, even when the retrieved column densities and slab radii are not.

The computation time to achieve true convergence of the MCMC chains for the full sample is prohibitively expensive at this time (see e.g., \citealt{romero-mirza24a, grant2024}); instead, we measure (or retrieve) each metric from the set of all unique slab models with $L2$-norm test statistic values within 5\% of the best-fit solution (consistent to within roughly 2$\sigma$; see Equation 1). We report the median values and standard deviations in Table \ref{tab:slabparams_C2H2}. 

We set a rough detection threshold for each of the three organic molecules considered in this work based on the integrated emission line luminosities measured from the best-fit slab models (see Figure \ref{fig:detection_limits}). For C$_2$H$_2$, the $Q$ branch is readily identified when the integrated emission line luminosity between 12-16 $\mu$m exceeds $\log (L/L_{\odot}) \sim -5.1 $. With this metric, C$_2$H$_2$ is detected in 22/31 sources, for a detection rate of 71\%. This is a significant increase over the 43-44\% detection rate reported for K and M type T Tauri stars observed with \emph{IRS} \citep{pontoppidan10b, carr11}; we note that our sample includes five of the disks with previous \emph{IRS} detections (AS 205, DoAr 25, IQ Tau, TW Cha, VZ Cha). The disks with C$_2$H$_2$ detections in JDISCS have a median slab temperature of $920^{+70}_{-130}$ K  --- see all retrieved temperatures in Figure \ref{fig:allmols_T}. The coolest temperatures are reported for SR 4 and DoAr 33 ($T \sim 700$ K; see also, \citealt{colmenares24}), and five disks have temperatures $>1000$ K. The group of sources with the highest C$_2$H$_2$ temperatures includes the four disks with the smallest sub-mm dust radii in our sample: GK Tau ($r_\mathrm{dust} = 13$ au), FZ Tau ($r_\mathrm{dust} = 15$ au), HP Tau ($r_\mathrm{dust} = 21$ au), and GQ Lup ($r_\mathrm{dust} = 22$ au). The final disk, Sz 129, has a larger sub-mm dust disk ($r_\mathrm{dust} = 76$ au) and the smallest inner disk C$_2$H$_2$ mass of the sample ($M = 5.8 \times 10^{-10} \, M_{\oplus}$). We compare these results to the HCN and CO$_2$ slab model parameters in the following sections.  

\subsubsection{A nitrogen carrier: HCN}

HCN is one of three expected nitrogen carriers within inner disk gas (including NH$_3$ and N$_2$; \citealt{pontoppidan19}), and it is as yet the only one that is readily detected in both \emph{IRS} and ground-based mid-infrared spectroscopy (see \citealt{najita21} for a TEXES detection of NH$_3$ in absorption, \citealt{kaeufer2024b} for a tentative detection of NH$_3$ with MRS). HCN emission is detected at a similar rate to C$_2$H$_2$ in our sample, with integrated emission line luminosities from the best-fit slab models exceeding $\log (L) \sim -5.1 \, L_{\odot}$ in 22/31 sources (71\%). Of the nine sources from which HCN was not observed, four are disks around Herbigs or intermediate mass T Tauri stars and four are transitional disks with $n_{13-26} > 0$. The exception is AS 209 (see \citealt{romero-mirza24a} for discussion of a marginal detection). This result is roughly consistent with the 60--70\% detection rate of HCN emission lines reported in analyses of solar-mass \emph{IRS} spectra \citep{pascucci09, pontoppidan10b, carr11}.

Figure \ref{fig:allmols_T} shows that the HCN slab temperatures are generally consistent with the C$_2$H$_2$ temperatures across the sample, with a median temperature of $820^{+70}_{-130}$ K. The integrated emission line luminosities and emitting masses of the two molecules are also generally similar (see Figure \ref{fig:allmols_Lemit_Memit}). The hottest temperature in the sample is retrieved from DoAr 25 ($T = 950 \pm 70$ K); however, the total HCN mass from this target falls near the median total HCN mass of the JDISCS sample ($1.4 \times 10^{-8} \, M_{\oplus}$, compared to the sample median of $1.2 \times 10^{-8} \, M_{\oplus}$), so it does not appear to be unusually HCN-rich. Although HCN is readily detected, we do not detect the HC$_3$N $Q$ branch that is identified in other sources with MIRI \citep{kanwar24, kaeufer2024b, arabhavi24, long25}, prominent emission from NH$_3$ \citep{kaeufer2024b}, or H$^{13}$CN \citep{salyk25}.  

\subsubsection{An oxygen carrier: CO$_2$}

\begin{figure*}
\epsscale{1.2}
\plotone{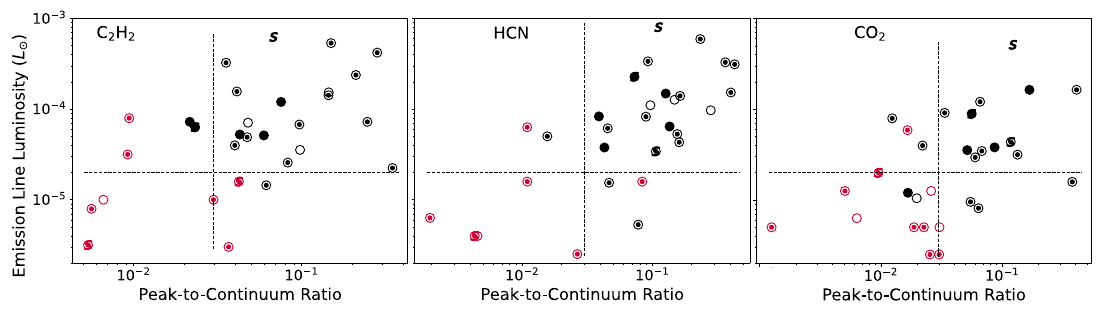}
\caption{Emission line luminosities from best-fit slab models (integrated between 12-16 $\mu$m) versus peak-to-continuum values from the slab models and fit to the continuum, with marker styles indicating the spatially resolved sub-mm structures in Table \ref{tab:diskprops}. Targets in the upper right quadrant show the clearest molecular gas detections, and we report non-detections (red markers) when peak-to-continuum ratios are $<0.03$ or emission line luminosities are smaller than $\log L < -5.1 \, L_{\odot}$ (black, dashed lines). While disks with resolved sub-mm dust cavities show C$_2$H$_2$ and HCN emission lines, none of these sources have CO$_2$ emission.}
\label{fig:detection_limits}
\end{figure*}

\begin{figure*}
\epsscale{1.05}
\plotone{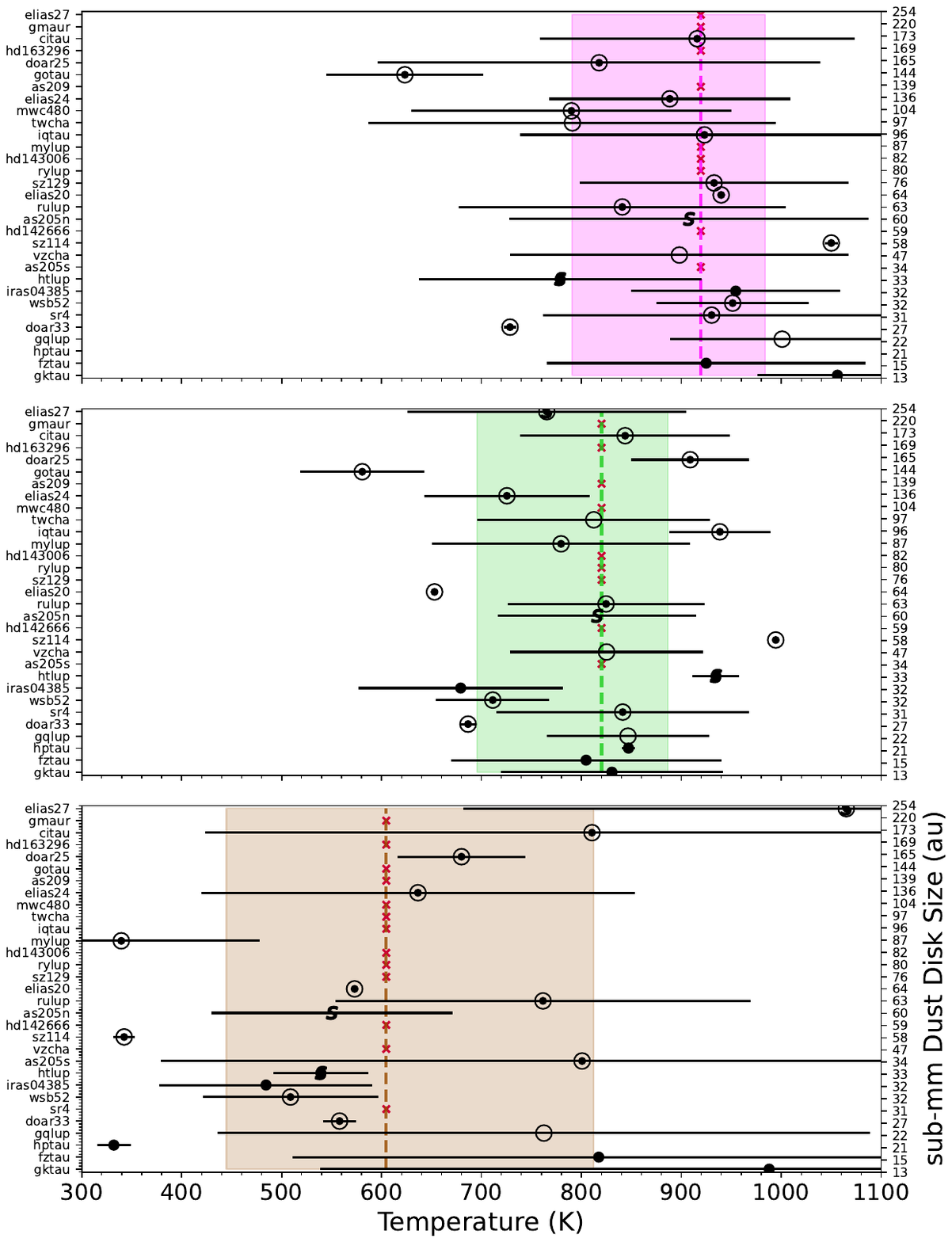}
\caption{Retrieved temperatures from slab model fits to C$_2$H$_2$ \emph{(top, magenta)}, HCN \emph{(middle, green)}, and CO$_2$ \emph{(bottom, brown)} emission lines from JDISCS sources in order of decreasing sub-mm disk size from top to bottom of each panel. Vertical dashed lines and shaded regions represent the sample median $\pm 1 \sigma$, with red x's along the median lines denoting non-detections for each molecule. Black markers identify the sub-mm substructures, as listed in Table \ref{tab:diskprops}. The C$_2$H$_2$ and HCN temperatures are generally consistent, while the CO$_2$ is cooler.}
\label{fig:allmols_T}
\end{figure*}

Like H$_2$O, CO$_2$ is a primary carrier of oxygen in the inner disk. MRS observations have revealed reservoirs of warm CO$_2$ that are abundant enough for robust detection of the $^{13}$CO$_2$ isotopologue $Q$ branch near 15.4 $\mu$m \citep{grant23}. MY Lup is the only JDISCS source in which the CO$_2$ isotopologues are unambiguously detected \citep{salyk25}, although Sz 114 has a marginal detection of $^{13}$CO$_2$ \citep{xie23}. The detection threshold of $^{12}$CO$_2$ from our slab model fits to the JDISCS sample is more ambiguous than the minimum peak-to-continuum ratios at which C$_2$H$_2$ and HCN are observed see Figure \ref{fig:detection_limits}), due to overlap with H I (16-10) emission at 14.962 $\mu$m that appears similar in shape to the CO$_2$ $Q$ branch (see e.g., GM Aur; Romero-Mirza et al., submitted). We identify 14/31 disks with integrated CO$_2$ emission line luminosities larger than $\log (L) \sim -5.1 \, L_{\odot}$. An additional four disks show CO$_2$ $Q$ branches upon visual inspection of the data, for a total detection rate of 18/31 (58\%). Since these targets have emission line luminosities and peak-to-continuum ratios that are similar to those disks in which CO$_2$ emission is not detected, any criterion that includes them will capture the non-detections as well. 

\begin{figure*}
\epsscale{1.2}
\plotone{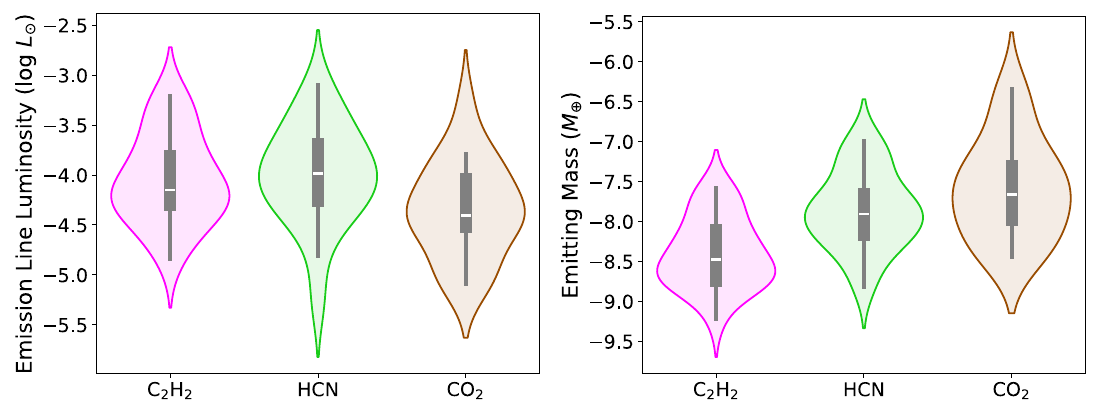}
\caption{Emission line luminosities \emph{(left)} and emitting masses \emph{(right)} integrated from slab model fits to C$_2$H$_2$, HCN, and CO$_2$ between 12-16 $\mu$m in JDISCS sources. Despite spanning a limited range in stellar masses, the targets display at least an order of magnitude variation in mid-infrared luminosity and mass of each molecule. Violin widths represent the number of disks with detections of each molecule, such that narrower violins correspond to lower detection rates.}
\label{fig:allmols_Lemit_Memit}
\end{figure*}

The $^{12}$CO$_2$ spectra in the JDISCS sample display the largest diversity in slab model parameters of all four molecules considered in this work, with best-fit temperatures ranging from 330--1050 K. AS 205N and MY Lup have the largest CO$_2$ masses in our sample ($3.7 \times 10^{-7} \, M_{\oplus}$), while IQ Tau has the smallest ($2.3 \times 10^{-10} \, M_{\oplus}$). The median temperature of $600^{+200}_{-160}$ K is significantly cooler than the median C$_2$H$_2$ and HCN temperatures of $920^{+70}_{-130}$ K and $820^{+70}_{-130}$ K, respectively. We note that this is likely an upper limit on the median CO$_2$ temperature, as the CO$_2$ slab models are slightly over-predicted in the $P$ and $R$ branch emission lines. Since the peak-to-continuum ratios in the $Q$ branches are similar to the overlapping H$_2$O emission lines, in addition to the aforementioned H I transition, CO$_2$ may be particularly dependent on the number of temperature components used to model the H$_2$O. We also find that the CO$_2$ emission line luminosities are slightly weaker than those from C$_2$H$_2$ and HCN and the emitting masses slightly higher (see Figure \ref{fig:allmols_Lemit_Memit}). These effects are discussed further in Section 4.1. 

\begin{deluxetable*}{lcccccc}
\tablecaption{Slab Model Fit Parameters: C$_2$H$_2$ \label{tab:slabparams_C2H2}}
\tablehead{
\colhead{Target} & \colhead{$\log N$} & \colhead{$r_{\text{slab}}$} & \colhead{$M_{\text{emit}}$} & \colhead{$T$} & \colhead{$L_{\text{emit}}$} & \colhead{P/C \tablenotemark{a}} \\
 & \colhead{(cm$^{-2}$)} & \colhead{(au)} & \colhead{$\log M$ ($M_{\oplus}$)} & \colhead{(K)} & \colhead{$\log L$ ($L_{\odot}$)} & }
\startdata
AS 205 N & 16.5 & 0.4 & -7.6 & $910 \pm 180$ & -3.2 & 0.068 \\
AS 205 S & \nodata & \nodata & $<-9.3$ & \nodata & $<-5.1$ & $<0.006$ \\
AS 209 & \nodata & \nodata & $<-8.7$ & \nodata & $<-4.5$ & $<0.009$\\
CI Tau & 15.6 & 0.6 & -8.2 & $920 \pm 160$ & -3.8 & 0.144 \\
DoAr 25 & 14.6 & 0.8 & $-8.9$ & $820 \pm 220$ & $-4.59$ & $0.083$ \\
DoAr 33\tablenotemark{b} & 15.0 & 0.9 & -8.4 & $730 \pm 10$ & -4.1 & 0.246 \\
Elias 20 & 15.4 & 1.1 & -7.8 & $940 \pm 100$ & -3.4 & 0.282 \\
Elias 24 & 15.8 & 0.9 & -7.6 & $890 \pm 120$ & -3.3 & 0.150 \\
Elias 27 & \nodata & \nodata & $<-9.2$ & \nodata & $<-4.8$ & $<0.042$ \\
FZ Tau & 15.2 & 0.9 & -8.3 & $920 \pm 160$ & -3.9 & 0.075 \\
GK Tau & 15.0 & 0.6 & -8.7 & $1060 \pm 80$ & -4.3 & 0.043 \\
GM Aur & \nodata & \nodata & $<-9.6$ & \nodata & $<-5.52$ & $<0.037$ \\
GO Tau & 15.2 & 0.5 & -8.8 & $620 \pm 80$ & -4.65 & 0.347 \\
GQ Lup & 14.9 & 0.7 & -8.6 & $1000 \pm 100$ & -4.2 & 0.048 \\
HD 142666 & \nodata & \nodata & $<-8.4$ & \nodata & $<-4.1$ & $<0.009$ \\
HD 143006 & \nodata & \nodata & $<-9.2$ & \nodata & $<-5.5$ & $<0.005$ \\
HD 163296 & \nodata & \nodata & \nodata & \nodata & \nodata & \nodata \\
HP Tau & 14.2 & 1.8 & -8.6 & $1130 \pm 100$ & -4.1 & 0.022 \\
HT Lup & 15.7 & 0.3 & -8.4 & $780 \pm 140$ & -4.2 & 0.023 \\
IQ Tau & 14.8 & 0.7 & -8.8 & $920 \pm 180$ & -4.4 & 0.060 \\
IRAS 04385 & 14.5 & 1.4 & -8.7 & $950 \pm 100$ & -4.3 & 0.040 \\
MWC 480 & 16.0 & 0.7 & -7.8 & $790 \pm 160$ & -3.5 & 0.04 \\
MY Lup & \nodata & \nodata & $<-9.3$ & \nodata & $<-5.0$ & $<0.030$ \\
RU Lup & 15.3 & 0.7 & $-8.1$ & $840 \pm 160$ & -3.8 & 0.041 \\
RY Lup & \nodata & \nodata & $<-9.4$ & \nodata & $<-5.0$ & $<0.007$ \\
SR 4 & 15.1 & 0.6 & -8.5 & $930 \pm 170$ & -4.2 & 0.047 \\
Sz 114\tablenotemark{c} & 14.5 & 1.1 & -8.8 & $1050 \pm 100$ & -4.3 & 0.097 \\
Sz 129 & 14.4 & 0.7 & $-9.2$ & $930 \pm 130$ & $<-4.8$ & $<0.062$ \\
TW Cha & 14.9 & 0.7 & -8.8 & $790 \pm 200$ & -4.5 & 0.098 \\
VZ Cha & 15.8 & 0.4 & -8.2 & $900 \pm 170$ & -3.8 & 0.145 \\
WSB 52 & 15.1 & 1.1 & -8.0 & $950 \pm 100$ & -3.6 & 0.210 \\
\enddata
\tablenotetext{a}{Peak/continuum ratio.}
\tablenotetext{b}{From \citet{colmenares24}: $\log N = 16.7$ cm$^{-2}$, $T = 550$ K, $r = 0.19$ au}
\tablenotetext{c}{From \citet{xie23}: $\log N = 15.5$ cm$^{-2}$, $T = 1400 K$, $r = 0.42$ au}
\tablecomments{Since $\log N$ and $r_{\rm{slab}}$ are degenerate, we do not include them in the analysis (see Section 3.3.3). The values reported here are those used to generate the slab models shown in Figures \ref{fig:organics_gallery_large} and \ref{fig:organics_gallery_small}. We use the set of slab model solutions within $<5\%$ of the minimum $L2$-norm test statistic for each disk to derive the median total number of molecules ($M_{\rm{emit}}$) and integrated emission line luminosities between 12-16 $\mu$m ($L_{\rm{emit}}$).}
\end{deluxetable*}

\begin{deluxetable*}{lcccccc}
\tablecaption{Slab Model Fit Parameters: HCN \label{tab:slabparams_HCN}}
\tablehead{
\colhead{Target} & \colhead{$\log N$} & \colhead{$r_{\text{slab}}$} & \colhead{$M_{\text{emit}}$} & \colhead{$T$} & \colhead{$L_{\text{emit}}$} & \colhead{P/C} \\
 & \colhead{(cm$^{-2}$)} & \colhead{(au)} & \colhead{$\log M$ ($M_{\oplus}$)} & \colhead{(K)} & \colhead{$\log L$ ($L_{\odot}$)} & }
\startdata
AS 205 N & 16.9 & 0.6 & -7.0 & $820 \pm 100$ & -3.1 & 0.112 \\
AS 205 S & \nodata & \nodata & $<-8.6$ & \nodata & $<-4.8$ & $<0.03$ \\
AS 209 & \nodata & \nodata & $<-8.1$ & \nodata & $<-4.3$ & $<0.01$ \\
CI Tau & 15.7 & 0.7 & -7.8 & $840 \pm 110$ & -3.9 & 0.163 \\
DoAr 25 & 17.0 & 0.2 & -7.8 & $910 \pm 100$ & -3.8 & 0.401 \\
DoAr 33\tablenotemark{a} & 14.8 & 1.4 & -8.1 & $690 \pm 100$ & -4.4 & 0.161 \\
Elias 20 & 15.3 & 2.3 & -7.2 & $650 \pm 100$ & -3.5 & 0.363 \\
Elias 24 & 15.3 & 2.6 & -7.1 & $730 \pm 100$ & -3.2 & 0.233 \\
Elias 27 & 15.3 & 0.6 & -8.3 & $770 \pm 140$ & -4.5 & 0.106 \\
FZ Tau & 15.0 & 2.3 & -7.7 & $800 \pm 140$ & -3.8 & 0.127 \\
GK Tau & 14.3 & 2.6 & -8.3 & $830 \pm 110$ & -4.4 & 0.043 \\
GM Aur & \nodata & \nodata & $<-9.4$ & \nodata & $<-5.6$ & $<0.03$ \\
GO Tau & 15.0 & 0.6 & -8.8 & $580 \pm 100$ & -5.3 & 0.078 \\
GQ Lup & 15.1 & 1.3 & -7.9 & $850 \pm 100$ & -4.0 & 0.097 \\
HD 142666 & \nodata & \nodata & $<-9.0$ & \nodata & $<-5.2$ & $<0.002$ \\
HD 143006 & \nodata & \nodata & $<-9.2$ & \nodata & $<-5.4$ & $<0.004$ \\
HD 163296 & \nodata & \nodata & \nodata & \nodata & \nodata & \nodata \\
HP Tau & 14.7 & 1.9 & -8.0 & $850 \pm 100$ & -4.1 & 0.039 \\
HT Lup & 16.6 & 0.4 & -7.7 & $930 \pm 100$ & -3.6 & 0.073 \\
IQ Tau & 15.1 & 1.1 & -8.1 & $940 \pm 100$ & -4.1 & 0.136 \\
IRAS 04385 & 14.6 & 2.7 & -7.9 & $680 \pm 100$ & -4.2 & 0.089 \\
MWC 480 & \nodata & \nodata & $<-8.0$ & \nodata & $<-4.2$ & $<0.01$ \\
MY Lup & 14.9 & 0.9 & -8.7 & $780 \pm 130$ & -4.8 & 0.046 \\
RU Lup & 15.1 & 2.5 & -7.4 & $820 \pm 100$ & -3.5 & 0.092 \\
RY Lup & \nodata & \nodata & $<-9.2$ & \nodata & $<-5.4$ & $<0.005$ \\
SR 4 & 15.0 & 1.1 & -8.2 & $840 \pm 130$ & -4.3 & 0.045 \\
Sz 114\tablenotemark{b} & 14.1 & 2.9 & -8.3 & $990 \pm 100$ & -4.2 & 0.155 \\
Sz 129 & \nodata & \nodata & $<-8.7$ & \nodata & $<-4.8$ & $<0.08$ \\
TW Cha & 15.1 & 1.5 & -7.9 & $810 \pm 120$ & -4.0 & 0.281 \\
VZ Cha & 17.5 & 0.2 & -7.8 & $830 \pm 100$ & -3.9 & 0.148 \\
WSB 52 & 16.1 & 0.8 & -7.3 & $710 \pm 100$ & -3.5 & 0.429 \\
\enddata
\tablenotetext{a}{From \citet{colmenares24}: $\log N = 14.0$ cm$^{-2}$, $T = 600$ K, $r = 3.95$ au}
\tablenotetext{b}{From \citet{xie23}: $\log N = 15.9$ cm$^{-2}$, $T = 870$ K, $r = 0.39$ au}
\tablecomments{Since $\log N$ and $r_{\rm{slab}}$ are degenerate, we do not include them in the analysis (see Section 3.3.3). The values reported here are those used to generate the slab models shown in Figures \ref{fig:organics_gallery_large} and \ref{fig:organics_gallery_small}. We use the set of slab model solutions within $<5\%$ of the minimum $L2$-norm test statistic for each disk to derive the median total number of molecules ($M_{\rm{emit}}$) and integrated emission line luminosities between 12-16 $\mu$m ($L_{\rm{emit}}$).}
\end{deluxetable*}

\begin{deluxetable*}{lcccccc}
\tablecaption{Slab Model Fit Parameters: CO$_2$ \label{tab:slabparams_CO2}}
\tablehead{
\colhead{Target} & \colhead{$\log N$} & \colhead{$r_{\text{slab}}$} & \colhead{$M_{\text{emit}}$} & \colhead{$T$} & \colhead{$L_{\text{emit}}$} & \colhead{P/C} \\
 & \colhead{(cm$^{-2}$)} & \colhead{(au)} & \colhead{$\log M$ ($M_{\oplus}$)} & \colhead{(K)} & \colhead{$\log L$ ($L_{\odot}$)} & }
\startdata
AS 205 N & 16.5 & 1.1 & -6.5 & $550 \pm 120$ & -3.3 & 0.122 \\
AS 205 S & 14.9 & 1.8 & -7.8 & $800 \pm 420$ & -4.4 & 0.022 \\
AS 209 & \nodata & \nodata & $<-7.9$ & \nodata & $<-4.1$ & $<0.01$ \\
CI Tau & 15.1 & 1.1 & -8.0 & $810 \pm 390$ & -4.5 & 0.060 \\
DoAr 25 & 14.2 & 1.7 & -8.4 & $680 \pm 100$ & -5.0 & 0.055 \\
DoAr 33\tablenotemark{a} & 14.8 & 0.9 & -8.3 & $560 \pm 100$ & -5.1 & 0.064 \\
Elias 20 & 14.9 & 1.6 & -7.7 & $570 \pm 100$ & -4.5 & 0.068 \\
Elias 24 & 16.0 & 0.6 & -7.3 & $640 \pm 220$ & -3.9 & 0.066 \\
Elias 27 & 15.7 & 0.6 & -8.0 & $1070 \pm 380$ & -4.4 & 0.118 \\
FZ Tau & 14.7 & 2.8 & -7.3 & $820 \pm 310$ & -3.8 & 0.168 \\
GK Tau & 14.3 & 2.5 & -8.4 & $990 \pm 450$ & -4.9 & 0.017 \\
GM Aur & \nodata & \nodata & $<-9.0$ & \nodata & $<-5.6$ & $<0.03$ \\
GO Tau & \nodata & \nodata & $<-9.1$ & \nodata & $<-5.6$ & $<0.03$ \\
GQ Lup & 15.6 & 0.6 & -8.3 & $760 \pm 330$ & -5.0 & 0.020 \\
HD 142666 & \nodata & \nodata & $<-8.8$ & \nodata & $<-5.3$ & $<0.001$ \\
HD 143006 & \nodata & \nodata & $<-8.4$ & \nodata & $<-4.7$ & $<0.01$ \\
HD 163296 & \nodata & \nodata & \nodata & \nodata & \nodata & \nodata \\
HP Tau & 15.7 & 1.3 & -7.1 & $330 \pm 100$ & -4.5 & 0.051 \\
HT Lup & 16.6 & 0.5 & -7.4 & $540 \pm 100$ & -4.1 & 0.056 \\
IQ Tau & \nodata & \nodata & $<-8.7$ & \nodata & $<-5.3$ & $<0.02$ \\
IRAS 04385 & 16.1 & 0.5 & -7.6 & $480 \pm 110$ & -4.4 & 0.086 \\
MWC 480 & \nodata & \nodata & $<-8.3$ & \nodata & $<-4.9$ & $<0.005$ \\
MY Lup & 19.9 & 0.2 & -7.8 & $340 \pm 140$ & -4.5 & 0.134 \\
RU Lup & 16.5 & 0.3 & -7.5 & $760 \pm 210$ & -4.0 & 0.034 \\
RY Lup & \nodata & \nodata & $<-8.8$ & \nodata & $<-5.2$ & $<0.006$ \\
SR 4 & \nodata & \nodata & $<-8.3$ & \nodata & $<-4.8$ & $<0.02$ \\
Sz 114\tablenotemark{b} & 17.3 & 0.5 & -6.3 & $340 \pm 100$ & -4.2 & 0.378 \\
Sz 129 & \nodata & \nodata & $<-9.0$ & \nodata & $<-5.3$ & $<0.02$ \\
TW Cha & \nodata & \nodata & $<-8.7$ & \nodata & $<-5.3$ & $<0.03$ \\
VZ Cha & \nodata & \nodata & $<-8.4$ & \nodata & $<-4.9$ & $<0.03$ \\
WSB 52 & 15.5 & 2.4 & -6.8 & $510 \pm 100$ & -3.8 & 0.410 \\
\enddata
\tablenotetext{a}{From \citet{colmenares24}: $\log N = 18.0$ cm$^{-2}$, $T = 225$ K, $r = 0.51$ au}
\tablenotetext{b}{From \citet{xie23}: $\log N = 17.6$ cm$^{-2}$, $T = 500$ K, $r = 0.18$ au}
\tablecomments{Since $\log N$ and $r_{\rm{slab}}$ are degenerate, we do not include them in the analysis (see Section 3.3.3). The values reported here are those used to generate the slab models shown in Figures \ref{fig:organics_gallery_large} and and \ref{fig:organics_gallery_small}. We use the set of slab model solutions within $<5\%$ of the minimum $L2$-norm test statistic for each disk to derive the median total number of molecules ($M_{\rm{emit}}$) and integrated emission line luminosities between 12-16 $\mu$m ($L_{\rm{emit}}$).}
\end{deluxetable*}

\subsection{A Comparison of Molecule Detection Rates from Spitzer and JWST}

\begin{figure}
\centering
\epsscale{1.1}
\plotone{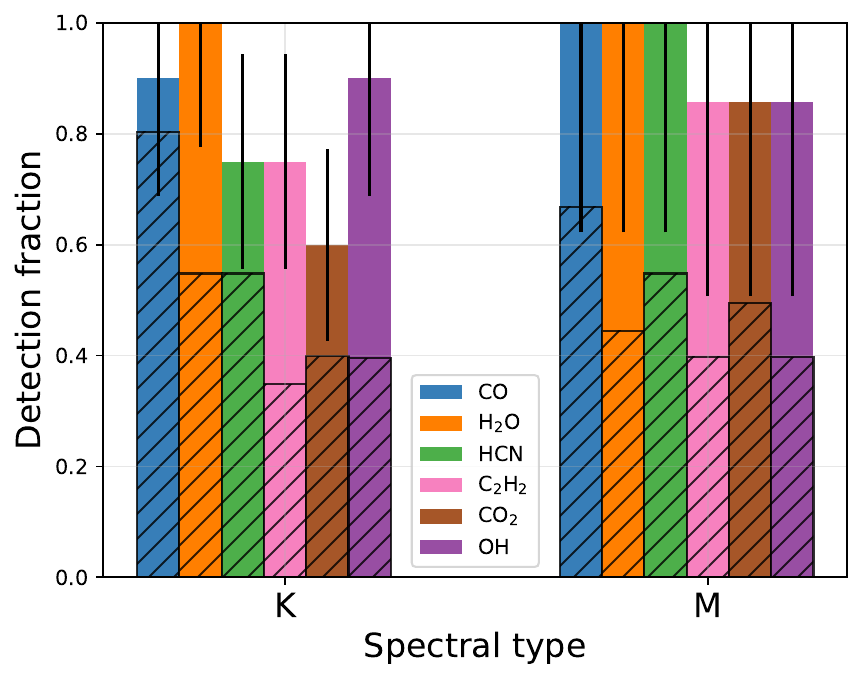}
\caption{Detection rates of CO, H$_2$O, HCN, C$_2$H$_2$, CO$_2$ and OH
for K and M stars in this sample of MRS spectra.  Hatches show IRS or ground-based (for CO) detection rates from \citet{pontoppidan10b}.  Earlier spectral types are excluded from this plot due to low number statistics in the JDISCS sample.}
\label{fig:histograms}
\end{figure}

Figure \ref{fig:histograms} compares the detection rates of H$_2$O, OH, C$_2$H$_2$, HCN, and CO$_2$ in K and M stars from this work to the \emph{Spitzer-IRS} sample from \citet{pontoppidan10b}. The detection rates with MIRI-MRS are significantly higher for all five molecules; however, we note that there are only eight targets in this paper that were also included in \citet{pontoppidan10b} and \citet{salyk11}. The disks with new detections that we report here, which were non-detections with \emph{Spitzer}, are HT Lup (H$_2$O, HCN, C$_2$H$_2$), DoAr 25 (H$_2$O, CO$_2$), RY Lup (H$_2$O), and RU Lup (C$_2$H$_2$). 

DoAr 25 and RY Lup both have positive infrared spectral indices ($n_{13-31} > 0$), which classified them as transition disks in the \emph{Spitzer} era. While transition disks were expected to be molecule-poor, the detections in both sources indicate that MIRI-MRS is sensitive to smaller line-to-continuum ratios than the IRS (see also, \citealt{perotti23}). The remaining disks have detections and non-detections that are consistent between \emph{JWST} and \emph{Spitzer}; notably, CO$_2$ emission is still not detected in RY Lup, TW Cha, or VZ Cha. The overall factor of two increase in CO$_2$ detection rates may be attributed to the improvement in spectral resolution and signal-to-noise ratios provided by the MRS. Since fewer disks in Figure \ref{fig:detection_limits} have CO$_2$ peak-to-continuum ratios $>0.1$ than C$_2$H$_2$ or HCN peak-to-continuum ratios $>0.1$, the CO$_2$ $Q$ branch lines that are now readily detected were likely too weak to resolve with \emph{Spitzer}.    

\subsection{An Inventory of Atomic Jet, Wind, and Accretion Tracers}

Figure \ref{fig:CO_OH_NeII_NeIII_gallery} shows a gallery of [Ne II] and [Ne III] atomic emission lines from JDISCS targets. The other forbidden emission lines all overlap with transitions from water and other organics in the molecule-rich JDISCS sources. In Channel 1, emission is detected near the [Fe II] transition at 5.340 $\mu$m in 22/31 JDISCS sources, but the feature is masked by CO $v = 1-0$ P(57) emission \citep[see Figure 1 in][]{banzatti24}. The [Fe II] emission line at 24.519 $\mu$m also overlaps with a H$_2$O transition \citep[see Appendix D in][]{banzatti24}, and we find that 22/31 targets have a broad emission feature at this wavelength ($FWHM \sim 350$ km s$^{-1}$). The H I Pf$\alpha$ (7.456 $\mu$m), Humphreys $\beta$ (7.503 $\mu$m), and Humphreys $\alpha$ (12.372 $\mu$m) emission lines are all blended with ro-vibrational ($6_{3 \, 4}-7_{4\, 3}$ at 7.460 $\mu$m; $7_{2 \, 5}-7_{5 \, 2}$ at 7.503 $\mu$m) and rotational H$_2$O features ($16_{4 \, 13}-15_{1 \, 14}$ at 12.375 $\mu$m; see Figure \ref{fig:example_retrieval} and Figures 1--4 in \citet{banzatti24}). Some HI lines are extracted and used to estimate the accretion luminosity in another work (Tofflemire et al. 2025, in press). Since a full treatment of the ro-vibrational and rotational water emission lines is beyond the scope of this work \citep[see e.g.,][]{romero-mirza24b, banzatti24}, our analysis is focused on the [Ne II] and [Ne III] emission lines, as they can be ``cleaned" of contaminating emission lines using the slab model fits described above. 

Figure \ref{fig:example_retrieval} shows that the [Ne II] 12.814 $\mu$m feature overlaps with weak C$_2$H$_2$ and HCN $R$ branch emission lines, which are generally well subtracted from the data using the best-fit slab models described in Section 3.2.2. Clean [Ne II] emission lines are then detected in nearly all sources included in this work, with the exception of three of the four Herbig disks (HD 143006, HD 142666, HD 163296). We use the interactive fitting tool iSLAT \citep{jellison2024} to fit Gaussian emission line profiles to the atomic lines; the corresponding emission line fluxes, velocity centroids, and FWHMs are reported in Table \ref{tab:atomicfluxes}. Across the sample, we find an average FWHM of $170 \pm 50$ km s$^{-1}$ and velocity centroid of $\sim 10$ km s$^{-1}$. Although the [Ne II] emission line is a critical tracer of outflowing disk material \citep{pascucci07, pascucci20, najita09}, higher spectral resolution than provided by MIRI-MRS is required to extract kinematic information. We note that the aperture used to extract the 1-D spectra also may not include spatially extended emission from jets or winds (see Section 2.2), which is likely the reason the average velocity centroid is not blue-shifted (see e.g., \citealt{xie23, pontoppidan24, bajaj24, arulanantham2024, schwarz24}). MWC 480 and RU Lup, which has a velocity-resolved MHD disk wind \citep{whelan21} and variability in outflow- and accretion-tracing UV and optical emission lines \citep{Herczeg05, stock22}, show the only significant blueshifted velocity centroids at $v_c = -175$ km s$^{-1}$ and $v_c = -140$ km s$^{-1}$, respectively, indicating that these two targets have significant outflowing emission originating close to the stars (within the aperture used to extract the 1-D spectra). Indeed, RU Lup only shows a [Ne II] high-velocity component (HVC) in high-resolution ground-based spectroscopy \citep{pascucci20}. The same effect is likely responsible for the difference of $\sim$50 km s$^{-1}$ between the velocity centroids measured for Sz 114 in this work and \citet{xie23}; we note that the difference is well within the spectral resolution limit at 12.814 $\mu$m.  

The [Ne III] 15.555 $\mu$m emission line overlaps with a strong rotational H$_2$O feature ($17_{10 \, 7}-16_{9 \, 8}$ at 15.554 $\mu$m). We subtract the best-fit single temperature component water model from each spectrum (see Section 3.2.1) and fit the residual spectra with Gaussian emission line profiles using iSLAT. Ten disks show clean [Ne III] detections: DoAr 33, GK Tau, GM Aur, GO Tau, GQ Lup, IQ Tau, MY Lup, RY Lup, Sz 129, and WSB 52. No significant blueshifts are detected, and we find an average Gaussian emission line width of 115 km s$^{-1}$. Notably, the only three disks in our sample with clear C$_2$H$_2$ (DoAr 33, GO Tau; \citealt{colmenares24}) and CO$_2$ (MY Lup; \citealt{salyk25}) isotopologue detections all have [Ne III] emission that dominates over the water transition. The integrated emission line fluxes, velocity centroids, and FWHMs are reported in Table \ref{tab:atomicfluxes}.

Lastly, we report the detection of strong [Ar II] emission at 6.985 $\mu$m in HD 143006, MWC 480, and MY Lup and marginal detections in IQ Tau and RY Lup (see Figure \ref{fig:herbig_cavity_water}), which is also expected to trace disk winds \citep{bajaj24, sellek24} or jets \citep{arulanantham2024}. In the rest of the sample, the H$_2$O $9_{1 \, 9}-9_{2 \, 8}$ ro-vibrational emission line dominates any weak emission from [Ar II]. As with the [Ne III], no significant blueshifts are detected, although again the aperture may not include spatially extended emission (see e.g., \citealt{worthen2024, bajaj24, arulanantham2024}). We do not detect [Ar III] emission in any of the targets included in this work (see e.g., \citealt{bajaj24} for a detection in the transition disk T Cha).  

\begin{deluxetable*}{l|ccc|ccc|ccc}
\tablecaption{Atomic Emission Line Fluxes ($10^{-14}$ erg s$^{-1}$ cm$^{-2}$) \label{tab:atomicfluxes}}
\tablehead{
\colhead{Target\tablenotemark{$\ast$}} & \colhead{} & \colhead{[Ne II]} & \colhead{} & \colhead{} & \colhead{[Ne III]} & \colhead{} & \colhead{} & \colhead{[Ar II]} & \colhead{} \\
\colhead{} & \colhead{$F$} & \colhead{$v_c$} & \colhead{FWHM} & \colhead{$F$} & \colhead{$v_c$} & \colhead{FWHM} & \colhead{$F$} & \colhead{$v_c$} & \colhead{FWHM}}
\startdata
AS 205 N\tablenotemark{a} & 1.77 & 3 & 225 & \nodata & \nodata & \nodata & \nodata & \nodata & \nodata \\
AS 205 S & 0.29 & 75 & 249 & $<0.22$ & 103 & 75 & \nodata & \nodata & \nodata \\
AS 209 & 2.38 & -100 & 166 & \nodata & \nodata & \nodata & \nodata & \nodata & \nodata \\
CI Tau & 0.32 & 2 & 193 & $<0.27$ & 99 & 104 & \nodata & \nodata & \nodata \\
DoAr 25 & 0.24 & 19 & 101 & \nodata & \nodata & \nodata & \nodata & \nodata & \nodata \\
DoAr 33 & 0.27 & 21 & 116 & 0.06 & 6 & 142 & \nodata & \nodata & \nodata \\
Elias 20 & 0.44 & -43 & 164 & $<0.17$ & 85 & 83 & \nodata & \nodata & \nodata \\
Elias 24 & 1.80 & -76 & 143 & $<0.72$ & 63 & 106 & \nodata & \nodata & \nodata \\
Elias 27 & 0.25 & -16 & 207 & $<0.19$ & 77 & 96 & \nodata & \nodata & \nodata \\
FZ Tau & 1.04 & 107 & 214 & \nodata & \nodata & \nodata & \nodata & \nodata & \nodata \\
GK Tau & 0.49 & 67 & 184 & 0.21 & 74 & 119 & \nodata & \nodata & \nodata \\
GM Aur & 0.83 & 7.3 & 120 & 0.13 & -32 & 149 & 0.26 & 29 & 118 \\
GO Tau & 0.10 & 43 & 95 & 0.02 & 13 & 146 & \nodata & \nodata & \nodata \\
GQ Lup & 0.26 & -4 & 188 & 0.20 & 49 & 134 & \nodata & \nodata & \nodata \\
HD 142666 & 0.41 & 65 & 127 & \nodata & \nodata & \nodata & \nodata & \nodata & \nodata \\
HD 143006 & 0.48 & -7 & 108 & \nodata & \nodata & \nodata & 0.43 & 5 & 93 \\
HD 163296 & \nodata & \nodata & \nodata & \nodata & \nodata & \nodata & \nodata & \nodata & \nodata \\
HP Tau & 0.38 & 38 & 108 & $<0.1$ & 66 & 29 & \nodata & \nodata & \nodata \\
HT Lup & $0.62$ & 11 & 197 & $<0.20$ & 114 & 77 & \nodata & \nodata & \nodata \\
IQ Tau & 1.06 & 22 & 105 & 0.17 & 88 & 131 & $<0.28$ & -6 & 114 \\
IRAS 04385 & 0.45 & -25 & 221 & $<0.25$ & -10 & 202 & \nodata & \nodata & \nodata \\
MWC 480 & 2.03 & -175 & 181 & \nodata & \nodata & \nodata & 3.35 & 46 & 126 \\
MY Lup & 1.84 & 32 & 99 & 0.21 & -10 & 124 & 0.23 & 11 & 98 \\
RU Lup & 3.18 & -142 & 230 & \nodata & \nodata & \nodata & \nodata & \nodata & \nodata \\
RY Lup & 0.42 & 37 & 109 & 0.08 & 18 & 196 & $<0.43$ & -6 & 161 \\
SR 4 & 1.39 & -45 & 196 & $<0.12$ & 114 & 33 & \nodata & \nodata & \nodata \\
Sz 114 & 0.23 & -22 & 164 & \nodata & \nodata & \nodata & \nodata & \nodata & \nodata \\
Sz 129 & 0.25 & 50 & 142 & 0.07 & 75 & 114 & \nodata & \nodata & \nodata \\
TW Cha & 0.32 & 76 & 244 & \nodata & \nodata & \nodata & \nodata & \nodata & \nodata \\
VZ Cha & 0.19 & 15 & 198 & \nodata & \nodata & \nodata & \nodata & \nodata & \nodata \\
WSB 52 & 2.24 & 69 & 123 & 0.92 & 42 & 126 & \nodata & \nodata & \nodata \\
\enddata
\tablenotetext{\ast}{All emission line properties were measured using the iSLAT tool \citep{jellison2024}, after subtracting the best-fit H$_2$O, C$_2$H$_2$, HCN, and CO$_2$ slab models. Emission line fluxes ($F$) are reported in units of $10^{-14}$ erg s$^{-1}$ cm$^{-2}$. Velocity centroids ($v_c$) and Gaussian FWHMs are reported in units of km s$^{-1}$, where negative $v_c$ corresponds to a blueshift.}
\tablenotetext{a}{Upper limits are reported for emission lines where residual flux was detected after subtracting the best-fit water model; no values are reported for targets where residual flux was not detected.}
\end{deluxetable*}

\section{Analysis} \label{sec:analysis}

In this section, we explore the connection between the observed molecular gas emission lines and the underlying chemical conditions within the warm inner disks. Thermochemical modeling work suggests that the total emission line fluxes are sensitive to both the temperatures and sizes of the emitting areas, along with the C/O elemental abundance ratios \citep[see e.g.][]{najita11,walsh15,woitke18,anderson21} and the hydrocarbon chemical network \citep{kanwar24a}. As discussed extensively in Section \ref{sec:results}, significant overlap between the $P$ and $R$ branch transitions of C$_2$H$_2$, HCN, and CO$_2$ makes it challenging to measure total emission line fluxes for the individual molecules directly from the data. Instead, we explore these trends using the emission line luminosities reported in Tables \ref{tab:slabparams_C2H2}, \ref{tab:slabparams_HCN}, and \ref{tab:slabparams_CO2}, which are the median values across the set of slab models with $L2$-norm test statistics within $<5\%$ of the ``best-fit" solution for each disk. This captures all degenerate parameter combinations, which contribute the largest source of uncertainty to the retrievals. The model emission line luminosities were derived by integrating the spectrum between $12-16$ $\mu$m, so we note that this is a wavelength-dependent luminosity for the mid-infrared component produced by each molecule.

\subsection{Dependence of Emission Line Luminosities on Slab Model Parameters}
\begin{figure*}
\centering
\epsscale{1.2}
\plotone{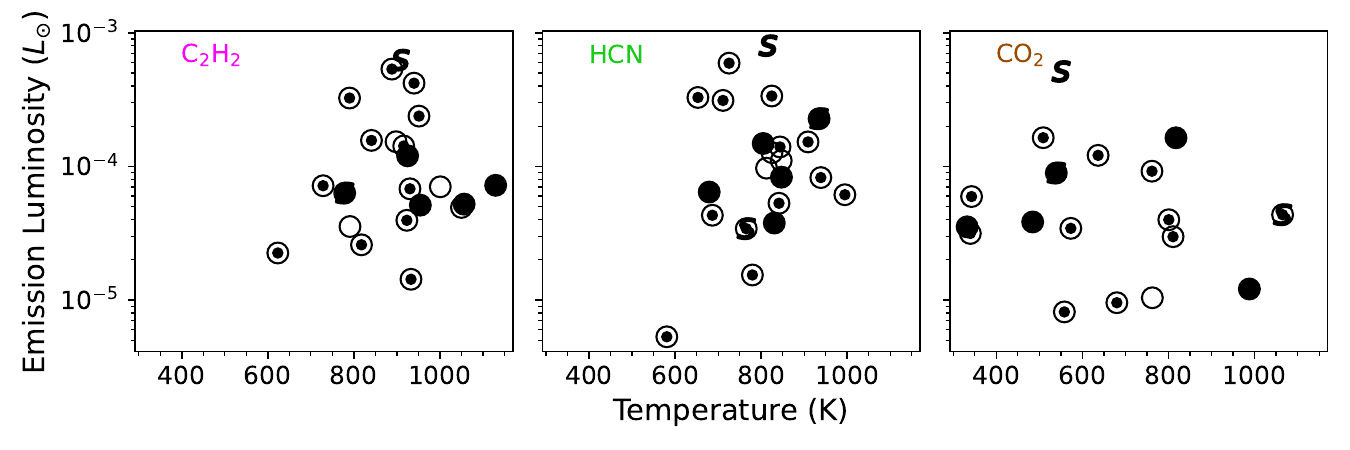}
\caption{Median emission line luminosities from C$_2$H$_2$ (left), HCN (middle), and CO$_2$ (right) versus the median slab temperatures measured from the 100 best-fit slab models for each disk, with markers representing the sub-mm dust substructures listed in Table \ref{tab:diskprops}. The slab temperatures and emission line luminosities are not correlated for any of the three molecules (Spearman $\rho = 0.006, 0.07, -0.1$; $p = 0.9, 0.8, 0.6$, respectively), implying that brighter emission lines do not necessarily originate in hotter gas.}
\label{fig:Lemit_vs_T}
\end{figure*}

Figure \ref{fig:Lemit_vs_T} compares the integrated emission line luminosities to the retrieved slab model temperatures for C$_2$H$_2$, HCN, and CO$_2$. We use the Spearman rank coefficient $\left( \rho \right)$ to identify monotonic relationships between pairs of parameters, where values closer to $\pm 1$ indicate that the values increase/decrease together (or inversely) and values closer to 0 are measured when no relationship is detected between the parameters. A threshold of $p < 0.05$ is typically indicative of a statistically significant (i.e., real) relationship. We find that the emission line luminosities and slab model temperatures are tentatively correlated for C$_2$H$_2$ (Spearman $\rho = 0.4$; $p = 0.04$) but not correlated for HCN or CO$_2$ (Spearman $\rho = 0.2, -0.3$; $p = 0.3, 0.1$). This indicates that physically warmer gas alone does not necessarily produce brighter mid-infrared emission lines. 

\begin{figure*}
\centering
\epsscale{1.2}
\plotone{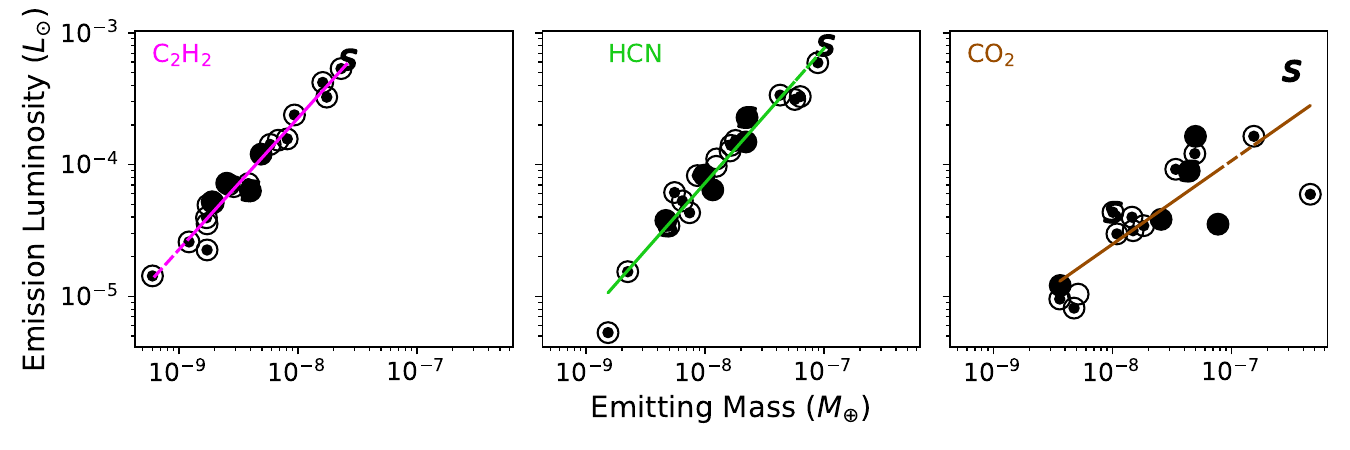}
\caption{Median emission line luminosities from C$_2$H$_2$ (left), HCN (middle), and CO$_2$ (right) versus the median emitting masses measured from the best-fit slab models for each disk, with marker styles indicating the sub-mm dust substructures listed in Table \ref{tab:diskprops}. We report statistically significant, positive correlations and linear relationships between the two parameters for all three molecules (see Equation 1), implying that the emission is optically thin to moderately optically thick (Spearman $\rho = 0.97, 0.98, 0.82$).}
\label{fig:Lemit_vs_Memit}
\end{figure*}

\begin{figure*}
\centering
\epsscale{1.2}
\plotone{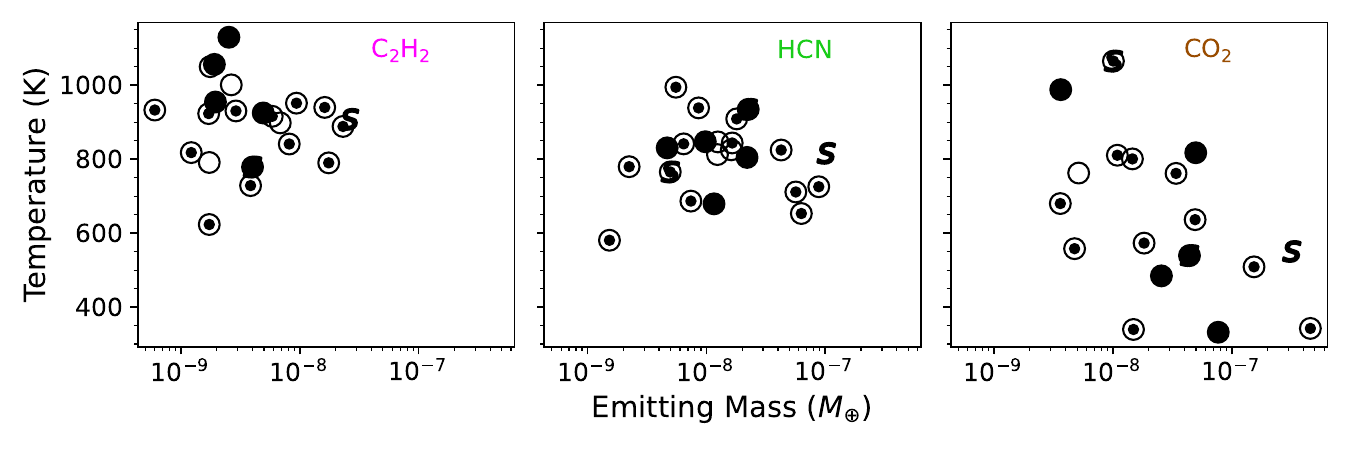}
\caption{Retrieved temperatures from C$_2$H$_2$ (left), HCN (middle), and CO$_2$ (right) versus the median emitting masses measured from the best-fit slab models for each disk, with marker styles indicating the sub-mm dust substructures listed in Table \ref{tab:diskprops}. No significant correlations are observed for C$_2$H$_2$ or HCN (Spearman $\rho = -0.1, -0.04$; $p = 0.6, 0.8$), but we report a significant negative correlation between CO$_2$ temperature and emitting mass (Spearman $\rho = -0.55$; $p = 0.02$).}
\label{fig:T_vs_Memit}
\end{figure*}

Instead, we find strong statistically significant positive correlations between the model emission line luminosities and emitting masses for all three molecules (see Figure \ref{fig:Lemit_vs_Memit}; Spearman $\rho = 0.98, 0.99, 0.93$). A linear regression analysis of the two parameters in log-space for each molecule returns:
\begin{eqnarray}
\log L \left(\rm{C_2H_2}\right) = 1.0 \times \log M \left(\rm{C_2H_2}\right) + 4.3 \\
\log L \left(\rm{HCN}\right) = 1.0 \times \log M \left(\rm{HCN}\right) + 4.0 \nonumber \\
\log L \left(\rm{CO_2}\right) = 0.6 \times \log M \left(\rm{CO_2}\right) + 0.5 \nonumber
\end{eqnarray}
The relationships suggest that the emitting regions are optically thin, as an increase in emitting mass leads directly to an increase in luminosity. This effect has been reported for previous studies of protoplanetary disks with both \emph{Spitzer} and MIRI \citep{salyk11, carr11, najita13, ramireztannus23, xie23, grant2024}. We note that the slope of the linear relationship is shallower for CO$_2$, as may be expected if the lines originate in cooler, more optically thick gas (see also, \citealt{lahuis06}). CO$_2$ is also the only molecule that shows an anti-correlation between the slab temperatures and emitting masses (see Figure \ref{fig:T_vs_Memit}; Spearman $\rho = -0.55$; $p = 0.003$), which may indicate that the emission originates in a deeper, more optically thick layer than the C$_2$H$_2$ and HCN (see e.g., \citealt{salyk25}).  

\begin{figure}
\epsscale{1.25}
\plotone{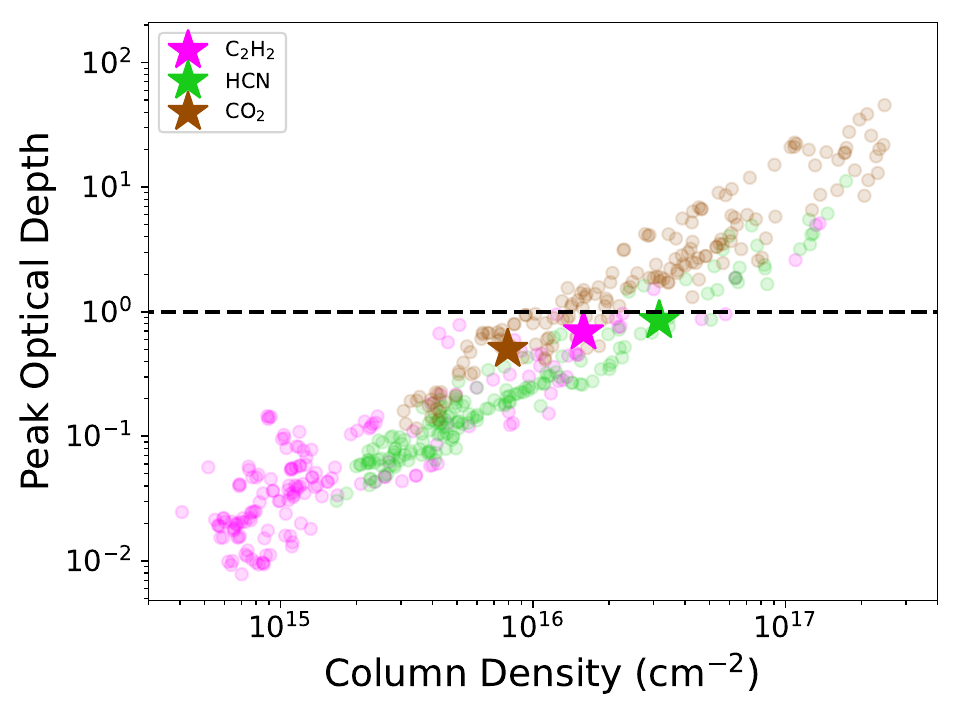}
\caption{Optical depths versus column densities for the combinations of C$_2$H$_2$ (magenta), HCN (green), and CO$_2$ (brown) slab model parameters that best reproduce the spectrum of AS 205N; markers with stars correspond to the column densities used in Figure \ref{fig:example_retrieval}, and the black, dashed line shows $\tau = 1$. In general, the C$_2$H$_2$ and HCN slab models are optically thin to larger column densities than the CO$_2$. We note that the $^{13}$CO$_2$ $Q$ branch near 15.4 $\mu$m is not detected in this target (see e.g., \citealt{grant23, xie23, salyk25}), or the $^{13}$C$^{12}$CH$_2$ $Q$ branch seen in DoAr 33 and GO Tau \citep{tabone23, kanwar24, arabhavi24, colmenares24}.}
\label{fig:optdepth_demo}
\end{figure}

Figure \ref{fig:optdepth_demo} shows the optical depth versus column density for all combinations of C$_2$H$_2$, HCN, and CO$_2$ slab model parameters that best reproduce the spectrum of AS 205N, confirming that the gas is optically thin and that the emission line strength is directly tracing the underlying warm molecular mass. The optical depths of C$_2$H$_2$ and HCN are always $\tau < 1$, up to column densities of $\sim 10^{16}$ cm$^{-2}$. The CO$_2$ solutions exceed $\tau > 1$ beyond $\sim 2 \times 10^{16}$ cm$^{-2}$, tentatively indicating more optically thick emission. However, we note that the $^{13}$CO$_2$ $Q$ branch is not detected in AS 205N, indicating that the true column densities are likely more consistent with the smaller values (see e.g., \citealt{xie23, salyk25}). By contrast, $R > 30000$ spectroscopy of H$_2$O in disks demonstrates that the gas is optically thick (see e.g., \citealt{banzatti23a}). In this work, we also report a weaker correlation between the H$_2$O emission line luminosities and emitting masses (Spearman $\rho = 0.66$) than is observed for the C$_2$H$_2$, HCN, and CO$_2$ emission (Spearman $\rho = 0.97, 0.98, 0.82$, respectively), which further supports this picture of optically thicker water and optically thinner organics.  

\begin{deluxetable*}{ccccccccc}
\tablecaption{Correlations Between Retrieval Parameters and Disk/Stellar Properties \label{tab:correlations}}
\tablehead{\colhead{Parameter\tablenotemark{$\ast$}} & \colhead{Molecule\tablenotemark{$\ast \ast$}} & \colhead{$r_{\rm{dust}}$} & \colhead{$r_{\rm{inner}}$} & \colhead{$n_{13-26}$} & \colhead{$M_{\ast}$} & \colhead{$\dot{M}_{\rm{acc}}$} & \colhead{[Ne II] Flux} & \colhead{[Ne III]/[Ne II]}  }
\startdata
Line Luminosity & C$_2$H$_2$ & -0.084 & 0.143 & \textbf{-0.538} & 0.367 & \textbf{0.642} & \textbf{0.583} & 0.250 \\
 & HCN & 0.053 & -0.015 & -0.331 & 0.259 & \textbf{0.504} & 0.400 & 0.392 \\
& CO$_2$ & 0.034 & 0.221 & -0.081 & -0.106 & \textbf{0.503} & \textbf{0.569} & -0.120 \\
\hline
Temperature & C$_2$H$_2$ & \textbf{-0.540} & -0.403 & 0.149 & -0.207 & 0.064 & 0.005 & \textbf{0.821} \\
 & HCN & -0.107 & 0.009 & 0.136 & -0.112 & -0.240 & -0.102 & 0.107 \\
& CO$_2$ & 0.109 & 0.200 & -0.262 & -0.156 & 0.479 & -0.181 & 0.800 \\
\hline
Emitting Mass & C$_2$H$_2$ & -0.036 & 0.181 & \textbf{-0.531} & 0.381 & \textbf{0.675} & \textbf{0.590} & 0.321 \\
 & HCN & 0.033 & -0.033 & -0.381 & 0.299 & \textbf{0.523} & 0.397 & 0.393 \\
& CO$_2$ & -0.123 & 0.042 & -0.076 & -0.125 & 0.147 & 0.424 & -0.300  \\
\enddata
\tablenotetext{\ast}{Spearman rank coefficients; numbers in bold are correlations where $p < 0.05$.}
\tablenotetext{\ast \ast}{The reported correlation coefficients exclude targets with non-detections.}
\end{deluxetable*}

\subsection{Correlations Between Slab Model Parameters and Disk and Stellar Properties}
Table \ref{tab:correlations} presents the Spearman rank coefficients between the emission line luminosities, slab temperatures, emitting masses, and the disk and stellar properties and atomic line fluxes included in Tables \ref{tab:targprops}, \ref{tab:diskprops}, and \ref{tab:atomicfluxes}. We do not identify any statistically significant correlations between the slab model temperatures and the radii of the innermost sub-mm dust rings, the infrared spectral indices, the mass accretion rates, or the [Ne II] emission line fluxes. There is a significant negative correlation between the C$_2$H$_2$ temperatures and the sub-mm disk radii (Spearman $\rho = -0.540$), indicating a cooler emitting layer in larger disks. However, the correlation is not present with the HCN or CO$_2$ emission line luminosities. 

\begin{figure*}
\epsscale{1.2}
\plotone{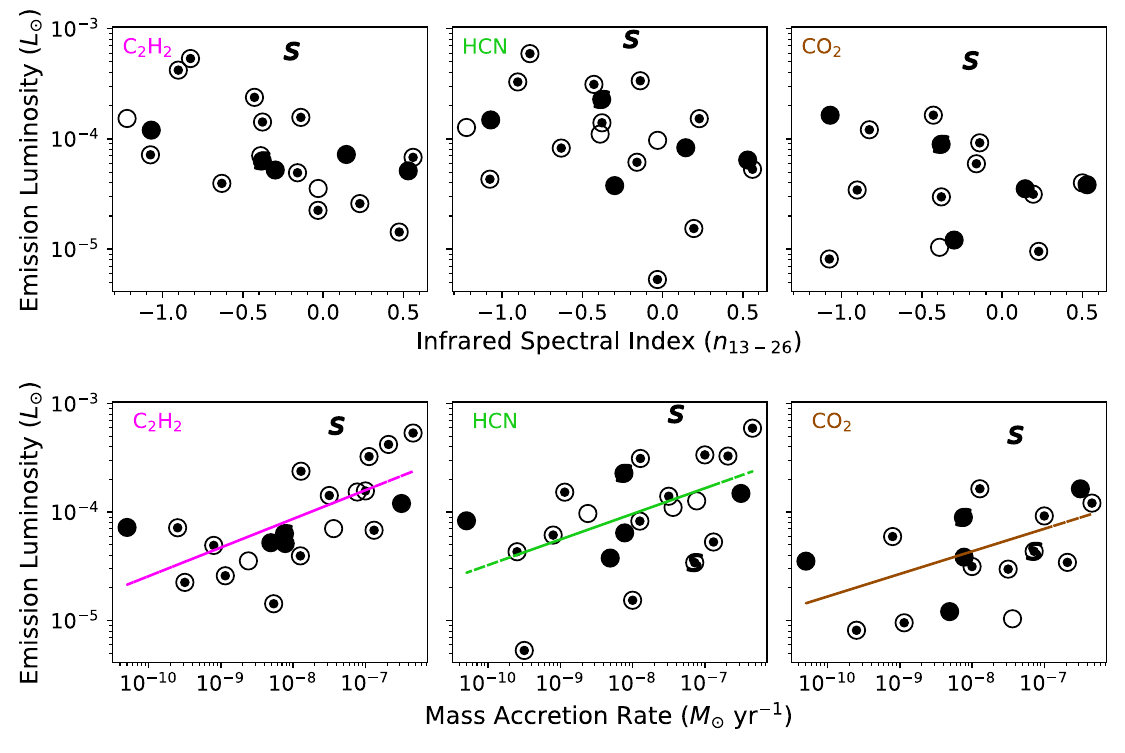}
\caption{Emission line luminosities measured from the best-fit slab models versus the infrared spectral indices measured between 13 and 26 $\mu$m \emph{(top)} and the literature mass accretion rates \emph{(bottom)}. Both trends are consistent with the \emph{Spitzer} results reported in \citet{banzatti2020}.}
\label{fig:correlations}
\end{figure*}

The most significant correlations are found between the emission line luminosities for all three molecules and the mass accretion rates, consistent with the \emph{Spitzer} findings reported by \citet{banzatti2020}. We also report significant anti-correlations between the C$_2$H$_2$, HCN, and CO$_2$ luminosities and the infrared spectral indices, again consistent with \emph{Spitzer} results reported in \citet{banzatti2020}, suggesting a depletion of warm molecular gas as the inner disk clears (see Figure \ref{fig:correlations}). Although the [Ne II] fluxes reported here do not include all spatially extended emission contained within the MRS field-of-view, they are also significantly positively correlated with the mass accretion rates (Spearman $\rho = 0.509$; $p = 0.005$), as expected if the bulk of the outflowing emission originates in the high-velocity component identified from spectrally resolved line profiles \citep{pascucci20}. The emission line luminosities of C$_2$H$_2$ and CO$_2$ are also correlated with the integrated [Ne II] fluxes, and the C$_2$H$_2$ temperatures are correlated with the [Ne III]/[Ne II] flux ratios. These trends are discussed further in the following section. 

\section{Discussion} \label{sec:discussion}

\subsection{Can we observe the influence of outer disk substructures on inner disk chemistry?}

As described in Section 3.1, sub-mm surveys at high angular resolution now provide a critical context for examining the impact of dust rings, gaps, cavities, and spiral arms on inner disk chemistry. Perhaps surprisingly, demographic studies with \emph{Spitzer} found a connection between the MIR spectroscopic properties of disks (i.e., arising from the inner au) and their properties at millimeter wavelengths (i.e., arising from $>10$ au). In particular, the flux ratio of HCN to H$_2$O is found to generally increase with disk mass \citep{carr11, najita13} and with the size of the dust disk at millimeter wavelengths \citep{banzatti2020}.

Some of the diversity in the MIR spectra of T Tauri disks could be driven by planet formation processes through their impact on water transport in disks \citep{najita13}. While the formation of large, non-migrating, icy objects (planetesimals and larger objects) can sequester oxygen beyond the snow line, their associated substructures, created either as a consequence of planet formation (spiral arms, gaps, rings, cavities) or as precursors to it (dust traps; see e.g., \citealt{bae2023} and references therein), can also inhibit the inward drift of small icy solids to the inner disk \citep{kalyaan21, kalyaan23}. In the case of multiple gaps within a single disk, the depth and radial location of the innermost gap remain the main factors that determine whether or not a disk is water-rich \citep{easterwood24}.  

Because the general consequence of these planet formation processes is to inhibit inward water transport, disks in which these processes along with carbon grain destruction inside the soot line are active will tend to have higher C/O ratios in their inner disks \citep{kress10, carr11, najita13, booth17, booth19, arabhavi24, colmenares24}. In contrast, disks that have not formed planets, contain leaky gaps, or otherwise have not trapped icy solids are likely to instead experience significant inward icy pebble drift, leading to an oxygen-rich inner disk \citep{banzatti2020,kalyaan21,kalyaan23,mah2024}. At the same time, this evolutionary effect may not be apparent from the measured H$_2$O column densities alone if dust grains remain coupled to the gas inside the snowline; CO$_2$ is also required to trace the time evolution of the oxygen content \citep{sellek25, houge25}. In other words, disks that have engaged in planet formation processes to a greater or lesser extent may therefore manifest different inner disk chemical ratios.   

\begin{figure*}
\epsscale{1.15}
\plotone{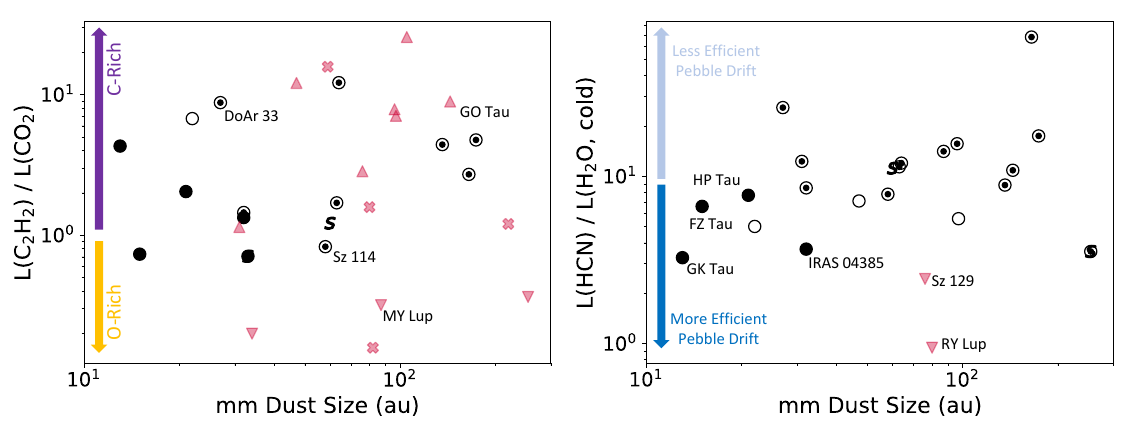}
\caption{\emph{Left:} Ratios of C$_2$H$_2$ to CO$_2$ integrated emission line luminosities versus sub-mm dust disk sizes from Table \ref{tab:diskprops}. Red markers denote lower limits for disks where CO$_2$ was not detected (VZ Cha, TW Cha, Sz 129, MWC 480), upper limits for disks where C$_2$H$_2$ was not detected (AS 205S, MY Lup, Elias 27), and Xs for disks where neither molecule was detected (RY Lup). \emph{Right:} Ratios of HCN integrated emission line luminosities and cold 23.85 $\mu$m H$_2$O emission from  \citep{banzatti24} versus sub-mm dust disk size. Black markers are used to identify sub-mm substructures, as listed in Table \ref{tab:diskprops}.}
\label{fig:organics_vs_rdust}
\end{figure*}

In Figure \ref{fig:organics_vs_rdust}, we compare the ratios of C$_2$H$_2$ and CO$_2$ integrated emission line luminosities to the sub-mm dust disk sizes reported in Table \ref{tab:diskprops}, using the markers listed in Table \ref{tab:diskprops} to indicate the detected outer disk substructures. The two parameters are uncorrelated (Spearman $\rho = 0.033$; $p = 0.8$), indicating that the outer disk size alone does not determine whether the inner disk is more carbon-rich or oxygen-rich. No clear division is observed as a function of resolved substructures, and the line luminosities and radii of the innermost dust rings are also not correlated (see Table \ref{tab:correlations}). We note that the closest resolved sub-mm dust rings in our sample are at $r \sim 6$ au \citep{huang2018, andrews21}, which is still well outside the expected zone of warm, mid-infrared molecular gas emission (see e.g., \citealt{anderson21, vlasblom24}). Further observations of smooth, compact disks, in addition to the four presented in this work, may help clarify this picture. Additional comparisons to 5 $\mu$m CO observations will also provide additional context for interpreting the carbon content within the inner regions of disks \citep{salyk11b, brown13}.       

Figure \ref{fig:organics_vs_rdust} also compares the ratios of HCN and cold H$_2$O luminosities \citep[the latter as adopted from][]{banzatti24} to the sub-mm dust disk sizes reported in Table \ref{tab:diskprops} (see also, \citealt{gasman25}). In multiple small disks in our sample, the observed H$_2$O emission from two transitions near $\sim 23.85$ $\mu$m is consistent with sublimation of icy pebbles \citep[$T \sim 170$--220~K, see][]{banzatti24}, which may produce excess flux in transitions with low rotational excitation energies \citep{zhang13,banzatti23b}. While the small, smooth disks show similar C$_2$H$_2$-to-CO$_2$ emission line luminosity ratios to the larger disks with structures, they generally show less HCN emission relative to the cold H$_2$O luminosity (Spearman $\rho = 0.36$; $p = 0.11$). The trend weakens to Spearman $\rho = 0.30$ and Spearman $\rho = 0.18$ when the denominator is replaced with the warm and hot H$_2$O luminosities adopted from \citet{banzatti24}, respectively. This again suggests that icy pebble drift is the mechanism driving the previously reported trend, producing an excess population of cold water vapor while leaving the warmer in situ water reservoirs unaffected.   

We note that the strong correlation between $L_{\rm{HCN}}/L_{\rm{H_2O}}$ reported in \citet{najita11, najita13} focused on a sample with a narrower span of infrared spectral indices, age, and stellar metallicity than the stellar and disk properties of our sample. Now that the sensitivity of MIRI-MRS has enabled the detection of a thermal gradient in the observed water vapor \citep{banzatti23b,banzatti24, romero-mirza24b}, it is possible to see the trend in a more diverse sample, although it is only marginally statistically significant. The prominent outliers in the $L_{\rm{HCN}}/L_{\rm{H_2O, \, cold}}$ correlation (Sz 129, Elias 27, RY Lup, two of which have an inner dust cavity) are disks that have relatively low HCN luminosities but still relatively high cold water fluxes \citep{banzatti24}. We also report that all disks smaller than $r < 25$ au show $L_{\rm{HCN}}/L_{\rm{H_2O, \, cold}} < 10$, again consistent with a picture where icy pebble drift proceeds more efficiently in compact systems and/or where ongoing planet formation has trapped oxygen in the outer disk. However, none of the four compact disks included in this work (HP Tau, FZ Tau, GK Tau, IRAS 04385) are particularly carbon-rich relative to the rest of the JDISCS sample, supporting the idea that the main driver of the correlation with disk radius is water itself \citep{banzatti2020,banzatti23b}.  

\subsection{How common is optically thick emission from warm organics?}

Optically thick molecular gas emission from deeper layers of the disk can produce a broad pseudo-continuum in the spectra, as the lines become wider than the instrument resolution and blend with adjacent transitions. This effect was recently confirmed in hydrocarbon emission for the first time, from a disk surrounding the very low mass star 2MASS J16053215-1933159 ($M_{\ast} = 0.14 \, M_{\odot}$; \citealt{tabone23}), as MIRI spectra were able to resolve the broadband feature first reported from \emph{Spitzer} observations \citep{pascucci13}. The emission is consistent with the superposition of optically thick and thin C$_2$H$_2$ components, one with column density $N \rm{\left( C_2H_2 \right)} = 2.4 \times 10^{20}$ cm$^{-2}$ and $T = 525$ K and the other with column density $N \rm{\left( C_2H_2 \right)} = 2.5 \times 10^{17}$ cm$^{-2}$ and $T = 400$ K \citep{tabone23, arabhavi24}. Such prominent, optically thick emission was attributed to increased destruction of carbon-bearing grains at the ``soot" line, which may proceed more efficiently as dust settling allows UV photons to travel deeper into the molecular gas layer (see also, \citealt{colmenares24}). However, in most C-rich disks around very low mass stars, the hydrocarbon excitation temperature is much lower than the temperature at the ``soot" line, indicating that this effect alone cannot fully explain the excess carbon content \citep{long25}. In any case, such features can be removed by iterative continuum fitting processes such as those described in Section \ref{sec:obs}, shifting the retrieved column densities to smaller values as a result. 

\begin{figure*}
\epsscale{1.1}
\plotone{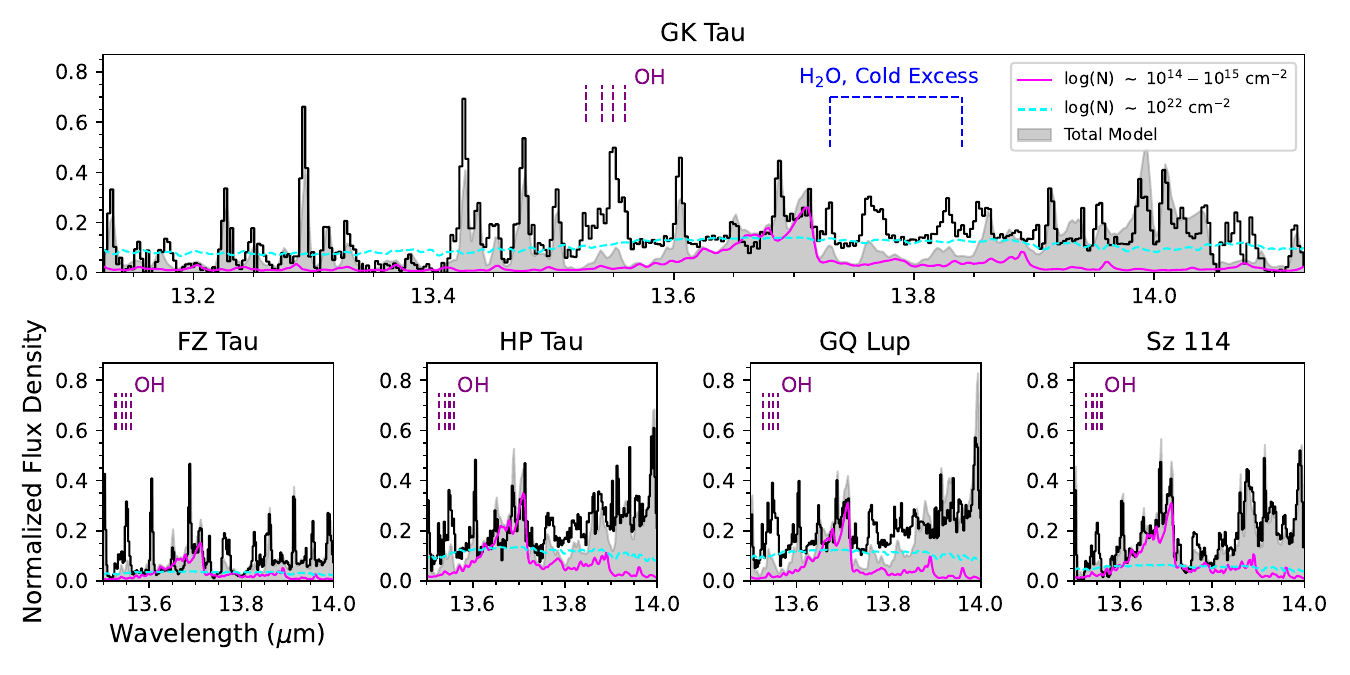}
\caption{Subset of JDISCS Cycle 1 targets with retrieved C$_2$H$_2$ slab temperatures of $\sim$1000 K, required to reproduce broader ``pedestal"-like emission than observed in the other disks from our sample (magenta; full models from Figure \ref{fig:organics_gallery_small} are filled in gray). Optically thick C$_2$H$_2$ components with $\log N = 22$ cm$^{-2}$ are over-plotted for comparison (cyan; \citealt{tabone23, arabhavi24}), and OH emission lines that were not included in our fits are marked with purple, dashed lines. While the optically thick gas models can reproduce the excess emission near the bandheads, they overpredict the emission from the $P$ and $R$ branch transitions and are therefore not consistent with the overall shapes of the pedestals.}
\label{fig:sz114bump}
\end{figure*}

While none of our sources show a broad pseudo-continuum near 13.7 $\mu$m that is as prominent as the features detected in the spectra of 2MASS J16053215-1933159 \citep{pascucci13, tabone23}, Sz 28 \citep{kanwar24}, ISO-ChaI 147 \citep{arabhavi24}, or the 30 Myr disk WISE J044634.16–262756.1B \citep{long25}, we identify a subset of disks in Figure \ref{fig:allmols_T} that require larger C$_2$H$_2$ slab temperatures to reproduce the spectra ($T \sim 1000$ K). Most are relatively compact, with sub-mm dust disk radii between $r = 13-58$ au). Figure \ref{fig:sz114bump} highlights these targets, showing ``pedestals" under the C$_2$H$_2$ and HCN $Q$-branch emission that are narrower than the broad pseudo-continuum observed in carbon-rich disks but still do not fall flat against the dust continuum (GK Tau, FZ Tau, HP Tau, GQ Lup, and Sz 114; see also \citealt{xie23}). Clear signatures of excess cold water emission at longer wavelengths that we do not model here are also detected \citep{banzatti23b, romero-mirza24b, banzatti24, gasman25}. We compare the residuals from our slab model fits to scaled, optically thick slab models with $\log N_{\rm{col}} = 22$ cm$^{-2}$ and $T = 525$ K \citep{tabone23, arabhavi24}, to explore whether an additional C$_2$H$_2$ component can fill in the missing flux. However, the optically thick models clearly over-predict the fluxes in the $P$ and $R$ branch emission lines at wavelengths further from the bandhead. 

Although some calibration issues complicated analysis between $\sim 13-15$ $\mu$m in early JWST observations \citep{chown2024}, the JDISCS pipeline has improved this considerably \citep{pontoppidan24}. Since only a subset of disks in our sample show the ``pedestals", it is also unlikely that the features can be attributed to the continuum subtraction routine described in Section 2.3; instead, we consider other mechanisms that may be responsible for the excess emission. There are no expected PAH features in the 13.7 $\mu$m region \citep{sloan2014,jones2023}, although this broad feature has long been noted in absorption toward embedded sources such as IRc2 in Orion \citep[e.g.][]{evans1991}. \citet{evans1991} attributed this to material in front of the hot protostellar core or the plateau feature in that region, in which emitting and absorbing material could have been co-located by line-of-sight. Alternatively, the excess emission also overlaps with double-peaked pyroxene features that were tentatively detected with \emph{Spitzer} at 13.6, 14.5, and 15.5 $\mu$m toward Class II disks \citep{chihara2002, carr11}, where the Fe concentration determined the peak wavelengths and shapes of the observed bands. However, a full treatment of the corresponding 10 $\mu$m silicate emission will be required to reproduce these features and is beyond the scope of this work. We conclude that the excess ``pedestal" under the organic emission in the small disks is not consistent with an optically thick gas pseudo-continuum emerging from deeper in the disk and may instead be attributed to dust emission, implying that the prominent bumps presented in previous works are unique features even among compact disks \citep{tabone23, kanwar24, arabhavi24, long25}. As current state-of-the-art techniques for simultaneous modeling of dust and molecular gas emission (see e.g., \citealt{kaeufer24}) have not yet been demonstrated on faint signatures such as those reported here, updated models of the dust continuum will be required to obtain the best constraints on the temperature structure of the mid-infrared disk layers.   

\subsection{How does inner disk chemistry evolve with accretion and disk dispersal?}

[Ne II], [Ne III], and [Ar II] emission lines are expected to trace either high-velocity winds/microjets in disks that are strong accretors ($M_{\rm{acc}} > 10^{-8} \, M_{\odot}$; \citealt{pascucci20}) or low-velocity winds otherwise \citep{hollenbach09, espaillat13, sellek24}. Photoevaporative wind models predict that [Ne III]/[Ne II] flux ratios will be $<1$ when X-ray irradiation dominates the disk ionization and dispersal processes, while ratios $>1$ are consistent with EUV driven wind models \citep{hollenbach09, espaillat13, bajaj24}. A [Ne II]/[Ar II] ratio $<2.5$ is expected under an EUV or soft X-ray radiation field, while [Ne II]/[Ar II] $>2.5$ points to a hard X-ray spectrum \citep{espaillat23}. Alternatively, a UV dominant spectrum would trigger gas-phase reactions in hotter layers; this effect may be particularly apparent in the OH-H$_2$O formation pathways \citep{agundez08, adamkovics14, adamkovics16, walsh15}. Since the C$_2$H$_2$, HCN, and CO$_2$ emission is generally optically thin (see Figure \ref{fig:Lemit_vs_Memit}) and therefore directly exposed to stellar and accretion-generated radiation, understanding the mechanisms driving the Ne and Ar ionization may also constrain the high-energy flux that catalyzes photochemistry in the warm molecular gas (see e.g., \citealt{agundez08, pascucci09, najita11, szulagyi12,  walsh15, gaidos25}).

\begin{figure*}
\epsscale{1.17}
\plottwo{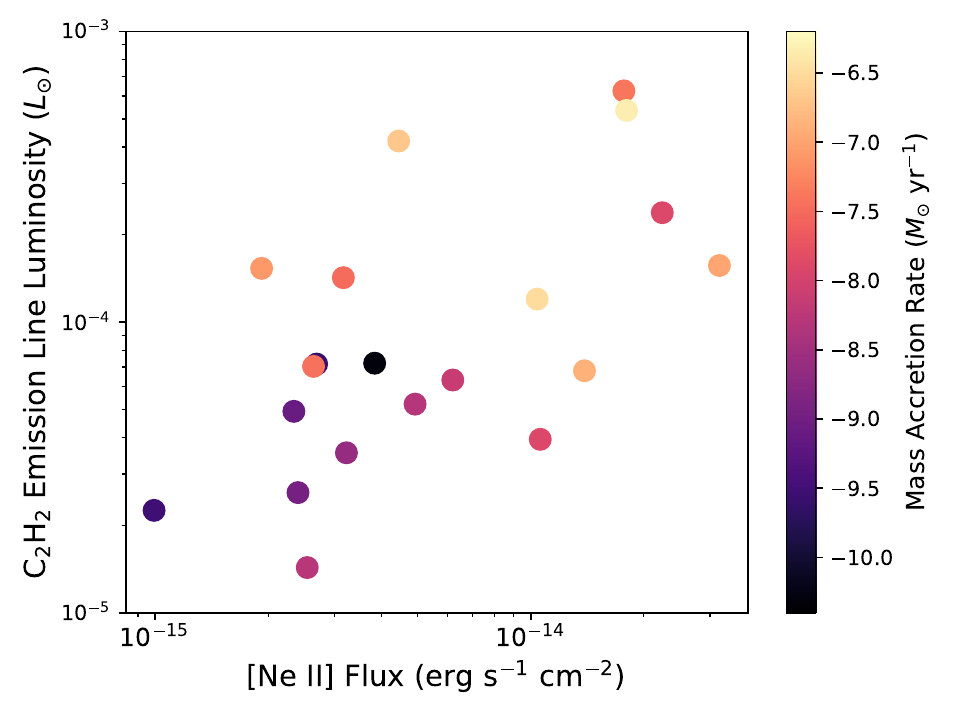}{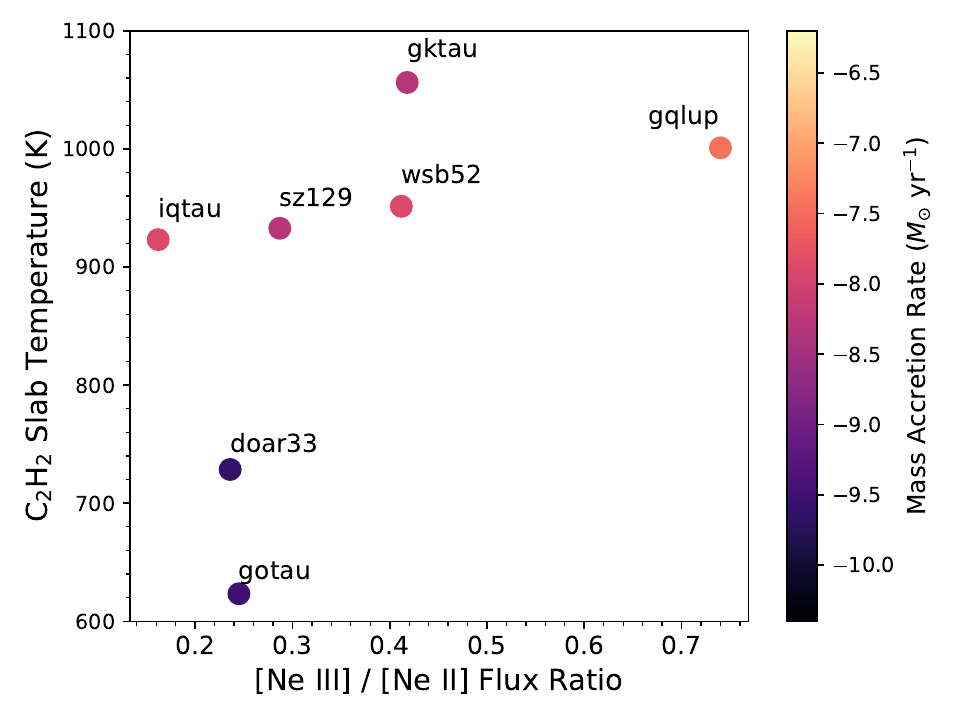}
\caption{\emph{Left:} C$_2$H$_2$ emission line luminosities versus [Ne II] emission line fluxes. \emph{Right:} C$_2$H$_2$ slab temperatures versus [Ne III] to [Ne II] flux ratios. In both panels, marker colors represent mass accretion rates (see also, \citealt{colmenares24}). Upper limits are omitted from the plot and correlation coefficient calculation (see Table \ref{tab:correlations}), due to the large uncertainty in the water removal at 15.555 $\mu$m.}
\label{fig:C2H2vsNeIINeIII}
\end{figure*}

All ten JDISCS targets with [Ne III] detections show [Ne III]/[Ne II] ratios $<1$, with a range from 0.1 (MY Lup) to 0.7 (GQ Lup), placing them all in the X-ray irradiation regime. This is consistent with results from \emph{Spitzer}, where [Ne III]/[Ne II] $>1$ was never detected \citep{najita10, espaillat13, bajaj24}. For the [Ar II] detections, we find that MY Lup, GM Aur, and IQ Tau have [Ne II]/[Ar II] $>2.5$ and HD 143006, MWC 480, and RY Lup $<2.5$. In MY Lup, GM Aur, and IQ Tau, the combination of $\rm{[Ne \, III]/[Ne \, II]} = 0.1-0.2$ and $\rm{[Ne \, II]/[Ar \, II]} > 2.5$ is consistent with hard X-ray spectra reaching the disks \citep{szulagyi12}. These radiation fields may drive gas-phase chemistry via ion-neutral reactions in cooler, UV-shielded disk layers (see e.g., \citealt{agundez08, najita11, walsh15}).

MY Lup and IQ Tau both have similar sub-mm dust disk sizes (see Figure \ref{fig:organics_gallery_large}; $r_{\rm{dust}} = 87, 96$ au for MY Lup, IQ Tau, respectively; \citealt{huang2018, long19}), and GM Aur has a much larger disk ($r_{\rm{dust}} = 220$ au; \citealt{huang2020a}) and a prominent dust cavity. Despite the similarity of their atomic features, the molecular emission spectra from all three targets do not show any clear grouping. MY Lup is one of the most distinctive targets in JDISCS, showing $^{12}$CO$_2$ with a peak/continuum ratio of $\sim 0.1$ and clear detections of C$^{18}$O$^{16}$O and H$^{13}$CN \citep{salyk25}. However, CO$_2$ is not detected at all in IQ Tau (peak/continuum $< 0.02$) or GM Aur (peak/continuum $<0.03$). Similarly, C$_2$H$_2$ is detected in IQ Tau with a peak/continuum ratio of $\sim 0.06$ but not detected in MY Lup or GM Aur (peak/continuum $<0.03$). It is possible that the line-of-sight toward IQ Tau probes a layer of gas with lower column densities than that toward MY Lup \citep{salyk25}. Identification of additional disks with organics-rich spectra and signatures of wind-driven inner disk clearing, i.e. pre-transitional disks, will be required to further explore this connection. 

Of the three organic molecules presented here, C$_2$H$_2$ shows the strongest evidence for a potential link between inner disk molecular gas emission and the MHD or photoevaporative wind-driven ionization structure across our sample (see Figure \ref{fig:C2H2vsNeIINeIII}). We find that the C$_2$H$_2$ emission line luminosities and emitting masses are significantly positively correlated with the [Ne II] fluxes ($\rho = 0.583, 0.590$, respectively; see Figure \ref{fig:C2H2vsNeIINeIII} for emission line luminosities and [Ne II] fluxes), which in turn are expected to scale with X-ray luminosity in static disk models with low EUV flux \citep{hollenbach09, pascucci14}. Meanwhile, the C$_2$H$_2$ slab temperatures are also significantly positively correlated with the [Ne III]/[Ne II] flux ratios ($\rho = 0.821$), indicating that the molecular layer becomes hotter as EUV heating becomes more significant. As reported by \citet{banzatti2017, banzatti2020} from analyses of \emph{Spitzer} spectra, we also find positive correlations between the mass accretion rates and C$_2$H$_2$, HCN, and CO$_2$ emission line luminosities ($\rho = 0.642, 0.504, 0.503$, respectively), as expected if UV photons generated at the accretion shocks are responsible for heating the gas. A new interesting aspect to explore in future work is emerging from studies of accretion variability and outbursts, which can increase/decrease the observed molecular fluxes in different phases of accretion due to disk heating and UV photo-dissociation (\citet{banzatti12}, Smith et al. 2025, in press).

However, only a small subset of targets have C$_2$H$_2$, [Ne III], and [Ne II] detections. Further clarification of this trend will require a more complete inventory of water emission lines across the full MIRI wavelength range, to place well-constrained upper limits on the [Ne III] emission. Complementary observations at FUV through X-ray wavelengths will also reveal the radiation fields generated by the accreting host stars, constraining the flux received at the disk surfaces. We note again that the aperture used to extract the 1-D spectra presented in this work does not capture spatially extended emission from the atomic wind tracers, which may in turn alter the reported flux ratios. An exploration of spatially extended emission across the JDISCS sample is forthcoming (Pontoppidan et al., in prep).  

\section{Conclusions} \label{sec:conclusions}

As observations of protoplanetary disks with \emph{JWST} MIRI are ongoing, JDISCS is providing a database of consistently reduced, high-quality mid-infrared spectra that enable inner disk population studies. From this database, we present an overview of H$_2$O, C$_2$H$_2$, HCN, CO$_2$, [Ne II], [Ne III], and [Ar II] emission lines from the 31 JDISCS systems observed in Cycle 1. Although this initial sample probes only a small range in stellar mass, spectra show approximately two orders of magnitude variation in molecular line luminosities. H$_2$O and OH emission is nearly ubiquitous, and detection rates are higher for MRS than they were for Spitzer-IRS. 
 
We fit LTE slab models to the spectra to retrieve column densities, temperatures, and emitting areas for three organic molecules and water at 12--16~$\mu$m, finding that:

\begin{itemize}
    \item Emission line luminosities are significantly correlated with the molecular emitting masses of C$_2$H$_2$, HCN, and CO$_2$, but they are not correlated with the slab temperatures. 
    \item CO$_2$ emission generally comes from cooler, more optically thick gas $\left(T = 600^{+200}_{-160} \, \rm{K} \right)$ than C$_2$H$_2$ and HCN emission ($T = 920^{+70}_{-130}, 820^{+70}_{-130}$ K, respectively). 
\end{itemize}
These results confirm that C$_2$H$_2$, HCN, and CO$_2$ emission lines are generally optically thin to moderately optically thick, originating in warm surface layers of the inner disk.   

All targets included in this work were previously observed with ALMA at high angular resolution, and we consider the impact of outer disk substructures on the molecular gas emission. We identify the following trends from our analysis: 
\begin{itemize}
\item Excess emission is visible as a ``pedestal" under the C$_2$H$_2$ and HCN $Q$ branches in the smallest disks, but it is not consistent with optically thick emission from deeper disk layers (see e.g., \citealt{tabone23, arabhavi24, long25}). This ``pedestal" emission drives the retrieved C$_2$H$_2$ temperatures higher than in other disks ($T > 1000$ K) and is likely due to dust features. 
\item The ratios of C$_2$H$_2$ to CO$_2$ emitting masses are not correlated with sub-mm disk sizes, indicating that disk size alone does not necessarily determine the ratios of carbon- and oxygen-bearing molecules \citep{colmenares24}. 
\item Smaller disks generally have larger fluxes from cold H$_2$O relative to HCN, consistent with a picture where either pebble drift and its impediment by dust traps regulates water delivery to the inner disk \citep{banzatti2020, banzatti23b, gasman25} or where advanced planet formation has sequestered oxygen in more massive disks \citep{najita13}.
\end{itemize}

Finally, we examine the brightest atomic forbidden emission lines, finding that:

\begin{itemize}

\item $\left[\rm{Ne \, II} \right]$ is detected in all but three disks included in this work, [Ne III] in ten sources, and [Ar II] in four systems.

\item The measured [Ne III]/[Ne II] emission line fluxes are all $<1$, consistent with X-ray irradiation as the primary ionization source \citep{espaillat23, bajaj24}. 

\item The [Ne II]/[Ar II] flux ratios indicate that the disks around MY Lup, IQ Tau, and GM Aur receive a hard X-ray spectrum, which may influence gas-phase chemistry in the optically thin emitting layers. 

\end{itemize}

While this study provides a first step toward a global analysis of atomic and molecular spectra from a {\it sample} of disks observed with MRS, the targets analyzed in this work are limited to stellar masses around solar and ages of a few Myr. Additional JDISCS programs are underway to better probe trends with stellar mass and disk evolutionary state. A more comprehensive treatment of the molecular gas across the entire MRS spectrum must also include the addition of less abundant species and isotopologues (e.g., $^{13}$C$^{12}$CH$_2$, C$_4$H$_2$, C$^{18}$O$^{16}$O, H$^{13}$CN; \citealt{tabone23, kanwar24, colmenares24, salyk25}), fit simultaneously to distinguish between overlapping transitions. Such efforts may require the adaptation of machine learning techniques that are not yet widely applied to spectra of protoplanetary disks, in order to capture the most relevant physics while minimizing computation time (see e.g., \citealt{diop24, kaeufer2024b}). However, the initial findings presented here highlight the potential of MIRI MRS to reveal demographic trends and links between the physics and chemistry of inner and outer protoplanetary disk zones. 

We thank the referee for a thorough review that greatly contributed to the clarity of the paper. A portion of this research was carried out at the Jet Propulsion Laboratory, California Institute of Technology, under a contract with the National Aeronautics and Space Administration (80NM0018D0004).

\vspace{5mm}
\facilities{JWST(MIRI)}

\software{astropy \citep{astropy:2013, astropy:2018, astropy:2022},  
spectools\_ir \citep{salykspectoolsir}, iSLAT \citep{iSLAT_code}
          }

\appendix
\restartappendixnumbering

\section{Impact of Water Removal on Slab Model Fits to Organics in MIRI-MRS Spectra \label{appendix:waterremoval}}

\begin{figure}
\epsscale{1.22}
\plotone{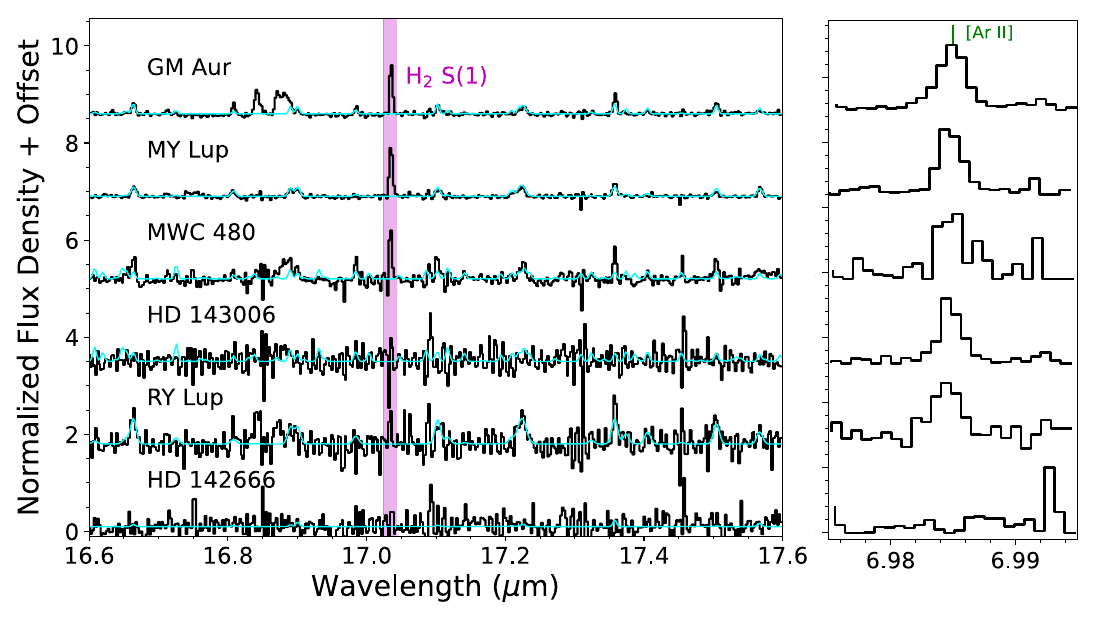}
\caption{Water emission lines between 16.6-17.6 $\mu$m and [Ar II] emission lines for JDISCS spectra that are dominated by atomic features, with the best-fit H$_2$O slab models overlaid (cyan). We place upper limits of $\log L_{\rm{emit}} < -5.3, -4.3 \, L_{\odot}$ for HD 142666 and HD 143006, respectively. The OH-rich spectrum of GM Aur will be presented in a forthcoming paper (Muñoz-Romero et al., in prep).}
\label{fig:herbig_cavity_water}
\end{figure}

The wavelength range in which HCN, C$_2$H$_2$, and CO$_2$ are detected in our MIRI-MRS spectra also includes a large number of rotational emission lines from the ground state and first excited state bending mode manifold of H$_2$O (see e.g., \citealt{banzatti23b, grant23, gasman23, xie23, pontoppidan24, romero-mirza24b}). These water lines must be subtracted in order to model the organics, to ensure that emission across the full rotational ladder within each vibrational band can be used to constrain the temperature of the emitting region. Previous work has achieved this by using scale factors to predict the water flux near the HCN $Q$ branch emission (see e.g., \citealt{najita13}) or by fitting the water emission lines in other regions of the spectra with single temperature slab models (see e.g., \citealt{salyk11, grant23}). 

The single temperature slab model fits to \emph{Spitzer}-IRS observations were generally consistent with a population of H$_2$O with $T \sim 450$ K and $N \sim 10^{18}$ cm$^{-2}$ \citep{carr08, salyk11}. However, at the higher spectral resolving power of the MRS ($R \sim 2250-2725$ between 13-14.5 $\mu$m; \citealt{pontoppidan24}), two ground state populations of H$_2$O with $T \sim 800$ and $T = 370-500$ K have been detected in several compact disks where efficient icy pebble drift produces excess emission from sublimated gas near the water snowline \citep{banzatti23a}. Notably, the cooler excess emission is not detected in larger disks where icy pebbles are likely trapped within dust rings \citep{kalyaan21, kalyaan23}. It is now possible to include multiple temperature components to reproduce the spectra (see e.g., \citealt{romero-mirza24b, grant2024}).

These two component slab model fits, or radial gradient retrievals \citep{romero-mirza24b}, typically assume LTE excitation of all ro-vibrational states. Building on this effort, \citet{banzatti24} reveals that the rovibrational band near 6 $\mu$m and the pure rotational hot band lines at longer wavelengths are suppressed compared to LTE. Here we do not consider the rovibrational emission, and note that the hot band $v=1-1$ rotational emission flux is weak compared to the ground state lines. Thus, we also do not consider these further in the present analysis. For more precise fits to individual sources, distinct but constant mutliplicative factors for the rovibrational and hot band rotational emission lines can be used as a first order characterization of the water vapor emission \citep{banzatti24}.

Since the water emission lines that show the strongest signatures of multiple temperature components are generally weaker in the 12-16 $\mu$m wavelength range where HCN, C$_2$H$_2$ and CO$_2$ emission is observed, we use single temperature slab models to reproduce and subtract the H$_2$O emission. All retrieved parameters are reported in Table \ref{tab:slabparams_H2O}.

\begin{deluxetable*}{lcccccc}
\tablecaption{Slab Model Fit Parameters: H$_2$O, Rotational lines at 12--16~$\mu$m \label{tab:slabparams_H2O}}
\tablehead{
\colhead{Target} & \colhead{$\log N$} & \colhead{$r_{\text{slab}}$} & \colhead{$M_{\text{emit}}$} & \colhead{$T$} & \colhead{$L_{\text{emit}}$} & \colhead{P/C} \\
 & \colhead{(cm$^{-2}$)} & \colhead{(au)} & \colhead{$\log M$ ($M_{\oplus}$)} & \colhead{(K)} & \colhead{$\log L$ ($L_{\odot}$)} & }
\startdata
AS 205 N & 18.7 & 1.6 & -4.4 & $610 \pm 5$ & -2.4 & 0.22 \\
AS 205 S & 17.6 & 1.3 & -5.6 & $530 \pm 10$ & -3.6 & 0.08 \\
AS 209 & 19.5 & 0.2 & -5.8 & $890 \pm 10$ & -3.3 & 0.04 \\
CI Tau & 18.3 & 0.4 & -6.1 & $880 \pm 10$ & -3.2 & 0.27 \\
DoAr 25 & 18.8 & 0.3 & -5.9 & $640 \pm 10$ & -3.8 & 0.14 \\
DoAr 33 & 18.8 & 0.2 & -6.0 & $610^{+20}_{-10}$ & -4.0 & 0.11 \\
Elias 20 & 18.5 & 0.7 & -5.3 & $640 \pm 10$ & -3.1 & 0.34 \\
Elias 24 & 18.4 & 1.2 & -4.9 & $600 \pm 10$ & -2.8 & 0.29 \\
Elias 27 & 18.5 & 0.5 & -5.6 & $640 \pm 10$ & -3.4 & 0.54 \\
FZ Tau & 18.5 & 0.8 & -5.1 & $690 \pm 10$ & -2.8 & 0.53 \\
GK Tau & 18.1 & 0.3 & -6.3 & $760 \pm 10$ & -3.6 & 0.10 \\
GM Aur & 17.8 & 0.2 & -6.6 & $480 \pm 50$ & -5.0 & 0.03 \\
GO Tau & 17.8 & 0.2 & -7.2 & $570 \pm 20$ & -5.0 & 0.06 \\
GQ Lup & 17.8 & 0.7 & -6.0 & $620 \pm 10$ & -3.6 & 0.12 \\
HD 142666 & \nodata & \nodata & $<-5.7$ & \nodata & $<-5.3$ & $<10^{-3}$ \\
HD 143006 & \nodata & \nodata & $<-6.8$ & \nodata & $<-4.3$ & $<0.01$ \\
HD 163296 & \nodata & \nodata & \nodata & \nodata & \nodata & \nodata \\
HP Tau & 18.2 & 0.3 & -6.2 & $760 \pm 10$ & -3.5 & 0.05 \\
HT Lup & 18.7 & 0.3 & -5.6 & $790 \pm 10$ & -3.2 & 0.06 \\
IQ Tau & 18.4 & 0.2 & -6.5 & $910^{+20}_{-10}$ & -3.6 & 0.15 \\
IRAS 04385 & 18.5 & 0.6 & -5.4 & $530 \pm 10$ & -3.6 & 0.15 \\
MWC 480 & 18.8 & 0.2 & -5.9 & $870^{+60}_{-10}$ & -3.4 & 0.01 \\
MY Lup & 18.5 & 0.3 & -5.9 & $390 \pm 10$ & -4.8 & 0.02 \\
RU Lup & 18.2 & 0.9 & -5.3 & $730 \pm 10$ & -2.7 & 0.19 \\
RY Lup & 17.6 & 0.4 & -6.7 & $530 \pm 10$ & -4.6 & 0.009 \\
SR 4 & 18.1 & 0.3 & -6.4 & $840 \pm 10$ & -3.5 & 0.10 \\
Sz 114 & 18.7 & 0.4 & -5.5 & $580 \pm 10$ & -3.7 & 0.26 \\
Sz 129 & 18.0 & 0.3 & -6.5 & $750 \pm 10$ & -3.8 & 0.24 \\
TW Cha & 18.1 & 0.5 & -6.0 & $740 \pm 10$ & -3.4 & 0.47 \\
VZ Cha & 18.1 & 0.5 & -6.1 & $840 \pm 10$ & -3.2 & 0.38 \\
WSB 52 & 18.8 & 1.4 & -4.3 & $500 \pm 5$ & -2.9 & 0.73 \\
\enddata
\end{deluxetable*}

\begin{figure*}
\epsscale{1.22}
\plotone{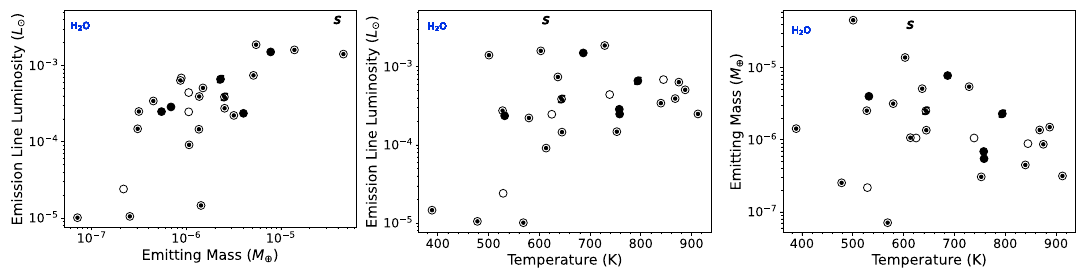}
\caption{Correlations between best-fit parameters retrieved from single temperature slab model fits to H$_2$O emission lines between 12-17.5 $\mu$m. Although statistically significant, the correlation between emission line luminosities and emitting masses \emph{(left)} is significantly weaker (Spearman $\rho = 0.66$; $p = 10^{-4}$) than what is observed for C$_2$H$_2$, HCN, and CO$_2$ (see Figure \ref{fig:detection_limits}). The single slab temperatures are not correlated with either the emission line luminosities (\emph{middle}; Spearman $\rho = 0.35$; $p = 0.064$) or emitting masses (\emph{right}; Spearman $\rho = -0.24$; $p = 0.21$).}
\label{fig:H2O_retrievalparam_correlations}
\end{figure*}

\bibliography{JDISCSoverview}{}

\begin{thebibliography}{}
\expandafter\ifx\csname natexlab\endcsname\relax\def\natexlab#1{#1}\fi
\providecommand{\url}[1]{\href{#1}{#1}}
\providecommand{\dodoi}[1]{doi:~\href{http://doi.org/#1}{\nolinkurl{#1}}}
\providecommand{\doeprint}[1]{\href{http://ascl.net/#1}{\nolinkurl{http://ascl.net/#1}}}
\providecommand{\doarXiv}[1]{\href{https://arxiv.org/abs/#1}{\nolinkurl{https://arxiv.org/abs/#1}}}

\bibitem[{{{\'A}d{\'a}mkovics} {et~al.}(2014){{\'A}d{\'a}mkovics}, {Glassgold}, \& {Najita}}]{adamkovics14}
{{\'A}d{\'a}mkovics}, M., {Glassgold}, A.~E., \& {Najita}, J.~R. 2014, \apj, 786, 135, \dodoi{10.1088/0004-637X/786/2/135}

\bibitem[{{{\'A}d{\'a}mkovics} {et~al.}(2016){{\'A}d{\'a}mkovics}, {Najita}, \& {Glassgold}}]{adamkovics16}
{{\'A}d{\'a}mkovics}, M., {Najita}, J.~R., \& {Glassgold}, A.~E. 2016, \apj, 817, 82, \dodoi{10.3847/0004-637X/817/1/82}

\bibitem[{{Ag{\'u}ndez} {et~al.}(2008){Ag{\'u}ndez}, {Cernicharo}, \& {Goicoechea}}]{agundez08}
{Ag{\'u}ndez}, M., {Cernicharo}, J., \& {Goicoechea}, J.~R. 2008, \aap, 483, 831, \dodoi{10.1051/0004-6361:20077927}

\bibitem[{{Alcal{\'a}} {et~al.}(2019){Alcal{\'a}}, {Manara}, {France}, {Schneider}, {Arulanantham}, {Miotello}, {G{\"u}nther}, \& {Brown}}]{alcala19}
{Alcal{\'a}}, J.~M., {Manara}, C.~F., {France}, K., {et~al.} 2019, \aap, 629, A108, \dodoi{10.1051/0004-6361/201935657}

\bibitem[{{Alcal{\'a}} {et~al.}(2017){Alcal{\'a}}, {Manara}, {Natta}, {Frasca}, {Testi}, {Nisini}, {Stelzer}, {Williams}, {Antoniucci}, {Biazzo}, {Covino}, {Esposito}, {Getman}, \& {Rigliaco}}]{alcala17}
{Alcal{\'a}}, J.~M., {Manara}, C.~F., {Natta}, A., {et~al.} 2017, \aap, 600, A20, \dodoi{10.1051/0004-6361/201629929}

\bibitem[{{Anderson} {et~al.}(2021){Anderson}, {Blake}, {Cleeves}, {Bergin}, {Zhang}, {Schwarz}, {Salyk}, \& {Bosman}}]{anderson21}
{Anderson}, D.~E., {Blake}, G.~A., {Cleeves}, L.~I., {et~al.} 2021, \apj, 909, 55, \dodoi{10.3847/1538-4357/abd9c1}

\bibitem[{{Andrews} {et~al.}(2018){Andrews}, {Huang}, {P{\'e}rez}, {Isella}, {Dullemond}, {Kurtovic}, {Guzm{\'a}n}, {Carpenter}, {Wilner}, {Zhang}, {Zhu}, {Birnstiel}, {Bai}, {Benisty}, {Hughes}, {{\"O}berg}, \& {Ricci}}]{andrews18}
{Andrews}, S.~M., {Huang}, J., {P{\'e}rez}, L.~M., {et~al.} 2018, \apjl, 869, L41, \dodoi{10.3847/2041-8213/aaf741}

\bibitem[{{Andrews} {et~al.}(2021){Andrews}, {Elder}, {Zhang}, {Huang}, {Benisty}, {Kurtovic}, {Wilner}, {Zhu}, {Carpenter}, {P{\'e}rez}, {Teague}, {Isella}, \& {Ricci}}]{andrews21}
{Andrews}, S.~M., {Elder}, W., {Zhang}, S., {et~al.} 2021, \apj, 916, 51, \dodoi{10.3847/1538-4357/ac00b9}

\bibitem[{{Ansdell} {et~al.}(2016){Ansdell}, {Williams}, {van der Marel}, {Carpenter}, {Guidi}, {Hogerheijde}, {Mathews}, {Manara}, {Miotello}, {Natta}, {Oliveira}, {Tazzari}, {Testi}, {van Dishoeck}, \& {van Terwisga}}]{ansdell16}
{Ansdell}, M., {Williams}, J.~P., {van der Marel}, N., {et~al.} 2016, \apj, 828, 46, \dodoi{10.3847/0004-637X/828/1/46}

\bibitem[{{Arabhavi} {et~al.}(2024){Arabhavi}, {Kamp}, {Henning}, {van Dishoeck}, {Christiaens}, {Gasman}, {Perrin}, {G{\"u}del}, {Tabone}, {Kanwar}, {Waters}, {Pascucci}, {Samland}, {Perotti}, {Bettoni}, {Grant}, {Lagage}, {Ray}, {Vandenbussche}, {Absil}, {Argyriou}, {Barrado}, {Boccaletti}, {Bouwman}, {Caratti o Garatti}, {Glauser}, {Lahuis}, {Mueller}, {Olofsson}, {Pantin}, {Scheithauer}, {Morales-Calder{\'o}n}, {Franceschi}, {Jang}, {Pawellek}, {Rodgers-Lee}, {Schreiber}, {Schwarz}, {Temmink}, {Vlasblom}, {Wright}, {Colina}, \& {{\"O}stlin}}]{arabhavi24}
{Arabhavi}, A.~M., {Kamp}, I., {Henning}, T., {et~al.} 2024, Science, 384, 1086, \dodoi{10.1126/science.adi8147}

\bibitem[{{Arulanantham} {et~al.}(2024){Arulanantham}, {McClure}, {Pontoppidan}, {Beck}, {Sturm}, {Harsono}, {Boogert}, {Cordiner}, {Dartois}, {Drozdovskaya}, {Espaillat}, {Melnick}, {Noble}, {Palumbo}, {Pendleton}, {Terada}, \& {van Dishoeck}}]{arulanantham2024}
{Arulanantham}, N., {McClure}, M.~K., {Pontoppidan}, K., {et~al.} 2024, \apjl, 965, L13, \dodoi{10.3847/2041-8213/ad35c9}

\bibitem[{{Astropy Collaboration} {et~al.}(2013){Astropy Collaboration}, {Robitaille}, {Tollerud}, {Greenfield}, {Droettboom}, {Bray}, {Aldcroft}, {Davis}, {Ginsburg}, {Price-Whelan}, {Kerzendorf}, {Conley}, {Crighton}, {Barbary}, {Muna}, {Ferguson}, {Grollier}, {Parikh}, {Nair}, {Unther}, {Deil}, {Woillez}, {Conseil}, {Kramer}, {Turner}, {Singer}, {Fox}, {Weaver}, {Zabalza}, {Edwards}, {Azalee Bostroem}, {Burke}, {Casey}, {Crawford}, {Dencheva}, {Ely}, {Jenness}, {Labrie}, {Lim}, {Pierfederici}, {Pontzen}, {Ptak}, {Refsdal}, {Servillat}, \& {Streicher}}]{astropy:2013}
{Astropy Collaboration}, {Robitaille}, T.~P., {Tollerud}, E.~J., {et~al.} 2013, \aap, 558, A33, \dodoi{10.1051/0004-6361/201322068}

\bibitem[{{Astropy Collaboration} {et~al.}(2018){Astropy Collaboration}, {Price-Whelan}, {Sip{\H{o}}cz}, {G{\"u}nther}, {Lim}, {Crawford}, {Conseil}, {Shupe}, {Craig}, {Dencheva}, {Ginsburg}, {Vand erPlas}, {Bradley}, {P{\'e}rez-Su{\'a}rez}, {de Val-Borro}, {Aldcroft}, {Cruz}, {Robitaille}, {Tollerud}, {Ardelean}, {Babej}, {Bach}, {Bachetti}, {Bakanov}, {Bamford}, {Barentsen}, {Barmby}, {Baumbach}, {Berry}, {Biscani}, {Boquien}, {Bostroem}, {Bouma}, {Brammer}, {Bray}, {Breytenbach}, {Buddelmeijer}, {Burke}, {Calderone}, {Cano Rodr{\'\i}guez}, {Cara}, {Cardoso}, {Cheedella}, {Copin}, {Corrales}, {Crichton}, {D'Avella}, {Deil}, {Depagne}, {Dietrich}, {Donath}, {Droettboom}, {Earl}, {Erben}, {Fabbro}, {Ferreira}, {Finethy}, {Fox}, {Garrison}, {Gibbons}, {Goldstein}, {Gommers}, {Greco}, {Greenfield}, {Groener}, {Grollier}, {Hagen}, {Hirst}, {Homeier}, {Horton}, {Hosseinzadeh}, {Hu}, {Hunkeler}, {Ivezi{\'c}}, {Jain}, {Jenness}, {Kanarek}, {Kendrew}, {Kern}, {Kerzendorf}, {Khvalko}, {King}, {Kirkby}, {Kulkarni},
  {Kumar}, {Lee}, {Lenz}, {Littlefair}, {Ma}, {Macleod}, {Mastropietro}, {McCully}, {Montagnac}, {Morris}, {Mueller}, {Mumford}, {Muna}, {Murphy}, {Nelson}, {Nguyen}, {Ninan}, {N{\"o}the}, {Ogaz}, {Oh}, {Parejko}, {Parley}, {Pascual}, {Patil}, {Patil}, {Plunkett}, {Prochaska}, {Rastogi}, {Reddy Janga}, {Sabater}, {Sakurikar}, {Seifert}, {Sherbert}, {Sherwood-Taylor}, {Shih}, {Sick}, {Silbiger}, {Singanamalla}, {Singer}, {Sladen}, {Sooley}, {Sornarajah}, {Streicher}, {Teuben}, {Thomas}, {Tremblay}, {Turner}, {Terr{\'o}n}, {van Kerkwijk}, {de la Vega}, {Watkins}, {Weaver}, {Whitmore}, {Woillez}, {Zabalza}, \& {Astropy Contributors}}]{astropy:2018}
{Astropy Collaboration}, {Price-Whelan}, A.~M., {Sip{\H{o}}cz}, B.~M., {et~al.} 2018, \aj, 156, 123, \dodoi{10.3847/1538-3881/aabc4f}

\bibitem[{{Astropy Collaboration} {et~al.}(2022){Astropy Collaboration}, {Price-Whelan}, {Lim}, {Earl}, {Starkman}, {Bradley}, {Shupe}, {Patil}, {Corrales}, {Brasseur}, {N{"o}the}, {Donath}, {Tollerud}, {Morris}, {Ginsburg}, {Vaher}, {Weaver}, {Tocknell}, {Jamieson}, {van Kerkwijk}, {Robitaille}, {Merry}, {Bachetti}, {G{"u}nther}, {Aldcroft}, {Alvarado-Montes}, {Archibald}, {B{'o}di}, {Bapat}, {Barentsen}, {Baz{'a}n}, {Biswas}, {Boquien}, {Burke}, {Cara}, {Cara}, {Conroy}, {Conseil}, {Craig}, {Cross}, {Cruz}, {D'Eugenio}, {Dencheva}, {Devillepoix}, {Dietrich}, {Eigenbrot}, {Erben}, {Ferreira}, {Foreman-Mackey}, {Fox}, {Freij}, {Garg}, {Geda}, {Glattly}, {Gondhalekar}, {Gordon}, {Grant}, {Greenfield}, {Groener}, {Guest}, {Gurovich}, {Handberg}, {Hart}, {Hatfield-Dodds}, {Homeier}, {Hosseinzadeh}, {Jenness}, {Jones}, {Joseph}, {Kalmbach}, {Karamehmetoglu}, {Ka{l}uszy{'n}ski}, {Kelley}, {Kern}, {Kerzendorf}, {Koch}, {Kulumani}, {Lee}, {Ly}, {Ma}, {MacBride}, {Maljaars}, {Muna}, {Murphy}, {Norman}, {O'Steen},
  {Oman}, {Pacifici}, {Pascual}, {Pascual-Granado}, {Patil}, {Perren}, {Pickering}, {Rastogi}, {Roulston}, {Ryan}, {Rykoff}, {Sabater}, {Sakurikar}, {Salgado}, {Sanghi}, {Saunders}, {Savchenko}, {Schwardt}, {Seifert-Eckert}, {Shih}, {Jain}, {Shukla}, {Sick}, {Simpson}, {Singanamalla}, {Singer}, {Singhal}, {Sinha}, {Sip{H{o}}cz}, {Spitler}, {Stansby}, {Streicher}, {{{S}}umak}, {Swinbank}, {Taranu}, {Tewary}, {Tremblay}, {Val-Borro}, {Van Kooten}, {Vasovi{'c}}, {Verma}, {de Miranda Cardoso}, {Williams}, {Wilson}, {Winkel}, {Wood-Vasey}, {Xue}, {Yoachim}, {Zhang}, {Zonca}, \& {Astropy Project Contributors}}]{astropy:2022}
{Astropy Collaboration}, {Price-Whelan}, A.~M., {Lim}, P.~L., {et~al.} 2022, \apj, 935, 167, \dodoi{10.3847/1538-4357/ac7c74}

\bibitem[{{Bae} {et~al.}(2023){Bae}, {Isella}, {Zhu}, {Martin}, {Okuzumi}, \& {Suriano}}]{bae2023}
{Bae}, J., {Isella}, A., {Zhu}, Z., {et~al.} 2023, in Astronomical Society of the Pacific Conference Series, Vol. 534, Protostars and Planets VII, ed. S.~{Inutsuka}, Y.~{Aikawa}, T.~{Muto}, K.~{Tomida}, \& M.~{Tamura}, 423, \dodoi{10.48550/arXiv.2210.13314}

\bibitem[{{Bajaj} {et~al.}(2024){Bajaj}, {Pascucci}, {Gorti}, {Alexander}, {Sellek}, {Morrison}, {Gaspar}, {Clarke}, {Xie}, {Ballabio}, \& {Deng}}]{bajaj24}
{Bajaj}, N.~S., {Pascucci}, I., {Gorti}, U., {et~al.} 2024, \aj, 167, 127, \dodoi{10.3847/1538-3881/ad22e1}

\bibitem[{{Banzatti} {et~al.}(2019){Banzatti}, {Pascucci}, {Edwards}, {Fang}, {Gorti}, \& {Flock}}]{banzatti19}
{Banzatti}, A., {Pascucci}, I., {Edwards}, S., {et~al.} 2019, \apj, 870, 76, \dodoi{10.3847/1538-4357/aaf1aa}

\bibitem[{{Banzatti} {et~al.}(2017){Banzatti}, {Pontoppidan}, {Salyk}, {Herczeg}, {van Dishoeck}, \& {Blake}}]{banzatti2017}
{Banzatti}, A., {Pontoppidan}, K.~M., {Salyk}, C., {et~al.} 2017, \apj, 834, 152, \dodoi{10.3847/1538-4357/834/2/152}

\bibitem[{{Banzatti} {et~al.}(2012){Banzatti}, {Meyer}, {Bruderer}, {Geers}, {Pascucci}, {Lahuis}, {Juh{\'a}sz}, {Henning}, \& {{\'A}brah{\'a}m}}]{banzatti12}
{Banzatti}, A., {Meyer}, M.~R., {Bruderer}, S., {et~al.} 2012, \apj, 745, 90, \dodoi{10.1088/0004-637X/745/1/90}

\bibitem[{{Banzatti} {et~al.}(2020){Banzatti}, {Pascucci}, {Bosman}, {Pinilla}, {Salyk}, {Herczeg}, {Pontoppidan}, {Vazquez}, {Watkins}, {Krijt}, {Hendler}, \& {Long}}]{banzatti2020}
{Banzatti}, A., {Pascucci}, I., {Bosman}, A.~D., {et~al.} 2020, \apj, 903, 124, \dodoi{10.3847/1538-4357/abbc1a}

\bibitem[{{Banzatti} {et~al.}(2022){Banzatti}, {Abernathy}, {Brittain}, {Bosman}, {Pontoppidan}, {Boogert}, {Jensen}, {Carr}, {Najita}, {Grant}, {Sigler}, {Sanchez}, {Kern}, \& {Rayner}}]{banzatti22}
{Banzatti}, A., {Abernathy}, K.~M., {Brittain}, S., {et~al.} 2022, \aj, 163, 174, \dodoi{10.3847/1538-3881/ac52f0}

\bibitem[{{Banzatti} {et~al.}(2023{\natexlab{a}}){Banzatti}, {Pontoppidan}, {Carr}, {Jellison}, {Pascucci}, {Najita}, {Romero-Mirza}, {{\"O}berg}, {Kalyaan}, {Pinilla}, {Krijt}, {Long}, {Lambrechts}, {Rosotti}, {Herczeg}, {Salyk}, {Zhang}, {Bergin}, {Ballering}, {Meyer}, {Bruderer}, \& {Jdiscs Collaboration}}]{banzatti23b}
{Banzatti}, A., {Pontoppidan}, K.~M., {Carr}, J.~S., {et~al.} 2023{\natexlab{a}}, \apjl, 957, L22, \dodoi{10.3847/2041-8213/acf5ec}

\bibitem[{{Banzatti} {et~al.}(2023{\natexlab{b}}){Banzatti}, {Pontoppidan}, {P{\'e}re Ch{\'a}vez}, {Salyk}, {Diehl}, {Bruderer}, {Herczeg}, {Carmona}, {Pascucci}, {Brittain}, {Jensen}, {Grant}, {van Dishoeck}, {Kamp}, {Bosman}, {{\"O}berg}, {Blake}, {Meyer}, {Gaidos}, {Boogert}, {Rayner}, \& {Wheeler}}]{banzatti23a}
{Banzatti}, A., {Pontoppidan}, K.~M., {P{\'e}re Ch{\'a}vez}, J., {et~al.} 2023{\natexlab{b}}, \aj, 165, 72, \dodoi{10.3847/1538-3881/aca80b}

\bibitem[{{Banzatti} {et~al.}(2025){Banzatti}, {Salyk}, {Pontoppidan}, {Carr}, {Zhang}, {Arulanantham}, {Krijt}, {{\"O}berg}, {Cleeves}, {Najita}, {Pascucci}, {Blake}, {Romero-Mirza}, {Bergin}, {Cieza}, {Pinilla}, {Long}, {Mallaney}, {Xie}, {Waggoner}, {Kaeufer}, \& {The Jdiscs Collaboration}}]{banzatti24}
{Banzatti}, A., {Salyk}, C., {Pontoppidan}, K.~M., {et~al.} 2025, \aj, 169, 165, \dodoi{10.3847/1538-3881/ada962}

\bibitem[{{Booth} {et~al.}(2017){Booth}, {Clarke}, {Madhusudhan}, \& {Ilee}}]{booth17}
{Booth}, R.~A., {Clarke}, C.~J., {Madhusudhan}, N., \& {Ilee}, J.~D. 2017, \mnras, 469, 3994, \dodoi{10.1093/mnras/stx1103}

\bibitem[{{Booth} \& {Ilee}(2019)}]{booth19}
{Booth}, R.~A., \& {Ilee}, J.~D. 2019, \mnras, 487, 3998, \dodoi{10.1093/mnras/stz1488}

\bibitem[{{Bosman} {et~al.}(2022){Bosman}, {Bergin}, {Calahan}, \& {Duval}}]{bosman22}
{Bosman}, A.~D., {Bergin}, E.~A., {Calahan}, J., \& {Duval}, S.~E. 2022, \apjl, 930, L26, \dodoi{10.3847/2041-8213/ac66ce}

\bibitem[{{Brown} {et~al.}(2013){Brown}, {Pontoppidan}, {van Dishoeck}, {Herczeg}, {Blake}, \& {Smette}}]{brown13}
{Brown}, J.~M., {Pontoppidan}, K.~M., {van Dishoeck}, E.~F., {et~al.} 2013, \apj, 770, 94, \dodoi{10.1088/0004-637X/770/2/94}

\bibitem[{{Brown} {et~al.}(2007){Brown}, {Blake}, {Dullemond}, {Mer{\'\i}n}, {Augereau}, {Boogert}, {Evans}, {Geers}, {Lahuis}, {Kessler-Silacci}, {Pontoppidan}, \& {van Dishoeck}}]{brown07}
{Brown}, J.~M., {Blake}, G.~A., {Dullemond}, C.~P., {et~al.} 2007, \apjl, 664, L107, \dodoi{10.1086/520808}

\bibitem[{{Brownlee} {et~al.}(2006){Brownlee}, {Tsou}, {Al{\'e}on}, {Alexander}, {Araki}, {Bajt}, {Baratta}, {Bastien}, {Bland}, {Bleuet}, {Borg}, {Bradley}, {Brearley}, {Brenker}, {Brennan}, {Bridges}, {Browning}, {Brucato}, {Bullock}, {Burchell}, {Busemann}, {Butterworth}, {Chaussidon}, {Cheuvront}, {Chi}, {Cintala}, {Clark}, {Clemett}, {Cody}, {Colangeli}, {Cooper}, {Cordier}, {Daghlian}, {Dai}, {D'Hendecourt}, {Djouadi}, {Dominguez}, {Duxbury}, {Dworkin}, {Ebel}, {Economou}, {Fakra}, {Fairey}, {Fallon}, {Ferrini}, {Ferroir}, {Fleckenstein}, {Floss}, {Flynn}, {Franchi}, {Fries}, {Gainsforth}, {Gallien}, {Genge}, {Gilles}, {Gillet}, {Gilmour}, {Glavin}, {Gounelle}, {Grady}, {Graham}, {Grant}, {Green}, {Grossemy}, {Grossman}, {Grossman}, {Guan}, {Hagiya}, {Harvey}, {Heck}, {Herzog}, {Hoppe}, {H{\"o}rz}, {Huth}, {Hutcheon}, {Ignatyev}, {Ishii}, {Ito}, {Jacob}, {Jacobsen}, {Jacobsen}, {Jones}, {Joswiak}, {Jurewicz}, {Kearsley}, {Keller}, {Khodja}, {Kilcoyne}, {Kissel}, {Krot}, {Langenhorst}, {Lanzirotti},
  {Le}, {Leshin}, {Leitner}, {Lemelle}, {Leroux}, {Liu}, {Luening}, {Lyon}, {MacPherson}, {Marcus}, {Marhas}, {Marty}, {Matrajt}, {McKeegan}, {Meibom}, {Mennella}, {Messenger}, {Messenger}, {Mikouchi}, {Mostefaoui}, {Nakamura}, {Nakano}, {Newville}, {Nittler}, {Ohnishi}, {Ohsumi}, {Okudaira}, {Papanastassiou}, {Palma}, {Palumbo}, {Pepin}, {Perkins}, {Perronnet}, {Pianetta}, {Rao}, {Rietmeijer}, {Robert}, {Rost}, {Rotundi}, {Ryan}, {Sandford}, {Schwandt}, {See}, {Schlutter}, {Sheffield-Parker}, {Simionovici}, {Simon}, {Sitnitsky}, {Snead}, {Spencer}, {Stadermann}, {Steele}, {Stephan}, {Stroud}, {Susini}, {Sutton}, {Suzuki}, {Taheri}, {Taylor}, {Teslich}, {Tomeoka}, {Tomioka}, {Toppani}, {Trigo-Rodr{\'\i}guez}, {Troadec}, {Tsuchiyama}, {Tuzzolino}, {Tyliszczak}, {Uesugi}, {Velbel}, {Vellenga}, {Vicenzi}, {Vincze}, {Warren}, {Weber}, {Weisberg}, {Westphal}, {Wirick}, {Wooden}, {Wopenka}, {Wozniakiewicz}, {Wright}, {Yabuta}, {Yano}, {Young}, {Zare}, {Zega}, {Ziegler}, {Zimmerman}, {Zinner}, \&
  {Zolensky}}]{Brownlee06}
{Brownlee}, D., {Tsou}, P., {Al{\'e}on}, J., {et~al.} 2006, Science, 314, 1711, \dodoi{10.1126/science.1135840}

\bibitem[{{Carr} \& {Najita}(2008)}]{carr08}
{Carr}, J.~S., \& {Najita}, J.~R. 2008, Science, 319, 1504, \dodoi{10.1126/science.1153807}

\bibitem[{{Carr} \& {Najita}(2011)}]{carr11}
---. 2011, \apj, 733, 102, \dodoi{10.1088/0004-637X/733/2/102}

\bibitem[{{Carr} \& {Najita}(2014)}]{carr14}
---. 2014, \apj, 788, 66, \dodoi{10.1088/0004-637X/788/1/66}

\bibitem[{{Chihara} {et~al.}(2002){Chihara}, {Koike}, {Tsuchiyama}, {Tachibana}, \& {Sakamoto}}]{chihara2002}
{Chihara}, H., {Koike}, C., {Tsuchiyama}, A., {Tachibana}, S., \& {Sakamoto}, D. 2002, \aap, 391, 267, \dodoi{10.1051/0004-6361:20020791}

\bibitem[{{Chown} {et~al.}(2024){Chown}, {Sidhu}, {Peeters}, {Tielens}, {Cami}, {Bern{\'e}}, {Habart}, {Alarc{\'o}n}, {Canin}, {Schroetter}, {Trahin}, {Van De Putte}, {Abergel}, {Bergin}, {Bernard-Salas}, {Boersma}, {Bron}, {Cuadrado}, {Dartois}, {Dicken}, {El-Yajouri}, {Fuente}, {Goicoechea}, {Gordon}, {Issa}, {Joblin}, {Kannavou}, {Khan}, {Lacinbala}, {Languignon}, {Le Gal}, {Maragkoudakis}, {Meshaka}, {Okada}, {Onaka}, {Pasquini}, {Pound}, {Robberto}, {R{\"o}llig}, {Schefter}, {Schirmer}, {Vicente}, {Wolfire}, {Zannese}, {Aleman}, {Allamandola}, {Auchettl}, {Baratta}, {Bejaoui}, {Bera}, {Black}, {Boulanger}, {Bouwman}, {Brandl}, {Brechignac}, {Br{\"u}nken}, {Buragohain}, {Burkhardt}, {Candian}, {Cazaux}, {Cernicharo}, {Chabot}, {Chakraborty}, {Champion}, {Colgan}, {Cooke}, {Coutens}, {Cox}, {Demyk}, {Meyer}, {Foschino}, {Garc{\'\i}a-Lario}, {Gavilan}, {Gerin}, {Gottlieb}, {Guillard}, {Gusdorf}, {Hartigan}, {He}, {Herbst}, {Hornekaer}, {J{\"a}ger}, {Janot-Pacheco}, {Kaufman}, {Kemper}, {Kendrew},
  {Kirsanova}, {Klaassen}, {Kwok}, {Labiano}, {Lai}, {Lee}, {Lefloch}, {Le Petit}, {Li}, {Linz}, {Mackie}, {Madden}, {Mascetti}, {McGuire}, {Merino}, {Micelotta}, {Misselt}, {Morse}, {Mulas}, {Neelamkodan}, {Ohsawa}, {Omont}, {Paladini}, {Palumbo}, {Pathak}, {Pendleton}, {Petrignani}, {Pino}, {Puga}, {Rangwala}, {Rapacioli}, {Ricca}, {Roman-Duval}, {Roser}, {Roueff}, {Rouill{\'e}}, {Salama}, {Sales}, {Sandstrom}, {Sarre}, {Sciamma-O'Brien}, {Sellgren}, {Shenoy}, {Teyssier}, {Thomas}, {Togi}, {Verstraete}, {Witt}, {Wootten}, {Zettergren}, {Zhang}, {Zhang}, \& {Zhen}}]{chown2024}
{Chown}, R., {Sidhu}, A., {Peeters}, E., {et~al.} 2024, \aap, 685, A75, \dodoi{10.1051/0004-6361/202346662}

\bibitem[{{Cieza} {et~al.}(2010){Cieza}, {Schreiber}, {Romero}, {Mora}, {Merin}, {Swift}, {Orellana}, {Williams}, {Harvey}, \& {Evans}}]{cieza2010}
{Cieza}, L.~A., {Schreiber}, M.~R., {Romero}, G.~A., {et~al.} 2010, \apj, 712, 925, \dodoi{10.1088/0004-637X/712/2/925}

\bibitem[{{Colmenares} {et~al.}(2024){Colmenares}, {Bergin}, {Salyk}, {Pontopiddan}, {Arulanantham}, {Calahan}, {Banzatti}, {Andrews}, {Blake}, {Ciesla}, {Green}, {Long}, {Lambrechts}, {Najita}, {Pascucci}, {Pinilla}, {Krijt}, {Trapman}, \& {the JDISCS Collaboration}}]{colmenares24}
{Colmenares}, M.~J., {Bergin}, E., {Salyk}, C., {et~al.} 2024, arXiv e-prints, arXiv:2410.18187.
\newblock \doarXiv{2410.18187}

\bibitem[{{Diop} {et~al.}(2024){Diop}, {Cleeves}, {Anderson}, {Pegues}, \& {Plunkett}}]{diop24}
{Diop}, A., {Cleeves}, L.~I., {Anderson}, D.~E., {Pegues}, J., \& {Plunkett}, A. 2024, \apj, 962, 90, \dodoi{10.3847/1538-4357/ad11ed}

\bibitem[{{Donati} {et~al.}(2024){Donati}, {Finociety}, {Cristofari}, {Alencar}, {Moutou}, {Delfosse}, {Fouqu{\'e}}, {Arnold}, {Baruteau}, {K{\'o}sp{\'a}l}, {M{\'e}nard}, {Carmona}, {Grankin}, {Takami}, {Artigau}, {Doyon}, {H{\'e}brard}, \& {the SPIRou science team}}]{donati24}
{Donati}, J.~F., {Finociety}, B., {Cristofari}, P.~I., {et~al.} 2024, \mnras, 530, 264, \dodoi{10.1093/mnras/stae675}

\bibitem[{{Dra{\.z}kowska} \& {Alibert}(2017)}]{drazkowska17}
{Dra{\.z}kowska}, J., \& {Alibert}, Y. 2017, \aap, 608, A92, \dodoi{10.1051/0004-6361/201731491}

\bibitem[{{Dullemond} {et~al.}(2018){Dullemond}, {Birnstiel}, {Huang}, {Kurtovic}, {Andrews}, {Guzm{\'a}n}, {P{\'e}rez}, {Isella}, {Zhu}, {Benisty}, {Wilner}, {Bai}, {Carpenter}, {Zhang}, \& {Ricci}}]{dullemond18}
{Dullemond}, C.~P., {Birnstiel}, T., {Huang}, J., {et~al.} 2018, \apjl, 869, L46, \dodoi{10.3847/2041-8213/aaf742}

\bibitem[{{Easterwood} {et~al.}(2024){Easterwood}, {Kalyaan}, \& {Banzatti}}]{easterwood24}
{Easterwood}, W., {Kalyaan}, A., \& {Banzatti}, A. 2024, \apj, 977, 21, \dodoi{10.3847/1538-4357/ad891d}

\bibitem[{{Eisner} {et~al.}(2005){Eisner}, {Hillenbrand}, {White}, {Akeson}, \& {Sargent}}]{eisner05}
{Eisner}, J.~A., {Hillenbrand}, L.~A., {White}, R.~J., {Akeson}, R.~L., \& {Sargent}, A.~I. 2005, \apj, 623, 952, \dodoi{10.1086/428828}

\bibitem[{{Espaillat} {et~al.}(2013){Espaillat}, {Ingleby}, {Furlan}, {McClure}, {Spatzier}, {Nieusma}, {Calvet}, {Bergin}, {Hartmann}, {Miller}, \& {Muzerolle}}]{espaillat13}
{Espaillat}, C., {Ingleby}, L., {Furlan}, E., {et~al.} 2013, \apj, 762, 62, \dodoi{10.1088/0004-637X/762/1/62}

\bibitem[{{Espaillat} {et~al.}(2023){Espaillat}, {Thanathibodee}, {Pittman}, {Sturm}, {McClure}, {Calvet}, {Walter}, {Franco-Hern{\'a}ndez}, \& {Muzerolle Page}}]{espaillat23}
{Espaillat}, C.~C., {Thanathibodee}, T., {Pittman}, C.~V., {et~al.} 2023, \apjl, 958, L4, \dodoi{10.3847/2041-8213/ad023d}

\bibitem[{{Evans} {et~al.}(1991){Evans}, {Lacy}, \& {Carr}}]{evans1991}
{Evans}, Neal~J., I., {Lacy}, J.~H., \& {Carr}, J.~S. 1991, \apj, 383, 674, \dodoi{10.1086/170824}

\bibitem[{{Fairlamb} {et~al.}(2015){Fairlamb}, {Oudmaijer}, {Mendigut{\'\i}a}, {Ilee}, \& {van den Ancker}}]{fairlamb15}
{Fairlamb}, J.~R., {Oudmaijer}, R.~D., {Mendigut{\'\i}a}, I., {Ilee}, J.~D., \& {van den Ancker}, M.~E. 2015, \mnras, 453, 976, \dodoi{10.1093/mnras/stv1576}

\bibitem[{{Fang} {et~al.}(2018){Fang}, {Pascucci}, {Edwards}, {Gorti}, {Banzatti}, {Flock}, {Hartigan}, {Herczeg}, \& {Dupree}}]{Fang18}
{Fang}, M., {Pascucci}, I., {Edwards}, S., {et~al.} 2018, \apj, 868, 28, \dodoi{10.3847/1538-4357/aae780}

\bibitem[{{Fedele} {et~al.}(2012){Fedele}, {Bruderer}, {van Dishoeck}, {Herczeg}, {Evans}, {Bouwman}, {Henning}, \& {Green}}]{fedele12}
{Fedele}, D., {Bruderer}, S., {van Dishoeck}, E.~F., {et~al.} 2012, \aap, 544, L9, \dodoi{10.1051/0004-6361/201219615}

\bibitem[{{Foreman-Mackey} {et~al.}(2013){Foreman-Mackey}, {Hogg}, {Lang}, \& {Goodman}}]{emcee}
{Foreman-Mackey}, D., {Hogg}, D.~W., {Lang}, D., \& {Goodman}, J. 2013, \pasp, 125, 306, \dodoi{10.1086/670067}

\bibitem[{{Francis} \& {van der Marel}(2020)}]{francis2020}
{Francis}, L., \& {van der Marel}, N. 2020, \apj, 892, 111, \dodoi{10.3847/1538-4357/ab7b63}

\bibitem[{{Frasca} {et~al.}(2017){Frasca}, {Biazzo}, {Alcal{\'a}}, {Manara}, {Stelzer}, {Covino}, \& {Antoniucci}}]{frasca2017}
{Frasca}, A., {Biazzo}, K., {Alcal{\'a}}, J.~M., {et~al.} 2017, \aap, 602, A33, \dodoi{10.1051/0004-6361/201630108}

\bibitem[{{Furlan} {et~al.}(2009){Furlan}, {Watson}, {McClure}, {Manoj}, {Espaillat}, {D'Alessio}, {Calvet}, {Kim}, {Sargent}, {Forrest}, \& {Hartmann}}]{furlan09}
{Furlan}, E., {Watson}, D.~M., {McClure}, M.~K., {et~al.} 2009, \apj, 703, 1964, \dodoi{10.1088/0004-637X/703/2/1964}

\bibitem[{{Gaia Collaboration} {et~al.}(2016){Gaia Collaboration}, {Prusti}, {de Bruijne}, {Brown}, {Vallenari}, {Babusiaux}, {Bailer-Jones}, {Bastian}, {Biermann}, {Evans}, {Eyer}, {Jansen}, {Jordi}, {Klioner}, {Lammers}, {Lindegren}, {Luri}, {Mignard}, {Milligan}, {Panem}, {Poinsignon}, {Pourbaix}, {Randich}, {Sarri}, {Sartoretti}, {Siddiqui}, {Soubiran}, {Valette}, {van Leeuwen}, {Walton}, {Aerts}, {Arenou}, {Cropper}, {Drimmel}, {H{\o}g}, {Katz}, {Lattanzi}, {O'Mullane}, {Grebel}, {Holland}, {Huc}, {Passot}, {Bramante}, {Cacciari}, {Casta{\~n}eda}, {Chaoul}, {Cheek}, {De Angeli}, {Fabricius}, {Guerra}, {Hern{\'a}ndez}, {Jean-Antoine-Piccolo}, {Masana}, {Messineo}, {Mowlavi}, {Nienartowicz}, {Ord{\'o}{\~n}ez-Blanco}, {Panuzzo}, {Portell}, {Richards}, {Riello}, {Seabroke}, {Tanga}, {Th{\'e}venin}, {Torra}, {Els}, {Gracia-Abril}, {Comoretto}, {Garcia-Reinaldos}, {Lock}, {Mercier}, {Altmann}, {Andrae}, {Astraatmadja}, {Bellas-Velidis}, {Benson}, {Berthier}, {Blomme}, {Busso}, {Carry}, {Cellino}, {Clementini},
  {Cowell}, {Creevey}, {Cuypers}, {Davidson}, {De Ridder}, {de Torres}, {Delchambre}, {Dell'Oro}, {Ducourant}, {Fr{\'e}mat}, {Garc{\'\i}a-Torres}, {Gosset}, {Halbwachs}, {Hambly}, {Harrison}, {Hauser}, {Hestroffer}, {Hodgkin}, {Huckle}, {Hutton}, {Jasniewicz}, {Jordan}, {Kontizas}, {Korn}, {Lanzafame}, {Manteiga}, {Moitinho}, {Muinonen}, {Osinde}, {Pancino}, {Pauwels}, {Petit}, {Recio-Blanco}, {Robin}, {Sarro}, {Siopis}, {Smith}, {Smith}, {Sozzetti}, {Thuillot}, {van Reeven}, {Viala}, {Abbas}, {Abreu Aramburu}, {Accart}, {Aguado}, {Allan}, {Allasia}, {Altavilla}, {{\'A}lvarez}, {Alves}, {Anderson}, {Andrei}, {Anglada Varela}, {Antiche}, {Antoja}, {Ant{\'o}n}, {Arcay}, {Atzei}, {Ayache}, {Bach}, {Baker}, {Balaguer-N{\'u}{\~n}ez}, {Barache}, {Barata}, {Barbier}, {Barblan}, {Baroni}, {Barrado y Navascu{\'e}s}, {Barros}, {Barstow}, {Becciani}, {Bellazzini}, {Bellei}, {Bello Garc{\'\i}a}, {Belokurov}, {Bendjoya}, {Berihuete}, {Bianchi}, {Bienaym{\'e}}, {Billebaud}, {Blagorodnova}, {Blanco-Cuaresma}, {Boch},
  {Bombrun}, {Borrachero}, {Bouquillon}, {Bourda}, {Bouy}, {Bragaglia}, {Breddels}, {Brouillet}, {Br{\"u}semeister}, {Bucciarelli}, {Budnik}, {Burgess}, {Burgon}, {Burlacu}, {Busonero}, {Buzzi}, {Caffau}, {Cambras}, {Campbell}, {Cancelliere}, {Cantat-Gaudin}, {Carlucci}, {Carrasco}, {Castellani}, {Charlot}, {Charnas}, {Charvet}, {Chassat}, {Chiavassa}, {Clotet}, {Cocozza}, {Collins}, {Collins}, {Costigan}, {Crifo}, {Cross}, {Crosta}, {Crowley}, {Dafonte}, {Damerdji}, {Dapergolas}, {David}, {David}, {De Cat}, {de Felice}, {de Laverny}, {De Luise}, {De March}, {de Martino}, {de Souza}, {Debosscher}, {del Pozo}, {Delbo}, {Delgado}, {Delgado}, {di Marco}, {Di Matteo}, {Diakite}, {Distefano}, {Dolding}, {Dos Anjos}, {Drazinos}, {Dur{\'a}n}, {Dzigan}, {Ecale}, {Edvardsson}, {Enke}, {Erdmann}, {Escolar}, {Espina}, {Evans}, {Eynard Bontemps}, {Fabre}, {Fabrizio}, {Faigler}, {Falc{\~a}o}, {Farr{\`a}s Casas}, {Faye}, {Federici}, {Fedorets}, {Fern{\'a}ndez-Hern{\'a}ndez}, {Fernique}, {Fienga}, {Figueras}, {Filippi},
  {Findeisen}, {Fonti}, {Fouesneau}, {Fraile}, {Fraser}, {Fuchs}, {Furnell}, {Gai}, {Galleti}, {Galluccio}, {Garabato}, {Garc{\'\i}a-Sedano}, {Gar{\'e}}, {Garofalo}, {Garralda}, {Gavras}, {Gerssen}, {Geyer}, {Gilmore}, {Girona}, {Giuffrida}, {Gomes}, {Gonz{\'a}lez-Marcos}, {Gonz{\'a}lez-N{\'u}{\~n}ez}, {Gonz{\'a}lez-Vidal}, {Granvik}, {Guerrier}, {Guillout}, {Guiraud}, {G{\'u}rpide}, {Guti{\'e}rrez-S{\'a}nchez}, {Guy}, {Haigron}, {Hatzidimitriou}, {Haywood}, {Heiter}, {Helmi}, {Hobbs}, {Hofmann}, {Holl}, {Holland}, {Hunt}, {Hypki}, {Icardi}, {Irwin}, {Jevardat de Fombelle}, {Jofr{\'e}}, {Jonker}, {Jorissen}, {Julbe}, {Karampelas}, {Kochoska}, {Kohley}, {Kolenberg}, {Kontizas}, {Koposov}, {Kordopatis}, {Koubsky}, {Kowalczyk}, {Krone-Martins}, {Kudryashova}, {Kull}, {Bachchan}, {Lacoste-Seris}, {Lanza}, {Lavigne}, {Le Poncin-Lafitte}, {Lebreton}, {Lebzelter}, {Leccia}, {Leclerc}, {Lecoeur-Taibi}, {Lemaitre}, {Lenhardt}, {Leroux}, {Liao}, {Licata}, {Lindstr{\o}m}, {Lister}, {Livanou}, {Lobel}, {L{\"o}ffler},
  {L{\'o}pez}, {Lopez-Lozano}, {Lorenz}, {Loureiro}, {MacDonald}, {Magalh{\~a}es Fernandes}, {Managau}, {Mann}, {Mantelet}, {Marchal}, {Marchant}, {Marconi}, {Marie}, {Marinoni}, {Marrese}, {Marschalk{\'o}}, {Marshall}, {Mart{\'\i}n-Fleitas}, {Martino}, {Mary}, {Matijevi{\v{c}}}, {Mazeh}, {McMillan}, {Messina}, {Mestre}, {Michalik}, {Millar}, {Miranda}, {Molina}, {Molinaro}, {Molinaro}, {Moln{\'a}r}, {Moniez}, {Montegriffo}, {Monteiro}, {Mor}, {Mora}, {Morbidelli}, {Morel}, {Morgenthaler}, {Morley}, {Morris}, {Mulone}, {Muraveva}, {Musella}, {Narbonne}, {Nelemans}, {Nicastro}, {Noval}, {Ord{\'e}novic}, {Ordieres-Mer{\'e}}, {Osborne}, {Pagani}, {Pagano}, {Pailler}, {Palacin}, {Palaversa}, {Parsons}, {Paulsen}, {Pecoraro}, {Pedrosa}, {Pentik{\"a}inen}, {Pereira}, {Pichon}, {Piersimoni}, {Pineau}, {Plachy}, {Plum}, {Poujoulet}, {Pr{\v{s}}a}, {Pulone}, {Ragaini}, {Rago}, {Rambaux}, {Ramos-Lerate}, {Ranalli}, {Rauw}, {Read}, {Regibo}, {Renk}, {Reyl{\'e}}, {Ribeiro}, {Rimoldini}, {Ripepi}, {Riva}, {Rixon},
  {Roelens}, {Romero-G{\'o}mez}, {Rowell}, {Royer}, {Rudolph}, {Ruiz-Dern}, {Sadowski}, {Sagrist{\`a} Sell{\'e}s}, {Sahlmann}, {Salgado}, {Salguero}, {Sarasso}, {Savietto}, {Schnorhk}, {Schultheis}, {Sciacca}, {Segol}, {Segovia}, {Segransan}, {Serpell}, {Shih}, {Smareglia}, {Smart}, {Smith}, {Solano}, {Solitro}, {Sordo}, {Soria Nieto}, {Souchay}, {Spagna}, {Spoto}, {Stampa}, {Steele}, {Steidelm{\"u}ller}, {Stephenson}, {Stoev}, {Suess}, {S{\"u}veges}, {Surdej}, {Szabados}, {Szegedi-Elek}, {Tapiador}, {Taris}, {Tauran}, {Taylor}, {Teixeira}, {Terrett}, {Tingley}, {Trager}, {Turon}, {Ulla}, {Utrilla}, {Valentini}, {van Elteren}, {Van Hemelryck}, {van Leeuwen}, {Varadi}, {Vecchiato}, {Veljanoski}, {Via}, {Vicente}, {Vogt}, {Voss}, {Votruba}, {Voutsinas}, {Walmsley}, {Weiler}, {Weingrill}, {Werner}, {Wevers}, {Whitehead}, {Wyrzykowski}, {Yoldas}, {{\v{Z}}erjal}, {Zucker}, {Zurbach}, {Zwitter}, {Alecu}, {Allen}, {Allende Prieto}, {Amorim}, {Anglada-Escud{\'e}}, {Arsenijevic}, {Azaz}, {Balm}, {Beck}, {Bernstein},
  {Bigot}, {Bijaoui}, {Blasco}, {Bonfigli}, {Bono}, {Boudreault}, {Bressan}, {Brown}, {Brunet}, {Bunclark}, {Buonanno}, {Butkevich}, {Carret}, {Carrion}, {Chemin}, {Ch{\'e}reau}, {Corcione}, {Darmigny}, {de Boer}, {de Teodoro}, {de Zeeuw}, {Delle Luche}, {Domingues}, {Dubath}, {Fodor}, {Fr{\'e}zouls}, {Fries}, {Fustes}, {Fyfe}, {Gallardo}, {Gallegos}, {Gardiol}, {Gebran}, {Gomboc}, {G{\'o}mez}, {Grux}, {Gueguen}, {Heyrovsky}, {Hoar}, {Iannicola}, {Isasi Parache}, {Janotto}, {Joliet}, {Jonckheere}, {Keil}, {Kim}, {Klagyivik}, {Klar}, {Knude}, {Kochukhov}, {Kolka}, {Kos}, {Kutka}, {Lainey}, {LeBouquin}, {Liu}, {Loreggia}, {Makarov}, {Marseille}, {Martayan}, {Martinez-Rubi}, {Massart}, {Meynadier}, {Mignot}, {Munari}, {Nguyen}, {Nordlander}, {Ocvirk}, {O'Flaherty}, {Olias Sanz}, {Ortiz}, {Osorio}, {Oszkiewicz}, {Ouzounis}, {Palmer}, {Park}, {Pasquato}, {Peltzer}, {Peralta}, {P{\'e}turaud}, {Pieniluoma}, {Pigozzi}, {Poels}, {Prat}, {Prod'homme}, {Raison}, {Rebordao}, {Risquez}, {Rocca-Volmerange}, {Rosen},
  {Ruiz-Fuertes}, {Russo}, {Sembay}, {Serraller Vizcaino}, {Short}, {Siebert}, {Silva}, {Sinachopoulos}, {Slezak}, {Soffel}, {Sosnowska}, {Strai{\v{z}}ys}, {ter Linden}, {Terrell}, {Theil}, {Tiede}, {Troisi}, {Tsalmantza}, {Tur}, {Vaccari}, {Vachier}, {Valles}, {Van Hamme}, {Veltz}, {Virtanen}, {Wallut}, {Wichmann}, {Wilkinson}, {Ziaeepour}, \& {Zschocke}}]{prusti2016}
{Gaia Collaboration}, {Prusti}, T., {de Bruijne}, J.~H.~J., {et~al.} 2016, \aap, 595, A1, \dodoi{10.1051/0004-6361/201629272}

\bibitem[{{Gaia Collaboration} {et~al.}(2018){Gaia Collaboration}, {Brown}, {Vallenari}, {Prusti}, {de Bruijne}, {Babusiaux}, {Bailer-Jones}, {Biermann}, {Evans}, {Eyer}, {Jansen}, {Jordi}, {Klioner}, {Lammers}, {Lindegren}, {Luri}, {Mignard}, {Panem}, {Pourbaix}, {Randich}, {Sartoretti}, {Siddiqui}, {Soubiran}, {van Leeuwen}, {Walton}, {Arenou}, {Bastian}, {Cropper}, {Drimmel}, {Katz}, {Lattanzi}, {Bakker}, {Cacciari}, {Casta{\~n}eda}, {Chaoul}, {Cheek}, {De Angeli}, {Fabricius}, {Guerra}, {Holl}, {Masana}, {Messineo}, {Mowlavi}, {Nienartowicz}, {Panuzzo}, {Portell}, {Riello}, {Seabroke}, {Tanga}, {Th{\'e}venin}, {Gracia-Abril}, {Comoretto}, {Garcia-Reinaldos}, {Teyssier}, {Altmann}, {Andrae}, {Audard}, {Bellas-Velidis}, {Benson}, {Berthier}, {Blomme}, {Burgess}, {Busso}, {Carry}, {Cellino}, {Clementini}, {Clotet}, {Creevey}, {Davidson}, {De Ridder}, {Delchambre}, {Dell'Oro}, {Ducourant}, {Fern{\'a}ndez-Hern{\'a}ndez}, {Fouesneau}, {Fr{\'e}mat}, {Galluccio}, {Garc{\'\i}a-Torres},
  {Gonz{\'a}lez-N{\'u}{\~n}ez}, {Gonz{\'a}lez-Vidal}, {Gosset}, {Guy}, {Halbwachs}, {Hambly}, {Harrison}, {Hern{\'a}ndez}, {Hestroffer}, {Hodgkin}, {Hutton}, {Jasniewicz}, {Jean-Antoine-Piccolo}, {Jordan}, {Korn}, {Krone-Martins}, {Lanzafame}, {Lebzelter}, {L{\"o}ffler}, {Manteiga}, {Marrese}, {Mart{\'\i}n-Fleitas}, {Moitinho}, {Mora}, {Muinonen}, {Osinde}, {Pancino}, {Pauwels}, {Petit}, {Recio-Blanco}, {Richards}, {Rimoldini}, {Robin}, {Sarro}, {Siopis}, {Smith}, {Sozzetti}, {S{\"u}veges}, {Torra}, {van Reeven}, {Abbas}, {Abreu Aramburu}, {Accart}, {Aerts}, {Altavilla}, {{\'A}lvarez}, {Alvarez}, {Alves}, {Anderson}, {Andrei}, {Anglada Varela}, {Antiche}, {Antoja}, {Arcay}, {Astraatmadja}, {Bach}, {Baker}, {Balaguer-N{\'u}{\~n}ez}, {Balm}, {Barache}, {Barata}, {Barbato}, {Barblan}, {Barklem}, {Barrado}, {Barros}, {Barstow}, {Bartholom{\'e} Mu{\~n}oz}, {Bassilana}, {Becciani}, {Bellazzini}, {Berihuete}, {Bertone}, {Bianchi}, {Bienaym{\'e}}, {Blanco-Cuaresma}, {Boch}, {Boeche}, {Bombrun}, {Borrachero},
  {Bossini}, {Bouquillon}, {Bourda}, {Bragaglia}, {Bramante}, {Breddels}, {Bressan}, {Brouillet}, {Br{\"u}semeister}, {Brugaletta}, {Bucciarelli}, {Burlacu}, {Busonero}, {Butkevich}, {Buzzi}, {Caffau}, {Cancelliere}, {Cannizzaro}, {Cantat-Gaudin}, {Carballo}, {Carlucci}, {Carrasco}, {Casamiquela}, {Castellani}, {Castro-Ginard}, {Charlot}, {Chemin}, {Chiavassa}, {Cocozza}, {Costigan}, {Cowell}, {Crifo}, {Crosta}, {Crowley}, {Cuypers}, {Dafonte}, {Damerdji}, {Dapergolas}, {David}, {David}, {de Laverny}, {De Luise}, {De March}, {de Martino}, {de Souza}, {de Torres}, {Debosscher}, {del Pozo}, {Delbo}, {Delgado}, {Delgado}, {Di Matteo}, {Diakite}, {Diener}, {Distefano}, {Dolding}, {Drazinos}, {Dur{\'a}n}, {Edvardsson}, {Enke}, {Eriksson}, {Esquej}, {Eynard Bontemps}, {Fabre}, {Fabrizio}, {Faigler}, {Falc{\~a}o}, {Farr{\`a}s Casas}, {Federici}, {Fedorets}, {Fernique}, {Figueras}, {Filippi}, {Findeisen}, {Fonti}, {Fraile}, {Fraser}, {Fr{\'e}zouls}, {Gai}, {Galleti}, {Garabato}, {Garc{\'\i}a-Sedano}, {Garofalo},
  {Garralda}, {Gavel}, {Gavras}, {Gerssen}, {Geyer}, {Giacobbe}, {Gilmore}, {Girona}, {Giuffrida}, {Glass}, {Gomes}, {Granvik}, {Gueguen}, {Guerrier}, {Guiraud}, {Guti{\'e}rrez-S{\'a}nchez}, {Haigron}, {Hatzidimitriou}, {Hauser}, {Haywood}, {Heiter}, {Helmi}, {Heu}, {Hilger}, {Hobbs}, {Hofmann}, {Holland}, {Huckle}, {Hypki}, {Icardi}, {Jan{\ss}en}, {Jevardat de Fombelle}, {Jonker}, {Juh{\'a}sz}, {Julbe}, {Karampelas}, {Kewley}, {Klar}, {Kochoska}, {Kohley}, {Kolenberg}, {Kontizas}, {Kontizas}, {Koposov}, {Kordopatis}, {Kostrzewa-Rutkowska}, {Koubsky}, {Lambert}, {Lanza}, {Lasne}, {Lavigne}, {Le Fustec}, {Le Poncin-Lafitte}, {Lebreton}, {Leccia}, {Leclerc}, {Lecoeur-Taibi}, {Lenhardt}, {Leroux}, {Liao}, {Licata}, {Lindstr{\o}m}, {Lister}, {Livanou}, {Lobel}, {L{\'o}pez}, {Managau}, {Mann}, {Mantelet}, {Marchal}, {Marchant}, {Marconi}, {Marinoni}, {Marschalk{\'o}}, {Marshall}, {Martino}, {Marton}, {Mary}, {Massari}, {Matijevi{\v{c}}}, {Mazeh}, {McMillan}, {Messina}, {Michalik}, {Millar}, {Molina}, {Molinaro},
  {Moln{\'a}r}, {Montegriffo}, {Mor}, {Morbidelli}, {Morel}, {Morris}, {Mulone}, {Muraveva}, {Musella}, {Nelemans}, {Nicastro}, {Noval}, {O'Mullane}, {Ord{\'e}novic}, {Ord{\'o}{\~n}ez-Blanco}, {Osborne}, {Pagani}, {Pagano}, {Pailler}, {Palacin}, {Palaversa}, {Panahi}, {Pawlak}, {Piersimoni}, {Pineau}, {Plachy}, {Plum}, {Poggio}, {Poujoulet}, {Pr{\v{s}}a}, {Pulone}, {Racero}, {Ragaini}, {Rambaux}, {Ramos-Lerate}, {Regibo}, {Reyl{\'e}}, {Riclet}, {Ripepi}, {Riva}, {Rivard}, {Rixon}, {Roegiers}, {Roelens}, {Romero-G{\'o}mez}, {Rowell}, {Royer}, {Ruiz-Dern}, {Sadowski}, {Sagrist{\`a} Sell{\'e}s}, {Sahlmann}, {Salgado}, {Salguero}, {Sanna}, {Santana-Ros}, {Sarasso}, {Savietto}, {Schultheis}, {Sciacca}, {Segol}, {Segovia}, {S{\'e}gransan}, {Shih}, {Siltala}, {Silva}, {Smart}, {Smith}, {Solano}, {Solitro}, {Sordo}, {Soria Nieto}, {Souchay}, {Spagna}, {Spoto}, {Stampa}, {Steele}, {Steidelm{\"u}ller}, {Stephenson}, {Stoev}, {Suess}, {Surdej}, {Szabados}, {Szegedi-Elek}, {Tapiador}, {Taris}, {Tauran}, {Taylor},
  {Teixeira}, {Terrett}, {Teyssandier}, {Thuillot}, {Titarenko}, {Torra Clotet}, {Turon}, {Ulla}, {Utrilla}, {Uzzi}, {Vaillant}, {Valentini}, {Valette}, {van Elteren}, {Van Hemelryck}, {van Leeuwen}, {Vaschetto}, {Vecchiato}, {Veljanoski}, {Viala}, {Vicente}, {Vogt}, {von Essen}, {Voss}, {Votruba}, {Voutsinas}, {Walmsley}, {Weiler}, {Wertz}, {Wevers}, {Wyrzykowski}, {Yoldas}, {{\v{Z}}erjal}, {Ziaeepour}, {Zorec}, {Zschocke}, {Zucker}, {Zurbach}, \& {Zwitter}}]{gaiaDR2}
{Gaia Collaboration}, {Brown}, A.~G.~A., {Vallenari}, A., {et~al.} 2018, \aap, 616, A1, \dodoi{10.1051/0004-6361/201833051}

\bibitem[{{Gaia Collaboration} {et~al.}(2021){Gaia Collaboration}, {Brown}, {Vallenari}, {Prusti}, {de Bruijne}, {Babusiaux}, {Biermann}, {Creevey}, {Evans}, {Eyer}, {Hutton}, {Jansen}, {Jordi}, {Klioner}, {Lammers}, {Lindegren}, {Luri}, {Mignard}, {Panem}, {Pourbaix}, {Randich}, {Sartoretti}, {Soubiran}, {Walton}, {Arenou}, {Bailer-Jones}, {Bastian}, {Cropper}, {Drimmel}, {Katz}, {Lattanzi}, {van Leeuwen}, {Bakker}, {Cacciari}, {Casta{\~n}eda}, {De Angeli}, {Ducourant}, {Fabricius}, {Fouesneau}, {Fr{\'e}mat}, {Guerra}, {Guerrier}, {Guiraud}, {Jean-Antoine Piccolo}, {Masana}, {Messineo}, {Mowlavi}, {Nicolas}, {Nienartowicz}, {Pailler}, {Panuzzo}, {Riclet}, {Roux}, {Seabroke}, {Sordo}, {Tanga}, {Th{\'e}venin}, {Gracia-Abril}, {Portell}, {Teyssier}, {Altmann}, {Andrae}, {Bellas-Velidis}, {Benson}, {Berthier}, {Blomme}, {Brugaletta}, {Burgess}, {Busso}, {Carry}, {Cellino}, {Cheek}, {Clementini}, {Damerdji}, {Davidson}, {Delchambre}, {Dell'Oro}, {Fern{\'a}ndez-Hern{\'a}ndez}, {Galluccio}, {Garc{\'\i}a-Lario},
  {Garcia-Reinaldos}, {Gonz{\'a}lez-N{\'u}{\~n}ez}, {Gosset}, {Haigron}, {Halbwachs}, {Hambly}, {Harrison}, {Hatzidimitriou}, {Heiter}, {Hern{\'a}ndez}, {Hestroffer}, {Hodgkin}, {Holl}, {Jan{\ss}en}, {Jevardat de Fombelle}, {Jordan}, {Krone-Martins}, {Lanzafame}, {L{\"o}ffler}, {Lorca}, {Manteiga}, {Marchal}, {Marrese}, {Moitinho}, {Mora}, {Muinonen}, {Osborne}, {Pancino}, {Pauwels}, {Petit}, {Recio-Blanco}, {Richards}, {Riello}, {Rimoldini}, {Robin}, {Roegiers}, {Rybizki}, {Sarro}, {Siopis}, {Smith}, {Sozzetti}, {Ulla}, {Utrilla}, {van Leeuwen}, {van Reeven}, {Abbas}, {Abreu Aramburu}, {Accart}, {Aerts}, {Aguado}, {Ajaj}, {Altavilla}, {{\'A}lvarez}, {{\'A}lvarez Cid-Fuentes}, {Alves}, {Anderson}, {Anglada Varela}, {Antoja}, {Audard}, {Baines}, {Baker}, {Balaguer-N{\'u}{\~n}ez}, {Balbinot}, {Balog}, {Barache}, {Barbato}, {Barros}, {Barstow}, {Bartolom{\'e}}, {Bassilana}, {Bauchet}, {Baudesson-Stella}, {Becciani}, {Bellazzini}, {Bernet}, {Bertone}, {Bianchi}, {Blanco-Cuaresma}, {Boch}, {Bombrun}, {Bossini},
  {Bouquillon}, {Bragaglia}, {Bramante}, {Breedt}, {Bressan}, {Brouillet}, {Bucciarelli}, {Burlacu}, {Busonero}, {Butkevich}, {Buzzi}, {Caffau}, {Cancelliere}, {C{\'a}novas}, {Cantat-Gaudin}, {Carballo}, {Carlucci}, {Carnerero}, {Carrasco}, {Casamiquela}, {Castellani}, {Castro-Ginard}, {Castro Sampol}, {Chaoul}, {Charlot}, {Chemin}, {Chiavassa}, {Cioni}, {Comoretto}, {Cooper}, {Cornez}, {Cowell}, {Crifo}, {Crosta}, {Crowley}, {Dafonte}, {Dapergolas}, {David}, {David}, {de Laverny}, {De Luise}, {De March}, {De Ridder}, {de Souza}, {de Teodoro}, {de Torres}, {del Peloso}, {del Pozo}, {Delbo}, {Delgado}, {Delgado}, {Delisle}, {Di Matteo}, {Diakite}, {Diener}, {Distefano}, {Dolding}, {Eappachen}, {Edvardsson}, {Enke}, {Esquej}, {Fabre}, {Fabrizio}, {Faigler}, {Fedorets}, {Fernique}, {Fienga}, {Figueras}, {Fouron}, {Fragkoudi}, {Fraile}, {Franke}, {Gai}, {Garabato}, {Garcia-Gutierrez}, {Garc{\'\i}a-Torres}, {Garofalo}, {Gavras}, {Gerlach}, {Geyer}, {Giacobbe}, {Gilmore}, {Girona}, {Giuffrida}, {Gomel}, {Gomez},
  {Gonzalez-Santamaria}, {Gonz{\'a}lez-Vidal}, {Granvik}, {Guti{\'e}rrez-S{\'a}nchez}, {Guy}, {Hauser}, {Haywood}, {Helmi}, {Hidalgo}, {Hilger}, {H{\l}adczuk}, {Hobbs}, {Holland}, {Huckle}, {Jasniewicz}, {Jonker}, {Juaristi Campillo}, {Julbe}, {Karbevska}, {Kervella}, {Khanna}, {Kochoska}, {Kontizas}, {Kordopatis}, {Korn}, {Kostrzewa-Rutkowska}, {Kruszy{\'n}ska}, {Lambert}, {Lanza}, {Lasne}, {Le Campion}, {Le Fustec}, {Lebreton}, {Lebzelter}, {Leccia}, {Leclerc}, {Lecoeur-Taibi}, {Liao}, {Licata}, {Lindstr{\o}m}, {Lister}, {Livanou}, {Lobel}, {Madrero Pardo}, {Managau}, {Mann}, {Marchant}, {Marconi}, {Marcos Santos}, {Marinoni}, {Marocco}, {Marshall}, {Martin Polo}, {Mart{\'\i}n-Fleitas}, {Masip}, {Massari}, {Mastrobuono-Battisti}, {Mazeh}, {McMillan}, {Messina}, {Michalik}, {Millar}, {Mints}, {Molina}, {Molinaro}, {Moln{\'a}r}, {Montegriffo}, {Mor}, {Morbidelli}, {Morel}, {Morris}, {Mulone}, {Munoz}, {Muraveva}, {Murphy}, {Musella}, {Noval}, {Ord{\'e}novic}, {Orr{\`u}}, {Osinde}, {Pagani}, {Pagano},
  {Palaversa}, {Palicio}, {Panahi}, {Pawlak}, {Pe{\~n}alosa Esteller}, {Penttil{\"a}}, {Piersimoni}, {Pineau}, {Plachy}, {Plum}, {Poggio}, {Poretti}, {Poujoulet}, {Pr{\v{s}}a}, {Pulone}, {Racero}, {Ragaini}, {Rainer}, {Raiteri}, {Rambaux}, {Ramos}, {Ramos-Lerate}, {Re Fiorentin}, {Regibo}, {Reyl{\'e}}, {Ripepi}, {Riva}, {Rixon}, {Robichon}, {Robin}, {Roelens}, {Rohrbasser}, {Romero-G{\'o}mez}, {Rowell}, {Royer}, {Rybicki}, {Sadowski}, {Sagrist{\`a} Sell{\'e}s}, {Sahlmann}, {Salgado}, {Salguero}, {Samaras}, {Sanchez Gimenez}, {Sanna}, {Santove{\~n}a}, {Sarasso}, {Schultheis}, {Sciacca}, {Segol}, {Segovia}, {S{\'e}gransan}, {Semeux}, {Shahaf}, {Siddiqui}, {Siebert}, {Siltala}, {Slezak}, {Smart}, {Solano}, {Solitro}, {Souami}, {Souchay}, {Spagna}, {Spoto}, {Steele}, {Steidelm{\"u}ller}, {Stephenson}, {S{\"u}veges}, {Szabados}, {Szegedi-Elek}, {Taris}, {Tauran}, {Taylor}, {Teixeira}, {Thuillot}, {Tonello}, {Torra}, {Torra}, {Turon}, {Unger}, {Vaillant}, {van Dillen}, {Vanel}, {Vecchiato}, {Viala}, {Vicente},
  {Voutsinas}, {Weiler}, {Wevers}, {Wyrzykowski}, {Yoldas}, {Yvard}, {Zhao}, {Zorec}, {Zucker}, {Zurbach}, \& {Zwitter}}]{brown2021}
---. 2021, \aap, 649, A1, \dodoi{10.1051/0004-6361/202039657}

\bibitem[{{Gaia Collaboration} {et~al.}(2023){Gaia Collaboration}, {Vallenari}, {Brown}, {Prusti}, {de Bruijne}, {Arenou}, {Babusiaux}, {Biermann}, {Creevey}, {Ducourant}, {Evans}, {Eyer}, {Guerra}, {Hutton}, {Jordi}, {Klioner}, {Lammers}, {Lindegren}, {Luri}, {Mignard}, {Panem}, {Pourbaix}, {Randich}, {Sartoretti}, {Soubiran}, {Tanga}, {Walton}, {Bailer-Jones}, {Bastian}, {Drimmel}, {Jansen}, {Katz}, {Lattanzi}, {van Leeuwen}, {Bakker}, {Cacciari}, {Casta{\~n}eda}, {De Angeli}, {Fabricius}, {Fouesneau}, {Fr{\'e}mat}, {Galluccio}, {Guerrier}, {Heiter}, {Masana}, {Messineo}, {Mowlavi}, {Nicolas}, {Nienartowicz}, {Pailler}, {Panuzzo}, {Riclet}, {Roux}, {Seabroke}, {Sordo}, {Th{\'e}venin}, {Gracia-Abril}, {Portell}, {Teyssier}, {Altmann}, {Andrae}, {Audard}, {Bellas-Velidis}, {Benson}, {Berthier}, {Blomme}, {Burgess}, {Busonero}, {Busso}, {C{\'a}novas}, {Carry}, {Cellino}, {Cheek}, {Clementini}, {Damerdji}, {Davidson}, {de Teodoro}, {Nu{\~n}ez Campos}, {Delchambre}, {Dell'Oro}, {Esquej},
  {Fern{\'a}ndez-Hern{\'a}ndez}, {Fraile}, {Garabato}, {Garc{\'\i}a-Lario}, {Gosset}, {Haigron}, {Halbwachs}, {Hambly}, {Harrison}, {Hern{\'a}ndez}, {Hestroffer}, {Hodgkin}, {Holl}, {Jan{\ss}en}, {Jevardat de Fombelle}, {Jordan}, {Krone-Martins}, {Lanzafame}, {L{\"o}ffler}, {Marchal}, {Marrese}, {Moitinho}, {Muinonen}, {Osborne}, {Pancino}, {Pauwels}, {Recio-Blanco}, {Reyl{\'e}}, {Riello}, {Rimoldini}, {Roegiers}, {Rybizki}, {Sarro}, {Siopis}, {Smith}, {Sozzetti}, {Utrilla}, {van Leeuwen}, {Abbas}, {{\'A}brah{\'a}m}, {Abreu Aramburu}, {Aerts}, {Aguado}, {Ajaj}, {Aldea-Montero}, {Altavilla}, {{\'A}lvarez}, {Alves}, {Anders}, {Anderson}, {Anglada Varela}, {Antoja}, {Baines}, {Baker}, {Balaguer-N{\'u}{\~n}ez}, {Balbinot}, {Balog}, {Barache}, {Barbato}, {Barros}, {Barstow}, {Bartolom{\'e}}, {Bassilana}, {Bauchet}, {Becciani}, {Bellazzini}, {Berihuete}, {Bernet}, {Bertone}, {Bianchi}, {Binnenfeld}, {Blanco-Cuaresma}, {Blazere}, {Boch}, {Bombrun}, {Bossini}, {Bouquillon}, {Bragaglia}, {Bramante}, {Breedt},
  {Bressan}, {Brouillet}, {Brugaletta}, {Bucciarelli}, {Burlacu}, {Butkevich}, {Buzzi}, {Caffau}, {Cancelliere}, {Cantat-Gaudin}, {Carballo}, {Carlucci}, {Carnerero}, {Carrasco}, {Casamiquela}, {Castellani}, {Castro-Ginard}, {Chaoul}, {Charlot}, {Chemin}, {Chiaramida}, {Chiavassa}, {Chornay}, {Comoretto}, {Contursi}, {Cooper}, {Cornez}, {Cowell}, {Crifo}, {Cropper}, {Crosta}, {Crowley}, {Dafonte}, {Dapergolas}, {David}, {David}, {de Laverny}, {De Luise}, {De March}, {De Ridder}, {de Souza}, {de Torres}, {del Peloso}, {del Pozo}, {Delbo}, {Delgado}, {Delisle}, {Demouchy}, {Dharmawardena}, {Di Matteo}, {Diakite}, {Diener}, {Distefano}, {Dolding}, {Edvardsson}, {Enke}, {Fabre}, {Fabrizio}, {Faigler}, {Fedorets}, {Fernique}, {Fienga}, {Figueras}, {Fournier}, {Fouron}, {Fragkoudi}, {Gai}, {Garcia-Gutierrez}, {Garcia-Reinaldos}, {Garc{\'\i}a-Torres}, {Garofalo}, {Gavel}, {Gavras}, {Gerlach}, {Geyer}, {Giacobbe}, {Gilmore}, {Girona}, {Giuffrida}, {Gomel}, {Gomez}, {Gonz{\'a}lez-N{\'u}{\~n}ez},
  {Gonz{\'a}lez-Santamar{\'\i}a}, {Gonz{\'a}lez-Vidal}, {Granvik}, {Guillout}, {Guiraud}, {Guti{\'e}rrez-S{\'a}nchez}, {Guy}, {Hatzidimitriou}, {Hauser}, {Haywood}, {Helmer}, {Helmi}, {Sarmiento}, {Hidalgo}, {Hilger}, {H{\l}adczuk}, {Hobbs}, {Holland}, {Huckle}, {Jardine}, {Jasniewicz}, {Jean-Antoine Piccolo}, {Jim{\'e}nez-Arranz}, {Jorissen}, {Juaristi Campillo}, {Julbe}, {Karbevska}, {Kervella}, {Khanna}, {Kontizas}, {Kordopatis}, {Korn}, {K{\'o}sp{\'a}l}, {Kostrzewa-Rutkowska}, {Kruszy{\'n}ska}, {Kun}, {Laizeau}, {Lambert}, {Lanza}, {Lasne}, {Le Campion}, {Lebreton}, {Lebzelter}, {Leccia}, {Leclerc}, {Lecoeur-Taibi}, {Liao}, {Licata}, {Lindstr{\o}m}, {Lister}, {Livanou}, {Lobel}, {Lorca}, {Loup}, {Madrero Pardo}, {Magdaleno Romeo}, {Managau}, {Mann}, {Manteiga}, {Marchant}, {Marconi}, {Marcos}, {Marcos Santos}, {Mar{\'\i}n Pina}, {Marinoni}, {Marocco}, {Marshall}, {Martin Polo}, {Mart{\'\i}n-Fleitas}, {Marton}, {Mary}, {Masip}, {Massari}, {Mastrobuono-Battisti}, {Mazeh}, {McMillan}, {Messina}, {Michalik},
  {Millar}, {Mints}, {Molina}, {Molinaro}, {Moln{\'a}r}, {Monari}, {Mongui{\'o}}, {Montegriffo}, {Montero}, {Mor}, {Mora}, {Morbidelli}, {Morel}, {Morris}, {Muraveva}, {Murphy}, {Musella}, {Nagy}, {Noval}, {Oca{\~n}a}, {Ogden}, {Ordenovic}, {Osinde}, {Pagani}, {Pagano}, {Palaversa}, {Palicio}, {Pallas-Quintela}, {Panahi}, {Payne-Wardenaar}, {Pe{\~n}alosa Esteller}, {Penttil{\"a}}, {Pichon}, {Piersimoni}, {Pineau}, {Plachy}, {Plum}, {Poggio}, {Pr{\v{s}}a}, {Pulone}, {Racero}, {Ragaini}, {Rainer}, {Raiteri}, {Rambaux}, {Ramos}, {Ramos-Lerate}, {Re Fiorentin}, {Regibo}, {Richards}, {Rios Diaz}, {Ripepi}, {Riva}, {Rix}, {Rixon}, {Robichon}, {Robin}, {Robin}, {Roelens}, {Rogues}, {Rohrbasser}, {Romero-G{\'o}mez}, {Rowell}, {Royer}, {Ruz Mieres}, {Rybicki}, {Sadowski}, {S{\'a}ez N{\'u}{\~n}ez}, {Sagrist{\`a} Sell{\'e}s}, {Sahlmann}, {Salguero}, {Samaras}, {Sanchez Gimenez}, {Sanna}, {Santove{\~n}a}, {Sarasso}, {Schultheis}, {Sciacca}, {Segol}, {Segovia}, {S{\'e}gransan}, {Semeux}, {Shahaf}, {Siddiqui}, {Siebert},
  {Siltala}, {Silvelo}, {Slezak}, {Slezak}, {Smart}, {Snaith}, {Solano}, {Solitro}, {Souami}, {Souchay}, {Spagna}, {Spina}, {Spoto}, {Steele}, {Steidelm{\"u}ller}, {Stephenson}, {S{\"u}veges}, {Surdej}, {Szabados}, {Szegedi-Elek}, {Taris}, {Taylor}, {Teixeira}, {Tolomei}, {Tonello}, {Torra}, {Torra}, {Torralba Elipe}, {Trabucchi}, {Tsounis}, {Turon}, {Ulla}, {Unger}, {Vaillant}, {van Dillen}, {van Reeven}, {Vanel}, {Vecchiato}, {Viala}, {Vicente}, {Voutsinas}, {Weiler}, {Wevers}, {Wyrzykowski}, {Yoldas}, {Yvard}, {Zhao}, {Zorec}, {Zucker}, \& {Zwitter}}]{GaiaDR3}
{Gaia Collaboration}, {Vallenari}, A., {Brown}, A.~G.~A., {et~al.} 2023, \aap, 674, A1, \dodoi{10.1051/0004-6361/202243940}

\bibitem[{{Gaidos} {et~al.}(2025){Gaidos}, {Gehrig}, \& {G{\"u}del}}]{gaidos25}
{Gaidos}, E., {Gehrig}, L., \& {G{\"u}del}, M. 2025, arXiv e-prints, arXiv:2502.16347, \dodoi{10.48550/arXiv.2502.16347}

\bibitem[{{Gasman} {et~al.}(2023){Gasman}, {van Dishoeck}, {Grant}, {Temmink}, {Tabone}, {Henning}, {Kamp}, {G{\"u}del}, {Lagage}, {Perotti}, {Christiaens}, {Samland}, {Arabhavi}, {Argyriou}, {Abergel}, {Absil}, {Barrado}, {Boccaletti}, {Bouwman}, {Caratti o Garatti}, {Geers}, {Glauser}, {Guadarrama}, {Jang}, {Kanwar}, {Lahuis}, {Morales-Calder{\'o}n}, {Mueller}, {Nehm{\'e}}, {Olofsson}, {Pantin}, {Pawellek}, {Ray}, {Rodgers-Lee}, {Scheithauer}, {Schreiber}, {Schwarz}, {Vandenbussche}, {Vlasblom}, {Waters}, {Wright}, {Colina}, {Greve}, \& {{\"O}stlin}}]{gasman23}
{Gasman}, D., {van Dishoeck}, E.~F., {Grant}, S.~L., {et~al.} 2023, \aap, 679, A117, \dodoi{10.1051/0004-6361/202347005}

\bibitem[{{Gasman} {et~al.}(2025){Gasman}, {Temmink}, {van Dishoeck}, {Kurtovic}, {Grant}, {Sellek}, {Tabone}, {Henning}, {Kamp}, {G{\"u}del}, {Barrado}, {Garatti}, {Glauser}, {Waters}, {Arabhavi}, {Jang}, {Kanwar}, {Lienert}, {Perotti}, {Schwarz}, \& {Vlasblom}}]{gasman25}
{Gasman}, D., {Temmink}, M., {van Dishoeck}, E.~F., {et~al.} 2025, arXiv e-prints, arXiv:2501.04587.
\newblock \doarXiv{2501.04587}

\bibitem[{{Gontcharov}(2006)}]{gontcharov06}
{Gontcharov}, G.~A. 2006, Astronomy Letters, 32, 759, \dodoi{10.1134/S1063773706110065}

\bibitem[{{Grant} {et~al.}(2023){Grant}, {van Dishoeck}, {Tabone}, {Gasman}, {Henning}, {Kamp}, {G{\"u}del}, {Lagage}, {Bettoni}, {Perotti}, {Christiaens}, {Samland}, {Arabhavi}, {Argyriou}, {Abergel}, {Absil}, {Barrado}, {Boccaletti}, {Bouwman}, {o Garatti}, {Geers}, {Glauser}, {Guadarrama}, {Jang}, {Kanwar}, {Lahuis}, {Morales-Calder{\'o}n}, {Mueller}, {Nehm{\'e}}, {Olofsson}, {Pantin}, {Pawellek}, {Ray}, {Rodgers-Lee}, {Scheithauer}, {Schreiber}, {Schwarz}, {Temmink}, {Vandenbussche}, {Vlasblom}, {Waters}, {Wright}, {Colina}, {Greve}, {Justannont}, \& {{\"O}stlin}}]{grant23}
{Grant}, S.~L., {van Dishoeck}, E.~F., {Tabone}, B., {et~al.} 2023, \apjl, 947, L6, \dodoi{10.3847/2041-8213/acc44b}

\bibitem[{{Grant} {et~al.}(2024){Grant}, {Kurtovic}, {van Dishoeck}, {Henning}, {Kamp}, {Nowacki}, {Perraut}, {Banzatti}, {Temmink}, {Christiaens}, {Samland}, {Gasman}, {Tabone}, {G{\"u}del}, {Lagage}, {Arabhavi}, {Barrado}, {Garatti}, {Glauser}, {Jang}, {Kanwar}, {Lahuis}, {Morales-Calder{\'o}n}, {Olofsson}, {Perotti}, {Schwarz}, {Vlasblom}, {Garcia Lopez}, \& {Long}}]{grant2024}
{Grant}, S.~L., {Kurtovic}, N.~T., {van Dishoeck}, E.~F., {et~al.} 2024, arXiv e-prints, arXiv:2406.10217, \dodoi{10.48550/arXiv.2406.10217}

\bibitem[{{Grossman}(1972)}]{grossman72}
{Grossman}, L. 1972, \gca, 36, 597, \dodoi{10.1016/0016-7037(72)90078-6}

\bibitem[{{Hayashi}(1981)}]{hayashi81}
{Hayashi}, C. 1981, Progress of Theoretical Physics Supplement, 70, 35, \dodoi{10.1143/PTPS.70.35}

\bibitem[{{Hendler} {et~al.}(2020){Hendler}, {Pascucci}, {Pinilla}, {Tazzari}, {Carpenter}, {Malhotra}, \& {Testi}}]{Hendler20}
{Hendler}, N., {Pascucci}, I., {Pinilla}, P., {et~al.} 2020, \apj, 895, 126, \dodoi{10.3847/1538-4357/ab70ba}

\bibitem[{{Henning} \& {Semenov}(2013)}]{henning13}
{Henning}, T., \& {Semenov}, D. 2013, Chemical Reviews, 113, 9016, \dodoi{10.1021/cr400128p}

\bibitem[{{Henning} {et~al.}(2024){Henning}, {Kamp}, {Samland}, {Arabhavi}, {Kanwar}, {van Dishoeck}, {G{\"u}del}, {Lagage}, {Waelkens}, {Abergel}, {Absil}, {Barrado}, {Boccaletti}, {Bouwman}, {Caratti o Garatti}, {Geers}, {Glauser}, {Lahuis}, {Mueller}, {Nehm{\'e}}, {Olofsson}, {Pantin}, {Ray}, {Scheithauer}, {Vandenbussche}, {Waters}, {Wright}, {Argyriou}, {Christiaens}, {Franceschi}, {Gasman}, {Grant}, {Guadarrama}, {Jang}, {Morales-Calder{\'o}n}, {Pawellek}, {Perotti}, {Rodgers-Lee}, {Schreiber}, {Schwarz}, {Tabone}, {Temmink}, {Vlasblom}, {Colina}, {Greve}, \& {{\"O}stlin}}]{henning24}
{Henning}, T., {Kamp}, I., {Samland}, M., {et~al.} 2024, \pasp, 136, 054302, \dodoi{10.1088/1538-3873/ad3455}

\bibitem[{{Herczeg} \& {Hillenbrand}(2014)}]{herczeg14}
{Herczeg}, G.~J., \& {Hillenbrand}, L.~A. 2014, \apj, 786, 97, \dodoi{10.1088/0004-637X/786/2/97}

\bibitem[{{Herczeg} {et~al.}(2005){Herczeg}, {Walter}, {Linsky}, {Gahm}, {Ardila}, {Brown}, {Johns-Krull}, {Simon}, \& {Valenti}}]{Herczeg05}
{Herczeg}, G.~J., {Walter}, F.~M., {Linsky}, J.~L., {et~al.} 2005, \aj, 129, 2777, \dodoi{10.1086/430075}

\bibitem[{{Hollenbach} \& {Gorti}(2009)}]{hollenbach09}
{Hollenbach}, D., \& {Gorti}, U. 2009, \apj, 703, 1203, \dodoi{10.1088/0004-637X/703/2/1203}

\bibitem[{{Houck} {et~al.}(2004){Houck}, {Roellig}, {van Cleve}, {Forrest}, {Herter}, {Lawrence}, {Matthews}, {Reitsema}, {Soifer}, {Watson}, {Weedman}, {Huisjen}, {Troeltzsch}, {Barry}, {Bernard-Salas}, {Blacken}, {Brandl}, {Charmandaris}, {Devost}, {Gull}, {Hall}, {Henderson}, {Higdon}, {Pirger}, {Schoenwald}, {Sloan}, {Uchida}, {Appleton}, {Armus}, {Burgdorf}, {Fajardo-Acosta}, {Grillmair}, {Ingalls}, {Morris}, \& {Teplitz}}]{houck04}
{Houck}, J.~R., {Roellig}, T.~L., {van Cleve}, J., {et~al.} 2004, \apjs, 154, 18, \dodoi{10.1086/423134}

\bibitem[{{Houge} {et~al.}(2025){Houge}, {Krijt}, {Banzatti}, {Blake}, {Pinilla}, {Pontoppidan}, {Trapman}, {Williams}, \& {Zhang}}]{houge25}
{Houge}, A., {Krijt}, S., {Banzatti}, A., {et~al.} 2025, \mnras, 537, 691, \dodoi{10.1093/mnras/staf057}

\bibitem[{{Hourihane} {et~al.}(2023{\natexlab{a}}){Hourihane}, {Fran{\c{c}}ois}, {Worley}, {Magrini}, {Gonneau}, {Casey}, {Gilmore}, {Randich}, {Sacco}, {Recio-Blanco}, {Korn}, {Allende Prieto}, {Smiljanic}, {Blomme}, {Bragaglia}, {Walton}, {Van Eck}, {Bensby}, {Lanzafame}, {Frasca}, {Franciosini}, {Damiani}, {Lind}, {Bergemann}, {Bonifacio}, {Hill}, {Lobel}, {Montes}, {Feuillet}, {Tautvai{\v{s}}ien{\.{e}}}, {Guiglion}, {Tabernero}, {Gonz{\'a}lez Hern{\'a}ndez}, {Gebran}, {Van der Swaelmen}, {Mikolaitis}, {Daflon}, {Merle}, {Morel}, {Lewis}, {Gonz{\'a}lez Solares}, {Murphy}, {Jeffries}, {Jackson}, {Feltzing}, {Prusti}, {Carraro}, {Biazzo}, {Prisinzano}, {Jofr{\'e}}, {Zaggia}, {Drazdauskas}, {Stonkut{\'e}}, {Marfil}, {Jim{\'e}nez-Esteban}, {Mahy}, {Guti{\'e}rrez Albarr{\'a}n}, {Berlanas}, {Santos}, {Morbidelli}, {Spina}, \& {Minkevi{\v{c}}i{\={u}}t{\.{e}}}}]{GaiaESO23}
{Hourihane}, A., {Fran{\c{c}}ois}, P., {Worley}, C.~C., {et~al.} 2023{\natexlab{a}}, \aap, 676, A129, \dodoi{10.1051/0004-6361/202345910}

\bibitem[{{Hourihane} {et~al.}(2023{\natexlab{b}}){Hourihane}, {Fran{\c{c}}ois}, {Worley}, {Magrini}, {Gonneau}, {Casey}, {Gilmore}, {Randich}, {Sacco}, {Recio-Blanco}, {Korn}, {Allende Prieto}, {Smiljanic}, {Blomme}, {Bragaglia}, {Walton}, {Van Eck}, {Bensby}, {Lanzafame}, {Frasca}, {Franciosini}, {Damiani}, {Lind}, {Bergemann}, {Bonifacio}, {Hill}, {Lobel}, {Montes}, {Feuillet}, {Tautvai{\v{s}}ien{\.{e}}}, {Guiglion}, {Tabernero}, {Gonz{\'a}lez Hern{\'a}ndez}, {Gebran}, {Van der Swaelmen}, {Mikolaitis}, {Daflon}, {Merle}, {Morel}, {Lewis}, {Gonz{\'a}lez Solares}, {Murphy}, {Jeffries}, {Jackson}, {Feltzing}, {Prusti}, {Carraro}, {Biazzo}, {Prisinzano}, {Jofr{\'e}}, {Zaggia}, {Drazdauskas}, {Stonkut{\'e}}, {Marfil}, {Jim{\'e}nez-Esteban}, {Mahy}, {Guti{\'e}rrez Albarr{\'a}n}, {Berlanas}, {Santos}, {Morbidelli}, {Spina}, \& {Minkevi{\v{c}}i{\={u}}t{\.{e}}}}]{hourihane23}
---. 2023{\natexlab{b}}, \aap, 676, A129, \dodoi{10.1051/0004-6361/202345910}

\bibitem[{{Huang} {et~al.}(2018{\natexlab{a}}){Huang}, {Andrews}, {Dullemond}, {Isella}, {P{\'e}rez}, {Guzm{\'a}n}, {{\"O}berg}, {Zhu}, {Zhang}, {Bai}, {Benisty}, {Birnstiel}, {Carpenter}, {Hughes}, {Ricci}, {Weaver}, \& {Wilner}}]{huang2018}
{Huang}, J., {Andrews}, S.~M., {Dullemond}, C.~P., {et~al.} 2018{\natexlab{a}}, \apjl, 869, L42, \dodoi{10.3847/2041-8213/aaf740}

\bibitem[{{Huang} {et~al.}(2018{\natexlab{b}}){Huang}, {Andrews}, {P{\'e}rez}, {Zhu}, {Dullemond}, {Isella}, {Benisty}, {Bai}, {Birnstiel}, {Carpenter}, {Guzm{\'a}n}, {Hughes}, {{\"O}berg}, {Ricci}, {Wilner}, \& {Zhang}}]{huang2018c}
{Huang}, J., {Andrews}, S.~M., {P{\'e}rez}, L.~M., {et~al.} 2018{\natexlab{b}}, \apjl, 869, L43, \dodoi{10.3847/2041-8213/aaf7a0}

\bibitem[{{Huang} {et~al.}(2020){Huang}, {Andrews}, {Dullemond}, {{\"O}berg}, {Qi}, {Zhu}, {Birnstiel}, {Carpenter}, {Isella}, {Mac{\'\i}as}, {McClure}, {P{\'e}rez}, {Teague}, {Wilner}, \& {Zhang}}]{huang2020a}
{Huang}, J., {Andrews}, S.~M., {Dullemond}, C.~P., {et~al.} 2020, \apj, 891, 48, \dodoi{10.3847/1538-4357/ab711e}

\bibitem[{{Jellison} {et~al.}(2024){Jellison}, {Banzatti}, {Johnson}, \& {Bruderer}}]{jellison2024}
{Jellison}, E.~G., {Banzatti}, A., {Johnson}, M.~B., \& {Bruderer}, S. 2024, \aj, 168, 99, \dodoi{10.3847/1538-3881/ad6142}

\bibitem[{Johnson {et~al.}(2024)Johnson, Banzatti, Fuller, \& Jellison}]{iSLAT_code}
Johnson, M., Banzatti, A., Fuller, J., \& Jellison, E. 2024, spexod/iSLAT: Second release, v4.03,  Zenodo, \dodoi{10.5281/zenodo.12167853}

\bibitem[{{Jones} {et~al.}(2023){Jones}, {{\'A}lvarez-M{\'a}rquez}, {Sloan}, {Kavanagh}, {Argyriou}, {Law}, {Labiano}, {Patapis}, {Mueller}, {Larson}, {Bright}, {Klaassen}, {Fox}, {Gasman}, {Geers}, {Glauser}, {Guillard}, {Nayak}, {Noriega-Crespo}, {Ressler}, {Sargent}, {Temim}, {Vandenbussche}, \& {Garc{\'\i}a Mar{\'\i}n}}]{jones2023}
{Jones}, O.~C., {{\'A}lvarez-M{\'a}rquez}, J., {Sloan}, G.~C., {et~al.} 2023, \mnras, 523, 2519, \dodoi{10.1093/mnras/stad1609}

\bibitem[{{J{\"o}nsson} {et~al.}(2020){J{\"o}nsson}, {Holtzman}, {Allende Prieto}, {Cunha}, {Garc{\'\i}a-Hern{\'a}ndez}, {Hasselquist}, {Masseron}, {Osorio}, {Shetrone}, {Smith}, {Stringfellow}, {Bizyaev}, {Edvardsson}, {Majewski}, {M{\'e}sz{\'a}ros}, {Souto}, {Zamora}, {Beaton}, {Bovy}, {Donor}, {Pinsonneault}, {Poovelil}, \& {Sobeck}}]{APOGEE20}
{J{\"o}nsson}, H., {Holtzman}, J.~A., {Allende Prieto}, C., {et~al.} 2020, \aj, 160, 120, \dodoi{10.3847/1538-3881/aba592}

\bibitem[{{Kaeufer} {et~al.}(2024{\natexlab{a}}){Kaeufer}, {Min}, {Woitke}, {Kamp}, \& {Arabhavi}}]{kaeufer24}
{Kaeufer}, T., {Min}, M., {Woitke}, P., {Kamp}, I., \& {Arabhavi}, A.~M. 2024{\natexlab{a}}, \aap, 687, A209, \dodoi{10.1051/0004-6361/202449936}

\bibitem[{{Kaeufer} {et~al.}(2024{\natexlab{b}}){Kaeufer}, {Woitke}, {Kamp}, {Kanwar}, \& {Min}}]{kaeufer2024b}
{Kaeufer}, T., {Woitke}, P., {Kamp}, I., {Kanwar}, J., \& {Min}, M. 2024{\natexlab{b}}, arXiv e-prints, arXiv:2408.06077.
\newblock \doarXiv{2408.06077}

\bibitem[{{Kalyaan} {et~al.}(2021){Kalyaan}, {Pinilla}, {Krijt}, {Mulders}, \& {Banzatti}}]{kalyaan21}
{Kalyaan}, A., {Pinilla}, P., {Krijt}, S., {Mulders}, G.~D., \& {Banzatti}, A. 2021, \apj, 921, 84, \dodoi{10.3847/1538-4357/ac1e96}

\bibitem[{{Kalyaan} {et~al.}(2023){Kalyaan}, {Pinilla}, {Krijt}, {Banzatti}, {Rosotti}, {Mulders}, {Lambrechts}, {Long}, \& {Herczeg}}]{kalyaan23}
{Kalyaan}, A., {Pinilla}, P., {Krijt}, S., {et~al.} 2023, \apj, 954, 66, \dodoi{10.3847/1538-4357/ace535}

\bibitem[{{Kanwar} {et~al.}(2024{\natexlab{a}}){Kanwar}, {Kamp}, {Woitke}, {Rab}, {Thi}, \& {Min}}]{kanwar24a}
{Kanwar}, J., {Kamp}, I., {Woitke}, P., {et~al.} 2024{\natexlab{a}}, \aap, 681, A22, \dodoi{10.1051/0004-6361/202346262}

\bibitem[{{Kanwar} {et~al.}(2024{\natexlab{b}}){Kanwar}, {Kamp}, {Jang}, {Waters}, {van Dishoeck}, {Christiaens}, {Arabhavi}, {Henning}, {G{\"u}del}, {Woitke}, {Absil}, {Barrado}, {Garatti}, {Glauser}, {Lahuis}, {Scheithauer}, {Vandenbussche}, {Gasman}, {Grant}, {Kurtovic}, {Perotti}, {Tabone}, \& {Temmink}}]{kanwar24}
{Kanwar}, J., {Kamp}, I., {Jang}, H., {et~al.} 2024{\natexlab{b}}, arXiv e-prints, arXiv:2407.14362, \dodoi{10.48550/arXiv.2407.14362}

\bibitem[{{Kounkel} {et~al.}(2019){Kounkel}, {Covey}, {Moe}, {Kratter}, {Su{\'a}rez}, {Stassun}, {Rom{\'a}n-Z{\'u}{\~n}iga}, {Hernandez}, {Kim}, {Pe{\~n}a Ram{\'\i}rez}, {Roman-Lopes}, {Stringfellow}, {Jaehnig}, {Borissova}, {Tofflemire}, {Krolikowski}, {Rizzuto}, {Kraus}, {Badenes}, {Longa-Pe{\~n}a}, {G{\'o}mez Maqueo Chew}, {Barba}, {Nidever}, {Brown}, {De Lee}, {Pan}, {Bizyaev}, {Oravetz}, \& {Oravetz}}]{kounkel2019}
{Kounkel}, M., {Covey}, K., {Moe}, M., {et~al.} 2019, \aj, 157, 196, \dodoi{10.3847/1538-3881/ab13b1}

\bibitem[{{Kress} {et~al.}(2010){Kress}, {Tielens}, \& {Frenklach}}]{kress10}
{Kress}, M.~E., {Tielens}, A. G.~G.~M., \& {Frenklach}, M. 2010, Advances in Space Research, 46, 44, \dodoi{10.1016/j.asr.2010.02.004}

\bibitem[{{Kurtovic} {et~al.}(2018){Kurtovic}, {P{\'e}rez}, {Benisty}, {Zhu}, {Zhang}, {Huang}, {Andrews}, {Dullemond}, {Isella}, {Bai}, {Carpenter}, {Guzm{\'a}n}, {Ricci}, \& {Wilner}}]{kurtovic2018}
{Kurtovic}, N.~T., {P{\'e}rez}, L.~M., {Benisty}, M., {et~al.} 2018, \apjl, 869, L44, \dodoi{10.3847/2041-8213/aaf746}

\bibitem[{{Lahuis} {et~al.}(2006){Lahuis}, {van Dishoeck}, {Boogert}, {Pontoppidan}, {Blake}, {Dullemond}, {Evans}, {Hogerheijde}, {J{\o}rgensen}, {Kessler-Silacci}, \& {Knez}}]{lahuis06}
{Lahuis}, F., {van Dishoeck}, E.~F., {Boogert}, A.~C.~A., {et~al.} 2006, \apjl, 636, L145, \dodoi{10.1086/500084}

\bibitem[{{Liu} {et~al.}(2019){Liu}, {Dipierro}, {Ragusa}, {Lodato}, {Herczeg}, {Long}, {Harsono}, {Boehler}, {Menard}, {Johnstone}, {Pascucci}, {Pinilla}, {Salyk}, {van der Plas}, {Cabrit}, {Fischer}, {Hendler}, {Manara}, {Nisini}, {Rigliaco}, {Avenhaus}, {Banzatti}, \& {Gully-Santiago}}]{liu2019}
{Liu}, Y., {Dipierro}, G., {Ragusa}, E., {et~al.} 2019, \aap, 622, A75, \dodoi{10.1051/0004-6361/201834157}

\bibitem[{{Long} {et~al.}(2020){Long}, {Zhang}, {Teague}, \& {Bergin}}]{long20}
{Long}, D.~E., {Zhang}, K., {Teague}, R., \& {Bergin}, E.~A. 2020, \apjl, 895, L46, \dodoi{10.3847/2041-8213/ab94a8}

\bibitem[{{Long} {et~al.}(2018){Long}, {Pinilla}, {Herczeg}, {Harsono}, {Dipierro}, {Pascucci}, {Hendler}, {Tazzari}, {Ragusa}, {Salyk}, {Edwards}, {Lodato}, {van de Plas}, {Johnstone}, {Liu}, {Boehler}, {Cabrit}, {Manara}, {Menard}, {Mulders}, {Nisini}, {Fischer}, {Rigliaco}, {Banzatti}, {Avenhaus}, \& {Gully-Santiago}}]{long18}
{Long}, F., {Pinilla}, P., {Herczeg}, G.~J., {et~al.} 2018, \apj, 869, 17, \dodoi{10.3847/1538-4357/aae8e1}

\bibitem[{{Long} {et~al.}(2019{\natexlab{a}}){Long}, {Herczeg}, {Harsono}, {Pinilla}, {Tazzari}, {Manara}, {Pascucci}, {Cabrit}, {Nisini}, {Johnstone}, {Edwards}, {Salyk}, {Menard}, {Lodato}, {Boehler}, {Mace}, {Liu}, {Mulders}, {Hendler}, {Ragusa}, {Fischer}, {Banzatti}, {Rigliaco}, {van de Plas}, {Dipierro}, {Gully-Santiago}, \& {Lopez-Valdivia}}]{long19}
{Long}, F., {Herczeg}, G.~J., {Harsono}, D., {et~al.} 2019{\natexlab{a}}, \apj, 882, 49, \dodoi{10.3847/1538-4357/ab2d2d}

\bibitem[{{Long} {et~al.}(2019{\natexlab{b}}){Long}, {Herczeg}, {Harsono}, {Pinilla}, {Tazzari}, {Manara}, {Pascucci}, {Cabrit}, {Nisini}, {Johnstone}, {Edwards}, {Salyk}, {Menard}, {Lodato}, {Boehler}, {Mace}, {Liu}, {Mulders}, {Hendler}, {Ragusa}, {Fischer}, {Banzatti}, {Rigliaco}, {van de Plas}, {Dipierro}, {Gully-Santiago}, \& {Lopez-Valdivia}}]{long2019}
---. 2019{\natexlab{b}}, \apj, 882, 49, \dodoi{10.3847/1538-4357/ab2d2d}

\bibitem[{{Long} {et~al.}(2025){Long}, {Pascucci}, {Houge}, {Banzatti}, {Pontoppidan}, {Najita}, {Krijt}, {Xie}, {Williams}, {Herczeg}, {Andrews}, {Bergin}, {Blake}, {Colmenares}, {Harsono}, {Romero-Mirza}, {Li}, {Lu}, {Pinilla}, {Wilner}, {Vioque}, {Zhang}, \& {The Jdiscs Collaboration}}]{long25}
{Long}, F., {Pascucci}, I., {Houge}, A., {et~al.} 2025, \apjl, 978, L30, \dodoi{10.3847/2041-8213/ad99d2}

\bibitem[{{Mah} {et~al.}(2023){Mah}, {Bitsch}, {Pascucci}, \& {Henning}}]{mah23}
{Mah}, J., {Bitsch}, B., {Pascucci}, I., \& {Henning}, T. 2023, \aap, 677, L7, \dodoi{10.1051/0004-6361/202347169}

\bibitem[{{Mah} {et~al.}(2024){Mah}, {Savvidou}, \& {Bitsch}}]{mah2024}
{Mah}, J., {Savvidou}, S., \& {Bitsch}, B. 2024, \aap, 686, L17, \dodoi{10.1051/0004-6361/202450322}

\bibitem[{{Manara} {et~al.}(2023){Manara}, {Ansdell}, {Rosotti}, {Hughes}, {Armitage}, {Lodato}, \& {Williams}}]{manara2023}
{Manara}, C.~F., {Ansdell}, M., {Rosotti}, G.~P., {et~al.} 2023, in Astronomical Society of the Pacific Conference Series, Vol. 534, Protostars and Planets VII, ed. S.~{Inutsuka}, Y.~{Aikawa}, T.~{Muto}, K.~{Tomida}, \& M.~{Tamura}, 539, \dodoi{10.48550/arXiv.2203.09930}

\bibitem[{{Manara} {et~al.}(2016){Manara}, {Fedele}, {Herczeg}, \& {Teixeira}}]{manara16}
{Manara}, C.~F., {Fedele}, D., {Herczeg}, G.~J., \& {Teixeira}, P.~S. 2016, \aap, 585, A136, \dodoi{10.1051/0004-6361/201527224}

\bibitem[{{Manara} {et~al.}(2014){Manara}, {Testi}, {Natta}, {Rosotti}, {Benisty}, {Ercolano}, \& {Ricci}}]{manara14}
{Manara}, C.~F., {Testi}, L., {Natta}, A., {et~al.} 2014, \aap, 568, A18, \dodoi{10.1051/0004-6361/201323318}

\bibitem[{{Manara} {et~al.}(2017){Manara}, {Testi}, {Herczeg}, {Pascucci}, {Alcal{\'a}}, {Natta}, {Antoniucci}, {Fedele}, {Mulders}, {Henning}, {Mohanty}, {Prusti}, \& {Rigliaco}}]{manara17}
{Manara}, C.~F., {Testi}, L., {Herczeg}, G.~J., {et~al.} 2017, \aap, 604, A127, \dodoi{10.1051/0004-6361/201630147}

\bibitem[{{Manara} {et~al.}(2021){Manara}, {Frasca}, {Venuti}, {Siwak}, {Herczeg}, {Calvet}, {Hernandez}, {Tychoniec}, {Gangi}, {Alcal{\'a}}, {Boffin}, {Nisini}, {Robberto}, {Briceno}, {Campbell-White}, {Sicilia-Aguilar}, {McGinnis}, {Fedele}, {K{\'o}sp{\'a}l}, {{\'A}brah{\'a}m}, {Alonso-Santiago}, {Antoniucci}, {Arulanantham}, {Bacciotti}, {Banzatti}, {Beccari}, {Benisty}, {Biazzo}, {Bouvier}, {Cabrit}, {Caratti o Garatti}, {Coffey}, {Covino}, {Dougados}, {Eisl{\"o}ffel}, {Ercolano}, {Espaillat}, {Erkal}, {Facchini}, {Fang}, {Fiorellino}, {Fischer}, {France}, {Gameiro}, {Garcia Lopez}, {Giannini}, {Ginski}, {Grankin}, {G{\"u}nther}, {Hartmann}, {Hillenbrand}, {Hussain}, {James}, {Koutoulaki}, {Lodato}, {Mauc{\'o}}, {Mendigut{\'\i}a}, {Mentel}, {Miotello}, {Oudmaijer}, {Rigliaco}, {Rosotti}, {Sanchis}, {Schneider}, {Spina}, {Stelzer}, {Testi}, {Thanathibodee}, {Vink}, {Walter}, {Williams}, \& {Zsidi}}]{manara21}
{Manara}, C.~F., {Frasca}, A., {Venuti}, L., {et~al.} 2021, \aap, 650, A196, \dodoi{10.1051/0004-6361/202140639}

\bibitem[{{McClure}(2019)}]{mcclure19}
{McClure}, M.~K. 2019, \aap, 632, A32, \dodoi{10.1051/0004-6361/201834361}

\bibitem[{{Meijerink} {et~al.}(2009){Meijerink}, {Pontoppidan}, {Blake}, {Poelman}, \& {Dullemond}}]{Meijerink09}
{Meijerink}, R., {Pontoppidan}, K.~M., {Blake}, G.~A., {Poelman}, D.~R., \& {Dullemond}, C.~P. 2009, \apj, 704, 1471, \dodoi{10.1088/0004-637X/704/2/1471}

\bibitem[{{Mendigut{\'\i}a} {et~al.}(2013){Mendigut{\'\i}a}, {Brittain}, {Eiroa}, {Meeus}, {Montesinos}, {Mora}, {Muzerolle}, {Oudmaijer}, \& {Rigliaco}}]{mendigutia13}
{Mendigut{\'\i}a}, I., {Brittain}, S., {Eiroa}, C., {et~al.} 2013, \apj, 776, 44, \dodoi{10.1088/0004-637X/776/1/44}

\bibitem[{{Miret-Roig} {et~al.}(2022){Miret-Roig}, {Galli}, {Olivares}, {Bouy}, {Alves}, \& {Barrado}}]{miretroig22}
{Miret-Roig}, N., {Galli}, P.~A.~B., {Olivares}, J., {et~al.} 2022, \aap, 667, A163, \dodoi{10.1051/0004-6361/202244709}

\bibitem[{{Najita} {et~al.}(2003){Najita}, {Carr}, \& {Mathieu}}]{najita03}
{Najita}, J., {Carr}, J.~S., \& {Mathieu}, R.~D. 2003, \apj, 589, 931, \dodoi{10.1086/374809}

\bibitem[{{Najita} {et~al.}(2011){Najita}, {{\'A}d{\'a}mkovics}, \& {Glassgold}}]{najita11}
{Najita}, J.~R., {{\'A}d{\'a}mkovics}, M., \& {Glassgold}, A.~E. 2011, \apj, 743, 147, \dodoi{10.1088/0004-637X/743/2/147}

\bibitem[{{Najita} {et~al.}(2021){Najita}, {Carr}, {Brittain}, {Lacy}, {Richter}, \& {Doppmann}}]{najita21}
{Najita}, J.~R., {Carr}, J.~S., {Brittain}, S.~D., {et~al.} 2021, \apj, 908, 171, \dodoi{10.3847/1538-4357/abcfc6}

\bibitem[{{Najita} {et~al.}(2013){Najita}, {Carr}, {Pontoppidan}, {Salyk}, {van Dishoeck}, \& {Blake}}]{najita13}
{Najita}, J.~R., {Carr}, J.~S., {Pontoppidan}, K.~M., {et~al.} 2013, \apj, 766, 134, \dodoi{10.1088/0004-637X/766/2/134}

\bibitem[{{Najita} {et~al.}(2018){Najita}, {Carr}, {Salyk}, {Lacy}, {Richter}, \& {DeWitt}}]{najita18}
{Najita}, J.~R., {Carr}, J.~S., {Salyk}, C., {et~al.} 2018, \apj, 862, 122, \dodoi{10.3847/1538-4357/aaca39}

\bibitem[{{Najita} {et~al.}(2010){Najita}, {Carr}, {Strom}, {Watson}, {Pascucci}, {Hollenbach}, {Gorti}, \& {Keller}}]{najita10}
{Najita}, J.~R., {Carr}, J.~S., {Strom}, S.~E., {et~al.} 2010, \apj, 712, 274, \dodoi{10.1088/0004-637X/712/1/274}

\bibitem[{{Najita} {et~al.}(2009){Najita}, {Doppmann}, {Carr}, {Graham}, \& {Eisner}}]{najita09}
{Najita}, J.~R., {Doppmann}, G.~W., {Carr}, J.~S., {Graham}, J.~R., \& {Eisner}, J.~A. 2009, \apj, 691, 738, \dodoi{10.1088/0004-637X/691/1/738}

\bibitem[{{Nguyen} {et~al.}(2012){Nguyen}, {Brandeker}, {van Kerkwijk}, \& {Jayawardhana}}]{nguyen12}
{Nguyen}, D.~C., {Brandeker}, A., {van Kerkwijk}, M.~H., \& {Jayawardhana}, R. 2012, \apj, 745, 119, \dodoi{10.1088/0004-637X/745/2/119}

\bibitem[{{{\"O}berg} {et~al.}(2021){{\"O}berg}, {Guzm{\'a}n}, {Walsh}, {Aikawa}, {Bergin}, {Law}, {Loomis}, {Alarc{\'o}n}, {Andrews}, {Bae}, {Bergner}, {Boehler}, {Booth}, {Bosman}, {Calahan}, {Cataldi}, {Cleeves}, {Czekala}, {Furuya}, {Huang}, {Ilee}, {Kurtovic}, {Le Gal}, {Liu}, {Long}, {M{\'e}nard}, {Nomura}, {P{\'e}rez}, {Qi}, {Schwarz}, {Sierra}, {Teague}, {Tsukagoshi}, {Yamato}, {van't Hoff}, {Waggoner}, {Wilner}, \& {Zhang}}]{Oberg2021}
{{\"O}berg}, K.~I., {Guzm{\'a}n}, V.~V., {Walsh}, C., {et~al.} 2021, \apjs, 257, 1, \dodoi{10.3847/1538-4365/ac1432}

\bibitem[{{Pascucci} {et~al.}(2009){Pascucci}, {Apai}, {Luhman}, {Henning}, {Bouwman}, {Meyer}, {Lahuis}, \& {Natta}}]{pascucci09}
{Pascucci}, I., {Apai}, D., {Luhman}, K., {et~al.} 2009, \apj, 696, 143, \dodoi{10.1088/0004-637X/696/1/143}

\bibitem[{{Pascucci} {et~al.}(2013){Pascucci}, {Herczeg}, {Carr}, \& {Bruderer}}]{pascucci13}
{Pascucci}, I., {Herczeg}, G., {Carr}, J.~S., \& {Bruderer}, S. 2013, \apj, 779, 178, \dodoi{10.1088/0004-637X/779/2/178}

\bibitem[{{Pascucci} {et~al.}(2014){Pascucci}, {Ricci}, {Gorti}, {Hollenbach}, {Hendler}, {Brooks}, \& {Contreras}}]{pascucci14}
{Pascucci}, I., {Ricci}, L., {Gorti}, U., {et~al.} 2014, \apj, 795, 1, \dodoi{10.1088/0004-637X/795/1/1}

\bibitem[{{Pascucci} {et~al.}(2007){Pascucci}, {Hollenbach}, {Najita}, {Muzerolle}, {Gorti}, {Herczeg}, {Hillenbrand}, {Kim}, {Carpenter}, {Meyer}, {Mamajek}, \& {Bouwman}}]{pascucci07}
{Pascucci}, I., {Hollenbach}, D., {Najita}, J., {et~al.} 2007, \apj, 663, 383, \dodoi{10.1086/518535}

\bibitem[{{Pascucci} {et~al.}(2016){Pascucci}, {Testi}, {Herczeg}, {Long}, {Manara}, {Hendler}, {Mulders}, {Krijt}, {Ciesla}, {Henning}, {Mohanty}, {Drabek-Maunder}, {Apai}, {Sz{\H{u}}cs}, {Sacco}, \& {Olofsson}}]{pascucci16}
{Pascucci}, I., {Testi}, L., {Herczeg}, G.~J., {et~al.} 2016, \apj, 831, 125, \dodoi{10.3847/0004-637X/831/2/125}

\bibitem[{{Pascucci} {et~al.}(2020){Pascucci}, {Banzatti}, {Gorti}, {Fang}, {Pontoppidan}, {Alexander}, {Ballabio}, {Edwards}, {Salyk}, {Sacco}, {Flaccomio}, {Blake}, {Carmona}, {Hall}, {Kamp}, {K{\"a}ufl}, {Meeus}, {Meyer}, {Pauly}, {Steendam}, \& {Sterzik}}]{pascucci20}
{Pascucci}, I., {Banzatti}, A., {Gorti}, U., {et~al.} 2020, \apj, 903, 78, \dodoi{10.3847/1538-4357/abba3c}

\bibitem[{{Perotti} {et~al.}(2023){Perotti}, {Christiaens}, {Henning}, {Tabone}, {Waters}, {Kamp}, {Olofsson}, {Grant}, {Gasman}, {Bouwman}, {Samland}, {Franceschi}, {van Dishoeck}, {Schwarz}, {G{\"u}del}, {Lagage}, {Ray}, {Vandenbussche}, {Abergel}, {Absil}, {Arabhavi}, {Argyriou}, {Barrado}, {Boccaletti}, {Caratti o Garatti}, {Geers}, {Glauser}, {Justannont}, {Lahuis}, {Mueller}, {Nehm{\'e}}, {Pantin}, {Scheithauer}, {Waelkens}, {Guadarrama}, {Jang}, {Kanwar}, {Morales-Calder{\'o}n}, {Pawellek}, {Rodgers-Lee}, {Schreiber}, {Colina}, {Greve}, {{\"O}stlin}, \& {Wright}}]{perotti23}
{Perotti}, G., {Christiaens}, V., {Henning}, T., {et~al.} 2023, \nat, 620, 516, \dodoi{10.1038/s41586-023-06317-9}

\bibitem[{{Pinilla} {et~al.}(2012){Pinilla}, {Birnstiel}, {Ricci}, {Dullemond}, {Uribe}, {Testi}, \& {Natta}}]{pinilla12}
{Pinilla}, P., {Birnstiel}, T., {Ricci}, L., {et~al.} 2012, \aap, 538, A114, \dodoi{10.1051/0004-6361/201118204}

\bibitem[{{Pollack} {et~al.}(1996){Pollack}, {Hubickyj}, {Bodenheimer}, {Lissauer}, {Podolak}, \& {Greenzweig}}]{pollack96}
{Pollack}, J.~B., {Hubickyj}, O., {Bodenheimer}, P., {et~al.} 1996, \icarus, 124, 62, \dodoi{10.1006/icar.1996.0190}

\bibitem[{{Pontoppidan} {et~al.}(2019){Pontoppidan}, {Salyk}, {Banzatti}, {Blake}, {Walsh}, {Lacy}, \& {Richter}}]{pontoppidan19}
{Pontoppidan}, K.~M., {Salyk}, C., {Banzatti}, A., {et~al.} 2019, \apj, 874, 92, \dodoi{10.3847/1538-4357/ab05d8}

\bibitem[{{Pontoppidan} {et~al.}(2014){Pontoppidan}, {Salyk}, {Bergin}, {Brittain}, {Marty}, {Mousis}, \& {{\"O}berg}}]{pontoppidan14}
{Pontoppidan}, K.~M., {Salyk}, C., {Bergin}, E.~A., {et~al.} 2014, in Protostars and Planets VI, ed. H.~{Beuther}, R.~S. {Klessen}, C.~P. {Dullemond}, \& T.~{Henning}, 363--385, \dodoi{10.2458/azu_uapress_9780816531240-ch016}

\bibitem[{{Pontoppidan} {et~al.}(2010){Pontoppidan}, {Salyk}, {Blake}, {Meijerink}, {Carr}, \& {Najita}}]{pontoppidan10b}
{Pontoppidan}, K.~M., {Salyk}, C., {Blake}, G.~A., {et~al.} 2010, \apj, 720, 887, \dodoi{10.1088/0004-637X/720/1/887}

\bibitem[{{Pontoppidan} {et~al.}(2024){Pontoppidan}, {Salyk}, {Banzatti}, {Zhang}, {Pascucci}, {{\"O}berg}, {Long}, {Mu{\~n}oz-Romero}, {Carr}, {Najita}, {Blake}, {Arulanantham}, {Andrews}, {Ballering}, {Bergin}, {Calahan}, {Cobb}, {Colmenares}, {Dickson-Vandervelde}, {Dignan}, {Green}, {Heretz}, {Herczeg}, {Kalyaan}, {Krijt}, {Pauly}, {Pinilla}, {Trapman}, \& {Xie}}]{pontoppidan24}
{Pontoppidan}, K.~M., {Salyk}, C., {Banzatti}, A., {et~al.} 2024, \apj, 963, 158, \dodoi{10.3847/1538-4357/ad20f0}

\bibitem[{{Ram{\'\i}rez-Tannus} {et~al.}(2023){Ram{\'\i}rez-Tannus}, {Bik}, {Cuijpers}, {Waters}, {G{\"o}ppl}, {Henning}, {Kamp}, {Preibisch}, {Getman}, {Chaparro}, {Cuartas-Restrepo}, {de Koter}, {Feigelson}, {Grant}, {Haworth}, {Hern{\'a}ndez}, {Kuhn}, {Perotti}, {Povich}, {Reiter}, {Roccatagliata}, {Sabbi}, {Tabone}, {Winter}, {McLeod}, {van Boekel}, \& {van Terwisga}}]{ramireztannus23}
{Ram{\'\i}rez-Tannus}, M.~C., {Bik}, A., {Cuijpers}, L., {et~al.} 2023, \apjl, 958, L30, \dodoi{10.3847/2041-8213/ad03f8}

\bibitem[{{Ribas} {et~al.}(2024){Ribas}, {Clarke}, \& {Zagaria}}]{ribas24}
{Ribas}, {\'A}., {Clarke}, C.~J., \& {Zagaria}, F. 2024, \mnras, 532, 1752, \dodoi{10.1093/mnras/stae1534}

\bibitem[{{Rieke} {et~al.}(2015){Rieke}, {Wright}, {B{\"o}ker}, {Bouwman}, {Colina}, {Glasse}, {Gordon}, {Greene}, {G{\"u}del}, {Henning}, {Justtanont}, {Lagage}, {Meixner}, {N{\o}rgaard-Nielsen}, {Ray}, {Ressler}, {van Dishoeck}, \& {Waelkens}}]{rieke15}
{Rieke}, G.~H., {Wright}, G.~S., {B{\"o}ker}, T., {et~al.} 2015, \pasp, 127, 584, \dodoi{10.1086/682252}

\bibitem[{{Romero-Mirza} {et~al.}(2024{\natexlab{a}}){Romero-Mirza}, {Banzatti}, {{\"O}berg}, {Pontoppidan}, {Salyk}, {Najita}, {Blake}, {Krijt}, {Arulanantham}, {Pinilla}, {Long}, {Rosotti}, {Andrews}, {Wilner}, {Calahan}, \& {The Jdiscs Collaboration}}]{romero-mirza24b}
{Romero-Mirza}, C.~E., {Banzatti}, A., {{\"O}berg}, K.~I., {et~al.} 2024{\natexlab{a}}, \apj, 975, 78, \dodoi{10.3847/1538-4357/ad769e}

\bibitem[{{Romero-Mirza} {et~al.}(2024{\natexlab{b}}){Romero-Mirza}, {{\"O}berg}, {Banzatti}, {Pontoppidan}, {Andrews}, {Wilner}, {Bergin}, {Czekala}, {Law}, {Salyk}, {Teague}, {Qi}, {Bergner}, {Huang}, {Walsh}, {Guzm{\'a}n}, {Cleeves}, {Aikawa}, {Bae}, {Booth}, {Cataldi}, {Ilee}, {Le Gal}, {Long}, {Loomis}, {Menard}, \& {Liu}}]{romero-mirza24a}
{Romero-Mirza}, C.~E., {{\"O}berg}, K.~I., {Banzatti}, A., {et~al.} 2024{\natexlab{b}}, \apj, 964, 36, \dodoi{10.3847/1538-4357/ad20e9}

\bibitem[{{Salyk}(2022)}]{salykspectoolsir}
{Salyk}, C. 2022, {csalyk/spectools\_ir: First release}, v1.0.0,  Zenodo, \dodoi{10.5281/zenodo.5818682}

\bibitem[{{Salyk} {et~al.}(2009){Salyk}, {Blake}, {Boogert}, \& {Brown}}]{salyk09}
{Salyk}, C., {Blake}, G.~A., {Boogert}, A.~C.~A., \& {Brown}, J.~M. 2009, \apj, 699, 330, \dodoi{10.1088/0004-637X/699/1/330}

\bibitem[{{Salyk} {et~al.}(2011{\natexlab{a}}){Salyk}, {Blake}, {Boogert}, \& {Brown}}]{salyk11b}
---. 2011{\natexlab{a}}, \apj, 743, 112, \dodoi{10.1088/0004-637X/743/2/112}

\bibitem[{{Salyk} {et~al.}(2013){Salyk}, {Herczeg}, {Brown}, {Blake}, {Pontoppidan}, \& {van Dishoeck}}]{salyk13}
{Salyk}, C., {Herczeg}, G.~J., {Brown}, J.~M., {et~al.} 2013, \apj, 769, 21, \dodoi{10.1088/0004-637X/769/1/21}

\bibitem[{{Salyk} {et~al.}(2015){Salyk}, {Lacy}, {Richter}, {Zhang}, {Blake}, \& {Pontoppidan}}]{salyk15}
{Salyk}, C., {Lacy}, J.~H., {Richter}, M.~J., {et~al.} 2015, \apjl, 810, L24, \dodoi{10.1088/2041-8205/810/2/L24}

\bibitem[{{Salyk} {et~al.}(2011{\natexlab{b}}){Salyk}, {Pontoppidan}, {Blake}, {Najita}, \& {Carr}}]{salyk11}
{Salyk}, C., {Pontoppidan}, K.~M., {Blake}, G.~A., {Najita}, J.~R., \& {Carr}, J.~S. 2011{\natexlab{b}}, \apj, 731, 130, \dodoi{10.1088/0004-637X/731/2/130}

\bibitem[{{Salyk} {et~al.}(2025){Salyk}, {Pontoppidan}, {Banzatti}, {Bergin}, {Arulanantham}, {Najita}, {Blake}, {Carr}, {Zhang}, \& {Xie}}]{salyk25}
{Salyk}, C., {Pontoppidan}, K.~M., {Banzatti}, A., {et~al.} 2025, \aj, 169, 184, \dodoi{10.3847/1538-3881/adb397}

\bibitem[{{Schwarz} {et~al.}(2024){Schwarz}, {Henning}, {Christiaens}, {Gasman}, {Samland}, {Perotti}, {Jang}, {Grant}, {Tabone}, {Morales-Calder{\'o}n}, {Kamp}, {van Dishoeck}, {G{\"u}del}, {Lagage}, {Barrado}, {Caratti o Garatti}, {Glauser}, {Ray}, {Vandenbussche}, {Waters}, {Arabhavi}, {Kanwar}, {Olofsson}, {Rodgers-Lee}, {Schreiber}, \& {Temmink}}]{schwarz24}
{Schwarz}, K.~R., {Henning}, T., {Christiaens}, V., {et~al.} 2024, \apj, 962, 8, \dodoi{10.3847/1538-4357/ad1393}

\bibitem[{{Sellek} {et~al.}(2025){Sellek}, {Vlasblom}, \& {van Dishoeck}}]{sellek25}
{Sellek}, A.~D., {Vlasblom}, M., \& {van Dishoeck}, E.~F. 2025, \aap, 694, A79, \dodoi{10.1051/0004-6361/202451137}

\bibitem[{{Sellek} {et~al.}(2024){Sellek}, {Bajaj}, {Pascucci}, {Clarke}, {Alexander}, {Xie}, {Ballabio}, {Deng}, {Gorti}, {Gaspar}, \& {Morrison}}]{sellek24}
{Sellek}, A.~D., {Bajaj}, N.~S., {Pascucci}, I., {et~al.} 2024, \aj, 167, 223, \dodoi{10.3847/1538-3881/ad34ae}

\bibitem[{{Sloan} {et~al.}(2014){Sloan}, {Lagadec}, {Zijlstra}, {Kraemer}, {Weis}, {Matsuura}, {Volk}, {Peeters}, {Duley}, {Cami}, {Bernard-Salas}, {Kemper}, \& {Sahai}}]{sloan2014}
{Sloan}, G.~C., {Lagadec}, E., {Zijlstra}, A.~A., {et~al.} 2014, \apj, 791, 28, \dodoi{10.1088/0004-637X/791/1/28}

\bibitem[{{Stock} {et~al.}(2022){Stock}, {McGinnis}, {Caratti o Garatti}, {Natta}, \& {Ray}}]{stock22}
{Stock}, C., {McGinnis}, P., {Caratti o Garatti}, A., {Natta}, A., \& {Ray}, T.~P. 2022, \aap, 668, A94, \dodoi{10.1051/0004-6361/202244315}

\bibitem[{{Sullivan} {et~al.}(2019){Sullivan}, {Wilking}, {Greene}, {Lisalda}, {Gibb}, \& {Ejeta}}]{sullivan2019}
{Sullivan}, T., {Wilking}, B.~A., {Greene}, T.~P., {et~al.} 2019, \aj, 158, 41, \dodoi{10.3847/1538-3881/ab24c0}

\bibitem[{{Szul{\'a}gyi} {et~al.}(2012){Szul{\'a}gyi}, {Pascucci}, {{\'A}brah{\'a}m}, {Apai}, {Bouwman}, \& {Mo{\'o}r}}]{szulagyi12}
{Szul{\'a}gyi}, J., {Pascucci}, I., {{\'A}brah{\'a}m}, P., {et~al.} 2012, \apj, 759, 47, \dodoi{10.1088/0004-637X/759/1/47}

\bibitem[{{Tabone} {et~al.}(2024){Tabone}, {van Dishoeck}, \& {Black}}]{tabone24}
{Tabone}, B., {van Dishoeck}, E.~F., \& {Black}, J.~H. 2024, arXiv e-prints, arXiv:2406.14560, \dodoi{10.48550/arXiv.2406.14560}

\bibitem[{{Tabone} {et~al.}(2023){Tabone}, {Bettoni}, {van Dishoeck}, {Arabhavi}, {Grant}, {Gasman}, {Henning}, {Kamp}, {G{\"u}del}, {Lagage}, {Ray}, {Vandenbussche}, {Abergel}, {Absil}, {Argyriou}, {Barrado}, {Boccaletti}, {Bouwman}, {Caratti o Garatti}, {Geers}, {Glauser}, {Justannont}, {Lahuis}, {Mueller}, {Nehm{\'e}}, {Olofsson}, {Pantin}, {Scheithauer}, {Waelkens}, {Waters}, {Black}, {Christiaens}, {Guadarrama}, {Morales-Calder{\'o}n}, {Jang}, {Kanwar}, {Pawellek}, {Perotti}, {Perrin}, {Rodgers-Lee}, {Samland}, {Schreiber}, {Schwarz}, {Colina}, {{\"O}stlin}, \& {Wright}}]{tabone23}
{Tabone}, B., {Bettoni}, G., {van Dishoeck}, E.~F., {et~al.} 2023, Nature Astronomy, 7, 805, \dodoi{10.1038/s41550-023-01965-3}

\bibitem[{{Temmink} {et~al.}(2024){Temmink}, {van Dishoeck}, {Gasman}, {Grant}, {Tabone}, {G{\"u}del}, {Henning}, {Barrado}, {Caratti o Garatti}, {Glauser}, {Kamp}, {Arabhavi}, {Jang}, {Kurtovic}, {Perotti}, {Schwarz}, \& {Vlasblom}}]{temmink24b}
{Temmink}, M., {van Dishoeck}, E.~F., {Gasman}, D., {et~al.} 2024, \aap, 689, A330, \dodoi{10.1051/0004-6361/202450355}

\bibitem[{{Testi} {et~al.}(2022){Testi}, {Natta}, {Manara}, {de Gregorio Monsalvo}, {Lodato}, {Lopez}, {Muzic}, {Pascucci}, {Sanchis}, {Miranda}, {Scholz}, {De Simone}, \& {Williams}}]{testi22}
{Testi}, L., {Natta}, A., {Manara}, C.~F., {et~al.} 2022, \aap, 663, A98, \dodoi{10.1051/0004-6361/202141380}

\bibitem[{{van der Marel} {et~al.}(2018){van der Marel}, {Williams}, {Ansdell}, {Manara}, {Miotello}, {Tazzari}, {Testi}, {Hogerheijde}, {Bruderer}, {van Terwisga}, \& {van Dishoeck}}]{vandermarel18}
{van der Marel}, N., {Williams}, J.~P., {Ansdell}, M., {et~al.} 2018, \apj, 854, 177, \dodoi{10.3847/1538-4357/aaaa6b}

\bibitem[{{Visser} {et~al.}(2011){Visser}, {Doty}, \& {van Dishoeck}}]{Visser11}
{Visser}, R., {Doty}, S.~D., \& {van Dishoeck}, E.~F. 2011, \aap, 534, A132, \dodoi{10.1051/0004-6361/201117249}

\bibitem[{{Vlasblom} {et~al.}(2024){Vlasblom}, {van Dishoeck}, {Tabone}, \& {Bruderer}}]{vlasblom24}
{Vlasblom}, M., {van Dishoeck}, E.~F., {Tabone}, B., \& {Bruderer}, S. 2024, \aap, 682, A91, \dodoi{10.1051/0004-6361/202348224}

\bibitem[{{Vorobyov} \& {Basu}(2010)}]{vorobyov10}
{Vorobyov}, E.~I., \& {Basu}, S. 2010, \apj, 719, 1896, \dodoi{10.1088/0004-637X/719/2/1896}

\bibitem[{{Walsh} {et~al.}(2015){Walsh}, {Nomura}, \& {van Dishoeck}}]{walsh15}
{Walsh}, C., {Nomura}, H., \& {van Dishoeck}, E. 2015, \aap, 582, A88, \dodoi{10.1051/0004-6361/201526751}

\bibitem[{{Wells} {et~al.}(2015){Wells}, {Pel}, {Glasse}, {Wright}, {Aitink-Kroes}, {Azzollini}, {Beard}, {Brandl}, {Gallie}, {Geers}, {Glauser}, {Hastings}, {Henning}, {Jager}, {Justtanont}, {Kruizinga}, {Lahuis}, {Lee}, {Martinez-Delgado}, {Mart{\'\i}nez-Galarza}, {Meijers}, {Morrison}, {M{\"u}ller}, {Nakos}, {O'Sullivan}, {Oudenhuysen}, {Parr-Burman}, {Pauwels}, {Rohloff}, {Schmalzl}, {Sykes}, {Thelen}, {van Dishoeck}, {Vandenbussche}, {Venema}, {Visser}, {Waters}, \& {Wright}}]{wells15}
{Wells}, M., {Pel}, J.~W., {Glasse}, A., {et~al.} 2015, \pasp, 127, 646, \dodoi{10.1086/682281}

\bibitem[{{Whelan} {et~al.}(2021){Whelan}, {Pascucci}, {Gorti}, {Edwards}, {Alexander}, {Sterzik}, \& {Melo}}]{whelan21}
{Whelan}, E.~T., {Pascucci}, I., {Gorti}, U., {et~al.} 2021, \apj, 913, 43, \dodoi{10.3847/1538-4357/abf55e}

\bibitem[{{White} \& {Hillenbrand}(2004)}]{white04}
{White}, R.~J., \& {Hillenbrand}, L.~A. 2004, \apj, 616, 998, \dodoi{10.1086/425115}

\bibitem[{{Woitke} {et~al.}(2018){Woitke}, {Min}, {Thi}, {Roberts}, {Carmona}, {Kamp}, {M{\'e}nard}, \& {Pinte}}]{woitke18}
{Woitke}, P., {Min}, M., {Thi}, W.~F., {et~al.} 2018, \aap, 618, A57, \dodoi{10.1051/0004-6361/201731460}

\bibitem[{{Worthen} {et~al.}(2024){Worthen}, {Chen}, {Law}, {Lu}, {Hoch}, {Chai}, {Sloan}, {Sargent}, {Kammerer}, {Hines}, {Rebollido}, {Balmer}, {Perrin}, {Watson}, {Pueyo}, {Girard}, {Lisse}, \& {Stark}}]{worthen2024}
{Worthen}, K., {Chen}, C.~H., {Law}, D.~R., {et~al.} 2024, \apj, 964, 168, \dodoi{10.3847/1538-4357/ad2354}

\bibitem[{{Wu} {et~al.}(2017){Wu}, {Sheehan}, {Males}, {Close}, {Morzinski}, {Teske}, {Haug-Baltzell}, {Merchant}, \& {Lyons}}]{wu2017}
{Wu}, Y.-L., {Sheehan}, P.~D., {Males}, J.~R., {et~al.} 2017, \apj, 836, 223, \dodoi{10.3847/1538-4357/aa5b96}

\bibitem[{{Xie} {et~al.}(2023){Xie}, {Pascucci}, {Long}, {Pontoppidan}, {Banzatti}, {Kalyaan}, {Salyk}, {Liu}, {Najita}, {Pinilla}, {Arulanantham}, {Herczeg}, {Carr}, {Bergin}, {Ballering}, {Krijt}, {Blake}, {Zhang}, {{\"O}berg}, {Green}, \& {Jdiscs Collaboration}}]{xie23}
{Xie}, C., {Pascucci}, I., {Long}, F., {et~al.} 2023, \apjl, 959, L25, \dodoi{10.3847/2041-8213/ad0ed9}

\bibitem[{{Zhang} {et~al.}(2013){Zhang}, {Pontoppidan}, {Salyk}, \& {Blake}}]{zhang13}
{Zhang}, K., {Pontoppidan}, K.~M., {Salyk}, C., \& {Blake}, G.~A. 2013, \apj, 766, 82, \dodoi{10.1088/0004-637X/766/2/82}

\bibitem[{{Zhu} {et~al.}(2009){Zhu}, {Hartmann}, \& {Gammie}}]{zhu09}
{Zhu}, Z., {Hartmann}, L., \& {Gammie}, C. 2009, \apj, 694, 1045, \dodoi{10.1088/0004-637X/694/2/1045}

\end{thebibliography}
\bibliographystyle{aasjournal}

\end{document}